\documentclass[a4paper,11pt]{article}

\usepackage{jheppub} 
\usepackage{lineno}
\usepackage{braket}
\usepackage{amsmath}
\usepackage{comment}
\usepackage[T1]{fontenc}
\newcommand{\be}{\begin{equation}}
\newcommand{\ee}{\end{equation}}

\title{\boldmath Electroweak Phase Transition and Bubble Wall Velocity in Local Thermal Equilibrium}

\author{Carlo Branchina$^{1,2}$, Angela Conaci$^{1,2}$, Stefania De Curtis$^{3}$, Luigi Delle Rose$^{1,2}$}
\affiliation{$^{1}$Dipartimento di Fisica, Università della Calabria, I-87036 Arcavacata di Rende, Cosenza, Italy}
\affiliation{$^{2}$INFN, Gruppo Collegato di Cosenza, Arcavacata di Rende, I-87036, Cosenza, Italy}
\affiliation{$^{3}$INFN, Sezione di Firenze, Via Sansone, 1, I-50019 Sesto Fiorentino (FI), Italy}

\emailAdd{carlo.branchina@unical.it}
\emailAdd{angela.conaci@unical.it}
\emailAdd{stefania.decurtis@fi.infn.it}
\emailAdd{luigi.dellerose@unical.it}

\abstract{
The dynamics of the electroweak phase transition in the early universe has profound implications for cosmology and particle physics.  We systematically study the steady-state dynamics of bubble walls in scenarios where the transition is first order within three representative beyond the Standard Model frameworks, characterised by the presence of an additional scalar in different electroweak representations. Focusing on the local thermal equilibrium regime, we numerically solve the coupled scalar and hydrodynamic equations to extract key properties of the phase transition front: the wall velocity, width, plasma and field profiles. We find a near-universal behaviour across models when expressed in terms of thermodynamic quantities, that can be captured by simple fitting functions, useful for phenomenological applications.  These results also provide an upper bound on the bubble velocity and represent the first necessary step for the full inclusion of out-of-equilibrium effects. As an application, we determine the gravitational wave signal and the amount of baryon asymmetry generated by the transition in local thermal equilibrium.} 

\makeatletter
\def\@fpheader{\relax}
\makeatother

\begin{document}
\maketitle
\flushbottom
\newpage
\section{Introduction}

\noindent
It has been known for a long time that interactions with a plasma at high temperature tend to stabilise scalar fields around the origin of field space \cite{{Kirzhnits:1972ut,Dolan:1973qd,Weinberg:1974hy}}, raising the question of how the electroweak (EW) vacuum was reached in the evolution of the universe from the initial hot state to the present day. It was shown in the late nineties that in the Standard Model (SM) the transition toward the EW vacuum is a smooth crossover \cite{Kajantie:1995kf,Kajantie:1996mn,Csikor:1998eu}, and does not generate any particularly interesting observational signature. This conclusion can be overturned in beyond the SM (BSM) scenarios with enlarged scalar sectors, where a first-order phase transition (PhT) can be easily accommodated. 

If a first-order EW phase transition (EWPhT) truly occurred in the early universe, it is expected to have produced a variety of cosmological signatures, such as a matter-antimatter asymmetry, primordial black holes, topological defects, dark matter remnants, and, most notably, a stochastic background of gravitational waves. The recent observation of the latter in astrophysical systems \cite{LIGOScientific:2016aoc,LIGOScientific:2016sjg} has fuelled renewed interest in the subject. The possibility to confront theories of the EWPhT with cosmological observables would make BSM scenarios  falsifiable, providing an extraordinarily rich arena to test new physics complementary to collider searches. Given a specific BSM model, an accurate description of the PhT dynamics is fundamental to give a quantitative assessment of its impact on cosmological observables. In particular, the expansion speed of the PhT front and its width are two fundamental quantities that characterise, for instance, the spectrum of emitted gravitational waves and the amount of matter asymmetry generated in scenarios of EW baryogenesis (see \cite{Hindmarsh:2020hop,Athron:2023xlk} for recent reviews of cosmological phase transitions).

A first-order PhT proceeds as follows: initially, scalar fields reside in a homogeneous configuration corresponding to the trivial vacuum, which is favoured at high temperatures. As the universe cools below the critical temperature $T_c$, where the minima of the potential become degenerate, new energetically deeper minima emerge. 
 When the probability to decay per Hubble volume exceeds the Hubble expansion rate, the transition proceeds through the nucleation and growth of bubbles of true vacuum, that eventually collide, merge, and fill up the whole space, leaving as an imprint the various cosmological signatures mentioned above. 

At a more fundamental level, bubbles correspond to non-trivial scalar field configurations that solve the scalar equations of motion with spherical symmetry. These solutions smoothly interpolate between the true vacuum (the deeper minimum of the potential) and the false vacuum (the higher minimum). When the potential difference between the two vacua is small, this transition occurs rapidly over a narrow spatial region. In this regime, the solution can be effectively interpreted as a spherical bubble of true vacuum, separated from the surrounding metastable phase by a domain wall (DW).

The potential difference between the phases generates an outward pressure on the wall that accelerates it. In turn, the presence of a moving front induces a back-reaction of the false vacuum background giving rise to a friction that hinders the acceleration and can eventually lead to the onset of a stable stationary solution, where the wall reaches a terminal velocity $v_w$. Given the importance of this quantity in the determination of gravitational wave signals, a huge effort has been put forward in recent years to develop methods for its calculation \cite{Moore:1995ua,Moore:1995si,John:2000zq,Moore:2000wx,Bodeker:2009qy,Megevand:2009gh,Espinosa:2010hh,Megevand:2013hwa,Huber:2013kj,Konstandin:2014zta,Megevand:2014dua,Kozaczuk:2015owa,Bodeker:2017cim,Dorsch:2018pat,DeCurtis:2019rxl,BarrosoMancha:2020fay,Hoche:2020ysm,Laurent:2020gpg,Friedlander:2020tnq,Azatov:2020ufh,Balaji:2020yrx,Wang:2020zlf,Lewicki:2021pgr,Dorsch:2021ubz,Ai:2021kak,Gouttenoire:2021kjv,Dorsch:2021nje,DeCurtis:2022hlx,Laurent:2022jrs,Lewicki:2022nba,Janik:2022wsx,Jiang:2022btc,DeCurtis:2023hil,Ai:2023see,Krajewski:2023clt,Dorsch:2023tss,Sanchez-Garitaonandia:2023zqz,Ai:2024shx,DeCurtis:2024hvh,Wang:2024wcs,Barni:2024lkj,Yuwen:2024hme,Ekstedt:2024fyq,Ai:2024btx,Krajewski:2024xuz,Krajewski:2024gma,Krajewski:2024zxg,Ramsey-Musolf:2025jyk,Ai:2025bjw,Branchina:2024rva,DeCurtis:2022llw,DeCurtis:2022djw,Cline:2000nw,Leitao:2010yw,Megevand:2013yua,Megevand:2014yua,Leitao:2014pda,Cline:2020jre,Cai:2020djd,Cline:2021iff,Bigazzi:2021ucw,Cline:2021dkf,Baldes:2023cih,Azatov:2023xem,Azatov:2024auq,Dorsch:2024jjl,Carena:2025flp,Ellis:2022lft}. This requires to simultaneously solve the equations of motion for the scalar fields, the continuity equations for the plasma, and the Boltzmann equation for the distribution functions of the particle species, making it a highly non-trivial task.  

The calculation of the bubble wall velocity, and more generally the determination of a stationary solution, has a long history, with the first full approach dating back to the seminal works of Moore and Prokopec \cite{Moore:1995ua,Moore:1995si}, where the computation was performed in the SM. In these works, the distribution functions are determined using a fluid ansatz and solving the (integro-differential) Boltzmann equation taking suitable weighted moments, that allow to express it as a system of ordinary differential equations.

In \cite{DeCurtis:2022hlx,DeCurtis:2023hil,DeCurtis:2024hvh} a significant progress was made in the development of a fully quantitative numerical method that allows to directly tackle the Boltzmann equation without resorting to any ansatz or arbitrary choice of decomposition. Technically, the collision integral represents the most challenging part of the computation. 
An iterative procedure was then devised, where, starting from the solution found in local thermal equilibrium (LTE), the out-of-equilibrium distribution functions are determined at each step inserting the solution found at the previous step in the source. At a more technical level, the great simplification allowed us to obtain for the first time a full solution to the Boltzmann equation, whereas previous works only dealt with a linearised (in the deviation from equilibrium of the distribution function) version of it. 
Important insights on the structure of the equation was also gained. 

In this work, we begin a systematic analysis of the EWPhT dynamics in BSM models. In \cite{DeCurtis:2022hlx,DeCurtis:2023hil,DeCurtis:2024hvh}, the numerical algorithm was tested on two benchmark points in the $\mathbb Z_2$-symmetric real scalar extension of the Standard Model. As noted there, the efficiency reached with the various simplifications of the collision integral opened the door to the possibility of performing scans of the parameter space of particle physics models. As a first step toward this goal, we present here the results for the bubble wall velocity and all the parameters entering in the steady-state dynamics obtained in local thermal equilibrium of three BSM models, namely the $\mathbb Z_2$-symmetric real singlet extension of the SM (SSM), the $\mathbb Z_2$-symmetric real triplet extension of the SM (RTSM), involving a $SU(2)_L$ triplet with vanishing hypercharge, and the inert doublet model (IDM), featuring a second Higgs doublet that is odd under a $\mathbb Z_2$ symmetry.  
All three models share a common feature: they allow for a two-step EW symmetry-breaking pattern in which the universe is first transitioning into a phase where a BSM scalar acquires a vacuum expectation value (VEV), followed by a first-order transition to the EW vacuum. Such scenarios generically give rise to tree-level potential barriers between vacua, circumventing the limitations of perturbation theory at finite temperature and allowing for a strong first-order EWPhT. \\
Our primary goal is to calculate the steady-state wall velocity $v_w$ and the (field and fluid) profiles in LTE across the relevant parameter space of these models. This represents the first step in a broader program aimed at eventually incorporating out-of-equilibrium dynamics, using iterative techniques, and will be used in subsequent works as the starting point for the full evaluation of out-of-equilibrium effects in the parameter space of BSM models. 

In the following analysis, we do not impose phenomenological constraints, such as those arising from collider searches, EW precision observables, or dark matter relic density and direct detection bounds, since our focus is on the accurate description of the wall dynamics. This choice allows us to explore the behaviour of the solutions and the structure of the equations governing them as functions of the parameters of the models, independently of external constraints. A detailed phenomenological analysis, incorporating these constraints and assessing the interplay between cosmological and experimental signatures, is beyond the scope of this work and can be addressed in future studies.
\\
It is also important to emphasize that the wall velocities computed within the LTE framework represent an upper bound on the actual wall velocities expected in a complete treatment. In realistic settings, deviations from equilibrium near the wall, captured by solving the Boltzmann equations, introduce additional sources of friction. These out-of-equilibrium effects tend to further slow down the motion of the wall, reducing the velocity relative to the LTE estimate. Therefore, the LTE results obtained here provide a crucial reference point, both for identifying the qualitative behaviour of solutions and for initialising more complete iterative procedures that include effects beyond (local) equilibrium.

In this paper, we solve the coupled scalar equations of motion and hydrodynamic conservation laws numerically. The resulting wall velocities, widths, and field displacements are obtained by imposing consistency conditions on the pressures, as well as on their gradients across the wall. Special attention is given to the classification of the hydrodynamic regimes (deflagration, detonation, hybrid) and their realisation within the LTE framework.
We find that stationary solutions exist throughout the parameter space corresponding to two-step transitions, except in regions where the wall experiences runaway behaviour. Actually, in the $v_w\to1$ limit a friction is known to arise that stops the bubble wall acceleration \cite{Bodeker:2009qy,Bodeker:2017cim,Ramsey-Musolf:2025jyk,Ai:2025bjw}, and these solutions become ultra-relativistic detonations. This effect can not be captured in LTE. Moreover, our results exhibit an approximate model-independence in the behaviour of $v_w$ when expressed in terms of thermodynamic quantities characterising the transition. 
Our analysis shows
the presence of common features across BSM scenarios. \\
To make this result of practical use, we have captured this universality in the form of simple fitting functions, which can be readily employed in independent phenomenological studies, such as gravitational wave signal modeling, baryogenesis estimates, or model parameter scans,
without going through the full computationally demanding numerical simulation, thus greatly facilitating phenomenological analyses. As a preliminary application, we determine the spectrum of gravitational waves and the baryon asymmetry generated by the EWPhT in the SSM augmented with a CP-violating wall-fermion interaction.  

The structure of the paper is as follows: In Section \ref{sec: hydro}, we review the hydrodynamic description of the plasma, derive the continuity equations and the equations
of motion for the scalar fields, and describe the numerical algorithm we used to solve them. Section \ref{sec: Models} details the BSM models under consideration, including renormalisation and thermal resummation schemes. Section \ref{sec: analytic} provides analytic insight into the structure of the constraint equations, aiding in the interpretation of numerical results. Section \ref{sec: numerical} presents our numerical results for the wall velocity and associated parameters in the three models. We apply our results to the gravitational wave spectrum and EW baryogenesis in Section \ref{sec: applications}. We conclude in Section \ref{sec: conclusions} with a discussion of the implications and future directions. In the Appendix we provide the analytic expressions of the thermal masses employed for the considered BSM scenarios.

\section{Hydrodynamics}
\label{sec: hydro}	
The determination of the wall dynamics relies, among others, on the hydrodynamic description of the plasma interacting with the DW. The expansion of the bubble drives the plasma out-of-equilibrium, the deviations from it being larger in regions close to the wall and for particles (more strongly) coupled to it. The plasma can thus be conveniently described as a mixture of two components: species directly coupled to the DW and out-of-equilibrium, as the top quark, and background particles, assumed to be in local equilibrium. The interaction between the plasma and the DW gives rise to position-dependent temperature and velocity plasma profiles, that must be accurately described to account for the friction exerted on the moving wall, and eventually determine the wall terminal velocity. 
	
\subsection{Hydrodynamic equations}
The dynamics of a relativistic plasma can be described using the stress-energy tensor, that can be written in terms of the distribution function $f_i(k,x) = f_i^{(0)}(k,x) + \delta f_i(k,x)$ of the particle species as
	\begin{equation}
		T_{\mu \nu}^{pl} = \sum_i\int \frac{d^3 k}{(2 \pi)^3 2 E_k}\,  k_\mu k_\nu f_i(k,x) = T^{lte}_{\mu \nu} + T^{out}_{\mu \nu},
	\end{equation}
where the sum extends to all the species in the plasma and $T^{lte}_{\mu \nu}$ describes the local thermal equilibrium contribution given by the position-dependent equilibrium distributions $f_i^{(0)}(k,x)$ while $T_{\mu \nu}^{out}$ is the out-of-equilibrium part determined by $\delta f_i(k,x)$, with the largest contribution coming from the species that are more strongly coupled to the DW, and in a region close to it. 

In the following, for the plasma in local equilibrium $T^{lte}_{\mu\nu}$ simplifies to 
	\begin{equation}
		T_{\mu \nu}^{lte} = w\, u_\mu u_\nu - g_{\mu \nu}\, p,
	\end{equation}
	with $w = e + p = T\,\partial_{_T} p $ the enthalpy, $u_\mu = \gamma(1, \vec{v_p})$ the plasma four-velocity, $p$ the pressure and $e$ the energy density.

The total stress-energy tensor of the system (the plasma and the DW) is thus given by the sum 
	\begin{equation}
T_{\mu\nu}=T^{pl}_{\mu \nu}+ T_{\mu \nu}^{\phi},
	\end{equation}
where $T_{\mu \nu}^{\phi}$ stands for the stress-energy tensor of the scalar fields driving the phase transition,
\begin{equation}
T^{\phi}_{\mu \nu}= \sum_{j} \left[\partial_\mu \phi_j \partial_\nu \phi_j -g_{\mu \nu}\frac{(\partial \phi_j)^2}{2}  \right] + g_{\mu\nu} V_{_{T=0}} , 
\end{equation}
with $V_{_{T=0}}\equiv V(\phi_1,\dots,\phi_n,T=0)$ the zero-temperature effective potential.
    
The hydrodynamic equations are obtained from the conservation of $T_{\mu\nu}$. 
Neglecting the expansion of the universe, and taking a planar approximation for the wall, the conservation equations can be simply written in the domain wall reference frame as
\begin{equation}
\partial_z T^{z0}= \partial_z T^{zz} = 0, 
\label{TensCons}
\end{equation}
where the $z$-axis is chosen to be aligned to the propagation direction of the wall. Integrating Eq.\,\eqref{TensCons} we obtain
	\begin{align}
		\label{hydro eq 1}
		T^{30} \equiv w\, \gamma^2 v_p + T^{30}_{out} = c_1, \\
		\label{hydro eq 2}
		T^{33} \equiv \sum_{i=1}^n\frac{(\partial_z \phi_i)^2}{2} - V(\phi_1,\dots,\phi_n, T) + w\, \gamma^2 v_p^2 + T^{33}_{out} = c_2,
	\end{align}
where $V(\phi_1,\dots,\phi_n,T)\equiv V_{_{T=0}}-p$ is the finite temperature effective potential.
The constants $c_1$ and $c_2$ are determined by the boundary conditions on the plasma velocity $v_{ \pm}$, the  plasma temperature $T_{\pm}$, and the scalar fields VEVs $\phi_i^{\pm}$, that can be imposed either inside or outside the bubble (to be consistent with the literature, we use the $+$ sign for quantities in front of the wall and the $-$ sign for quantities behind it).
They are more easily determined far away from the wall at $z \to \pm \infty$ where $\delta f \to 0$, and depend on the class of the hydrodynamic solution, as we will briefly review below.
	
Integrating the conservation equations \eqref{TensCons} across the DW, $v_p$ and $T$ in front and behind the wall are found to be related by the matching equations 
	\be
	\label{matching eq}
	v_+ v_- = \frac{p_+ - p_{-}}{e_+ - e_{-}}, \ \ \ \ \ \
	\frac{v_{+}}{v_-}= \frac{e_{-}+ p_{+}}{e_+ + p_{-}}, 
	\ee
where $p$ was redefined to include the contribution from the zero-temperature potential $p\to p - V_{_{T=0}}$, and $e$ is the total energy density associated to it, $e \to e + V_{_{T=0}}=T\,\partial_{_T} p - p$. The scalar VEVs $\phi_i^\pm$ are straightforwardly found by minimizing the potential $V(\phi,T)\equiv V(\phi_1,\dots,\phi_n,T)$ at $T_+$ and $T_-$. The dynamics of the system is then fully described by the equation  of state (EoS) for the plasma, and the equations of motion (EoM) for the scalar fields. 
	
Concerning the EoS, we take the generalized bag EoS (see \cite{Espinosa:2010hh}), where the quantities of interest are determined from the finite-temperature effective potential $V(\phi,T)=-p$ as 
	\begin{equation}
        \label{eq: EOS}
		p_\pm=\frac13 a_\pm T_{\pm}^4-\epsilon_\pm, \qquad e_\pm =a_\pm T_\pm^4+\epsilon_\pm 
	\end{equation}
	and 
	\begin{equation}
		a_\pm= \frac{3}{4T_\pm^3}\left(\frac{\partial p}{\partial T}\right)_{\pm}=\frac{3\omega_\pm}{4 T_\pm^4}, \qquad \epsilon_\pm = \frac14(e_\pm - 3p_\pm ),
	\end{equation}
where $\epsilon_\pm$ stands for the potential energy inside and outside the bubble. Inserting \eqref{eq: EOS}  in \eqref{matching eq}, $v_+$ and $v_-$ are expressed in terms of $T_+$ and $T_-$.

Far away from the wall, the scalar fields become constant and the solution must be matched to a self-similar dynamical one where the profiles of the plasma velocity and temperature are described by the conservation of $T^{\mu\nu}_{lte}$.
Projecting the equation $\partial_\mu T^{\mu\nu}_{lte}=0$ along the flow ($u_\nu \partial_\mu T^{\mu\nu}_{lte}=0$) and in the direction perpendicular to it ($\bar u_\nu \partial_\mu T^{\mu\nu}_{lte}=0$, with $\bar u_\mu u^\mu =0$ and $\bar u^2=1$), one gets, in the bubble frame,
	\begin{align}
		(\xi - v)\frac{\partial_\xi e}{w}&=2\frac{v}{\xi}+\left(1-\gamma^2v(\xi-v)\right)\partial_\xi v,\\
		(1-v\xi)\frac{\partial_\xi p}{w}&=\gamma^2(\xi-v)\partial_\xi v,
	\end{align}
	where a spherical symmetry was assumed, and $\xi$ is defined as $\xi\equiv r/t$, with $r$ the distance from the centre of the bubble and $t$ the time from nucleation. Using $c_s^2=dp/de$, with $c_s$ the speed of  sound, the equations above can be rearranged into 
	\begin{align}
		\label{eq: diff eq plasma 1}
		2\frac{v_p}{\xi}&=\gamma^2(1-v_p\,\xi)\left(\frac{\mu^2}{c_s^2}-1\right)\partial_\xi v_p,\\
		\label{eq: diff eq plasma 2}
		\partial_\xi T&=\gamma^2\mu\, T\,\partial_\xi v_p,
	\end{align}
	where $\mu$ is the Lorentz-transformed fluid velocity 
	\begin{equation}
		\mu(\xi, v_p)=\frac{\xi-v_p}{1-\xi v_p}.
	\end{equation}
	In these coordinates, the bubble wall is located  at $\xi=v_w$.

	\subsection{Hydrodynamic regimes}
	
The conservation equations give rise to different types of solutions, that can be classified by the wall velocity $v_w$. In terms of $v_+$ and $v_-$, Eq.\,\eqref{matching eq} determines two branches, one with $v_+>v_-$ (detonations), and another with $v_+<v_-$ (deflagrations). The lowest value $v_+$ can assume in the detonation regime 
is the so-called Jouguet velocity $v_{_J}$
and is obtained for a solution with $v_-=c_s$, where $c_s=\left(\partial_{_T}V/(T\partial_{_T}^2V)\right)_{-}$ is the speed of sound inside the bubble and the subscript indicates that the derivative must be evaluated inside the bubble. Below we briefly recall the most important features of the three types of solution (we refer to \cite{Espinosa:2010hh,Gyulassy:1983rq} for further details).

\textbf{Detonation $(v_w > v_{_J})$}. The wall moves faster than sound, leaving a rarefaction wave behind it. The plasma in front of the bubble is at rest, with perturbations occurring only behind it, so that $v_w=v_+$, and $T_n=T_+$. The plasma temperature and velocity have opposite behaviour\footnote{In local equilibrium, this is due to entropy conservation, that imposes $\gamma(z)T(z)={\rm \text{constant}}$\,\cite{Ai:2021kak}.}, so that $v_+>v_-$ implies $T_+ < T_-$.

\textbf{Deflagration $(v_w < c_s)$}. The wall velocity is lower than the speed of sound inside the bubble, generating a shock wave ahead of it. Perturbations are found in front of the wall, while the plasma is at rest behind it and in front of the shock wave. The wall velocity is then $v_w=v_-$, and  $T_n=T_+^{SW}$, where $T_+^{SW}$ is the temperature in front of the shock wave. Quantities at the shock front are related to one another as in a detonation solution, and one finds the following chain $T_+> T_-^{SW}>  T_+^{SW}>T_-$.
	
\textbf{Hybrid  $(c_s < v_w <v_{_J} )$}. These solutions are given by a combination of deflagration and detonation ones. They have both a shock front and a rarefaction wave, with $T_+^{SW} = T_n$ and $v_- = c_s$. Computationally, they are similar to deflagration solutions, and  we will generically refer to deflagrations for both. 
	
As the wall velocity approaches the Jouguet velocity from below, the wall and the shock front get closer and closer until the latter disappears at $v_w=v_{_J}$,  and the type of solution changes from an hybrid to a detonation. The theoretical set-up described above, and in particular equations \eqref{hydro eq 1} and \eqref{hydro eq 2}, ignores the presence of the shock front, and requires a sufficiently large distance between the latter and the bubble wall to give an accurate description of the region in between. As a consequence, the approach presented here is not valid when $v_w\to v_{_J}$ from below, and, more generally, the limit $v_w\to v_{_J}$ should be considered with care.

\subsection{Scalar equations of motion}
\label{sec: scalar EOMs}

As our main interest is in multi-step phase transitions driven by two scalar fields, in the following we consider a system of two scalars, $h$ and $s$. In the applications presented below, the first one will be the Higgs field, and the second one a CP-even, neutral, additional scalar. Assuming a steady-state expansion and a mean free path of particles in the plasma much smaller than the scale of variation of the scalars, the equations of motion for $h$ and $s$ read~\cite{Moore:1995ua,Moore:1995si} 
\begin{align}
\label{eq: scalar EOMs}
E_h \equiv - \partial^2_z h + \frac{\partial V(h,s,T)}{\partial h} + \frac{F_h^{out}(z)}{h'} = 0, \\
E_s \equiv - \partial^2_z s + \frac{\partial V (h,s,T)}{\partial s} +  \frac{F_s^{out}(z)}{s'} =0,
\end{align}
where ($\phi_j=h,s$)
\begin{equation}
F^{out}_j(z)=\sum_i \frac{n_i}{2}\frac{\partial m^2_i}{\partial \phi_j}\int\frac{d^3p}{(2\pi)^3 E_p} \,\delta f_i
\end{equation}
accounts for the out-of-equilibrium contributions and $n_i$ counts the number of degrees of freedom.
An approximate solution to these equations can be obtained taking a {\it tanh} ansatz for the fields\footnote{Having in mind the application to the EWPhT, we consider here a transition from the state $(h,s)=(0,s_+)$ to $(h,s)=(h_-,0)$.},
\begin{align}
h(z) = \frac{h_{-}}{2} \left( 1 + \tanh{\left( \frac{z}{L_h}\right)}\right), \\
s(z) = \frac{s_{+}}{2} \left( 1 - \tanh{\left( \frac{z}{L_s} - \delta_s \right)}\right),
\label{eq: ansatz scalar profiles}
\end{align}
where $L_h$ and $L_s$ are the thicknesses of the $h$ and $s$ walls (the $z$-axis is chosen here so that positive values of $z$ correspond to the interior of the ($h$-)bubble, while negative values correspond to the exterior).

Exploiting translation invariance, the first solution is centred in $z=0$, and $\delta_s$ is the displacement of the $s$ wall with respect to the $h$ one. The coefficients $h_{-}$ and $s_{+}$ are the VEVs of the $h$ and $s$ fields inside and outside the bubble, respectively
	\begin{equation}
		\frac{\partial V(h_{-}, 0, T_-)}{\partial h}=0, \ \ \ \ \frac{\partial V(s_{+}, 0, T_+)}{\partial s}=0.
	\end{equation}
	The EoM are then traded for four constraints, obtained taking suitable moments of them, 
	\begin{align}
		\label{eq: constraints}
		P_h &= \int dz E_h h' = 0, \qquad  G_h = \int dz E_h \left(  2\frac{h}{h_{-}} -1\right) h' = 0, \nonumber \\
		P_s &= \int dz E_s s' = 0, \qquad  G_s = \int dz E_s \left(  2\frac{s}{s_{+}} -1\right) s' = 0,
	\end{align}
where $h' = \partial_z h$ and $s' = \partial_z s$.
Here, $P_h$ and $P_s$ determine the pressure acting on the $h$ and $s$ walls, respectively, while $G_h$ and $G_s$ are the corresponding pressure gradients (see \cite{Moore:1995ua,Moore:1995si}). Vanishing of $P_h$ and $P_s$ is necessary for a steady-state regime to be reached, where the outward pressure, that tends to accelerate the wall, is balanced by the friction acting on it. The equations $G_h=0$ and $G_s=0$ correspond to the requirement that the pressure gradient vanishes, enforcing that the solution has fixed widths.  
	
When out-of-equilibrium effects are considered, the Boltzmann equation must be included to determine $\delta f$. An iterative procedure was recently proposed to get a complete (numerical) solution, and was tested for given benchmark points in a specific model \cite{DeCurtis:2022hlx, DeCurtis:2023hil, DeCurtis:2024hvh}. In this approach, the equilibrium solution is the first, crucial, step, needed to proceed to the iterative evaluation of the out-of-equilibrium effects and of their impact on the physical quantities of interest. 
The present work is only concerned with the first part of this program, {\it i.e.}\,the determination of the equilibrium solution.

\subsection{Numerical algorithm} \label{sec: algorithm}	
In this section, we describe the numerical method used to solve the hydrodynamic equations and the EoMs. This approach allows us to determine the bubble wall velocity $v_w$, the parameters $\delta_s$, $L_h$, $L_s$, and the profiles for the plasma, $T(z)$ and $v_p(z)$. The profiles of the fields $h(z)$ and $s(z)$ are then determined too.   
	
For a given point in the parameter space of a model, the Jouguet velocity $v_{_J}$ and the speed of sound inside the bubble $c_s$ are calculated once and for all, before the numerical loop starts, using a numerical technique for root-finding and function minimization. The algorithm then proceeds with the following steps.

\vskip 5pt
	
\textbf{1. Initial guess} 
	
\vskip2pt
\noindent
The algorithm starts with an initial guess for $v_w, \,\delta_s,\, L_h$ and $L_s$. When scanning the parameter space of a model, the region where the scan is performed is divided into a grid of points. For each point, the initial guess corresponds to the result that was obtained for one of its nearest neighbours.
	
\vskip 5pt
	
\textbf{2. Boundary conditions} 
	
\vskip2pt
\noindent
The first step is to calculate the four constraints $P_{tot}$, $\Delta P$, $G_h$ and $G_s$ using the initial guess, where $P_{tot}\equiv P_h+P_s$ and $\Delta P\equiv P_s-P_h$ (we will motivate the choice of $P_{tot}$ and $\Delta P$, rather than $P_h$ and $P_s$, later). To compute these quantities, the plasma temperature profile $T(z)$ must be known in the appropriate combustion regime. This is determined by comparing the bubble wall velocity $v_w$ from the initial guess to the Jouguet velocity $v_{_J}$, and then classifying the tentative solution.
\vskip 5pt
	
\textbf{3a. Detonations ($v_w > v_{_J}$)} 
	
\vskip2pt
\noindent
For detonations, $T_{+}= T_n$ ($s_+ = s_n$) and $v_{+}= v_w$. The two unknown $T_{-}$ and $v_{-}$ are then easily found using the matching equations \eqref{matching eq}.
Note that there is an intrinsic dependence on $T_-$ in $h_-$, so $T_-$ is found searching for the root of those equations once $h_{-}$ is determined as the minimum of the potential in the $h$ direction at the temperature $T=T_-$. 
    The velocity $v_-$ is then computed, and $h_-$ is used to construct the profile $h(z)$ (together with $s(z)$, that was already available), from which the plasma profiles $T(z)$ and $v_p(z)$ are in turn determined, using \eqref{hydro eq 1}, \eqref{hydro eq 2}.   
	
\vskip 5pt
\textbf{3b. Deflagrations / hybrids ($v_w < v_{_J}$)}
	
\vskip2pt
\noindent
The determination of the plasma profiles for deflagrations is more involved, as the boundary conditions $T^{SW}_{+}=T_n$ and $v_{-}=v_w$ (or $v_-=c_s$ for hybrids) are given at two different points. Before the matching equations can be used, the additional task is to find the value of $T_+$. This is done in two sequential steps. The first step calculates $T^{SW}_{+}$ for a given tentative value of $T_+$ with the following procedure. The tentative value of $T_+$, along with $v_w$, is then used to determine $v_+$ and $T_-$ from the matching conditions. Equations \eqref{eq: diff eq plasma 1} and \eqref{eq: diff eq plasma 2} are then integrated from the bubble wall to the shock front\footnote{A technical remark is in order here. To deal with single-valued functions, the equations are inverted and solved for $\xi(v_p)$ and $T(v_p)$ in the bubble frame. The integration is performed from $v_p=0$ to $v_p=\mu(v_w, v_+)$.} to find $T_{-}^{SW}$ and $v_{-}^{SW}$, from which $T_{+}^{SW}$ is obtained. The second step refines the value of $T_+$ until the condition $T_{+}^{SW} = T_n$ is satisfied. Once $T_+$ is found, the determination of the profiles proceeds as for detonations.

	\vskip 5pt
	
	\textbf{4. Root-finding}
	
	\vskip2pt
	\noindent
	The constraints are finally evaluated on the initial guess using an adaptive quadrature integration technique. Starting from the initial guess, the values of $v_w$, $\delta_s$, $L_h$ and $L_s$ are progressively refined until the solution is found. 
As a check, the constraints are calculated on the final result, which is accepted only if each of them is below a predefined threshold.

\section{Extensions of the Higgs sector}

In this section we present the three  BSM models used to test our numerical algorithm and calculate the bubble wall velocity, each characterised by an extra scalar field with different charge under the EW $SU(2)$ gauge group. We consider a real singlet extension of the Standard Model (SSM), a real triplet extension of the Standard Model (RTSM), and the inert two-Higgs doublet model (IDM). 
An enlarged scalar sector offers the possibility to realise a first-order EWPhT with a two-step pattern, where the EW minimum is reached from a configuration in which  the additional neutral, CP-even scalar, possesses a VEV. The potential barrier between the two minima receives sizeable contributions from the tree-level potential, so that the EWPhT is typically strong with observable GW signatures and more efficient mechanisms for EW baryogenesis.
On the contrary, when the EWPhT proceeds directly from the completely symmetric vacuum, this cannot happen in general. Indeed, as the Higgs field quadratic term must be positive at the nucleation temperature for the origin to be a minimum itself, the barrier can only be generated by a combination of loop and thermal contributions, exposing it more to the (well-known) issues of the perturbative expansion at finite temperature.  

\label{sec: Models}
\subsection{Singlet extension of the Standard Model (SSM)}

We consider a $\mathbb{Z}_2$-symmetric extension of the SM with a scalar field $s$. 
The tree-level Lagrangian of the scalar sector is 
\be
\mathcal{L} =  (D_{\mu} H)^{\dagger}(D^{\mu} H) + \frac{1}{2} \partial_{\mu} s\, \partial^{\mu} s - V_0(H, s),
\ee
where $H$ is the Higgs doublet 
\be
\label{eq: Higgs doublet param}
    H= \frac{1}{\sqrt 2}\begin{pmatrix}
        \chi_1+i\chi_2 \\ h + i \chi_3
    \end{pmatrix},
\ee
$h$ the Higgs field and $V_0$ the potential
\begin{equation}
   V_0(H,s) = \mu_h^2 |H|^2+\lambda_h |H|^4+\frac{\mu_s^2}{2} s^2 +\frac{\lambda_s}{4}s^4 +\lambda_{hs} |H|^2s^2. 
\end{equation}
In terms of $h$ and $s$, it reads
\begin{equation}
\label{eq: V0}
 V_0(h,s) = \frac{\mu_h^2}{2}  h^2 + \frac{1}{4} \lambda_h h^4 + \frac{\mu^2_s}{2} s^2 + \frac{\lambda_s}{4} s^4 + \frac{\lambda_{hs}}{2} h^2 s^2.
\end{equation}
The model consists of three free parameters, $\mu_s,\, \lambda_s$ and $\lambda_{hs}$, with the former that can be traded for the mass of the singlet in the EW vacuum $(h,s) = (v, 0)$, $m_s^2=\mu_s^2 +\lambda_{hs}  v^2$, while $\mu_h^2$ and $\lambda_h$ are fixed
by the usual relations $\lambda_h = m_h^2/2 v^2$ and $\mu_h^2 = - \lambda_h v^2$. 
To match the parameters in the tree-level potential directly to the measured values of the Higgs VEV and mass, we will use an on-shell renormalisation scheme.
 The singlet mass will also be fixed to its tree-level value by the renormalisation conditions given below. 

When $\mu_s^2<0$, a non-trivial minimum is found in the singlet direction. The viability of this scenario requires that, at $T=0$, the EW minimum is deeper than the minimum in the $s$-direction. A two-step phase transition (with a first-order transition towards the EW vacuum in the second step) can then occur if the non-trivial minimum in the $s$ direction appears at temperatures higher than the one in the $h$ direction, and conditions for nucleation toward the EW vacuum are met. The region of parameter space where this is verified is the one of interest for our analysis. A rough analytic estimate of this region can be obtained working with the Parwani-resummed tree-level potential \cite{DeCurtis:2019rxl}, {\it i.e.}\,the tree-level potential supplemented by thermal masses.     

We choose to adopt the Parwani resummation method \cite{Parwani:1991gq}, where daisy diagrams are resummed ab initio. In terms of the thermal eigenmasses $m_i^2(h,s,T)$, the one-loop effective potential in the $\overline{\hbox{MS}}$ scheme with renormalisation scale $\mu=v$ reads 
\begin{equation} \label{CW}
V_{_{CW}}(h,s,T)=\frac{1}{64 \pi^2}\sum_{i} n_i\,m_i^4(h,s,T)\left(\log\frac{m^2_i(h,s,T)}{v^2} - c_i\right),
\end{equation}
where $n_i$ is the number of degrees of freedom of the corresponding particle, $i\in\{h,s,\chi,t,$ $W_{L/T},Z_{L/T},\gamma\}$, with $t, W_{L/T}, Z_{L/T}$ and $\gamma$ the top quark, (longitudinal and transverse components of the) $W$ bosons, (longitudinal and transverse components of the) $Z$ boson and the photon, respectively. For the scalar fields, longitudinal $W/Z$ and the top quark, $c_i = 3/2$, while for transverse gauge bosons, $c_i = 1/2$. 
The expressions for the field-dependent thermal masses $m^2_i(h,s,T)$ can be found in Appendix \ref{section: thermal masses SSM}. 

As mentioned above, we compensate for radiative shifts of the Higgs VEV, Higgs mass and singlet mass in the EW vacuum by introducing finite counterterms
\begin{equation}
\label{eq: Vct}
    V_{_{CT}}(h, s)= \frac{1}{2} \delta m^2_h h^2 + \frac{1}{2} \delta m^2_{s} {s}^2 + \frac{1}{4} \delta \lambda_h h^4, 
    \end{equation}
and imposing the conditions 
\begin{equation}
\label{eq: reno condition 1}
    \frac{\partial (V_{_{CW}} + V_{_{CT}})}{\partial h} \bigg|^{T=0}_{h = v, s= 0} = 0, \\
\end{equation}
\begin{equation}
\label{eq: reno condition 2}
    \frac{\partial^2 (V_{_{CW}} + V_{_{CT}})}{\partial h^2} \bigg|^{T=0}_{h = v, s= 0} = 0, 
    \ \ \ \ 
\frac{\partial^2 (V_{_{CW}} + V_{_{CT}})}{\partial {s}^2} \bigg|^{T=0}_{h = v, s= 0} = 0.
\end{equation}
The second derivative with respect to $h$ of the Goldstone contribution is singular in the EW vacuum and needs to be regularised. We do so by shifting the field-dependent mass $m^2_\chi$, $m^2_\chi\to m^2_\chi+\mu_{_{IR}}^2$, in the singular terms. Specifically, we make the following substitution ($V_{_{CW}}^{\chi}$ indicates the Goldstone contribution to $V_{_{CW}}$)
\begin{equation}
	\label{sub1}
\frac{\partial^2 V_{_{CW}}^{\chi}}{\partial h^2}\Bigg|^{T=0}_{h=v,s=0} \to \frac{n_\chi}{8\pi^2}\lambda_h v^2\log\left(\frac{\mu_{_{IR}}^2}{v^2}\right),
\end{equation} 
and we choose $\mu_{_{IR}}=m_h$. 
The counterterms are then unambiguously determined as 
\begin{align}
\delta m_h^2 &= \left(\frac{1}{2} \frac{\partial^2 V_{_{CW}}}{\partial h^2} - \frac{3}{2 v} \frac{\partial V_{_{CW}}}{\partial h}\right)\Bigg|^{T=0}_{h=v,s=0} \\
\delta m_s^2 &= - \frac{\partial^2 V_{_{CW}}}{\partial s^2}\Bigg|^{T=0}_{h=v,s=0} \\ 
\delta\lambda_h &= \frac{1}{2 v^2}\left( \frac{1}{ v} \frac{\partial V_{_{CW}}}{\partial h} - \frac{\partial^2 V_{_{CW}}}{\partial h^2} \right)\Bigg|^{T=0}_{h=v,s=0}.
\end{align}

The thermal contribution to the one-loop potential is given by
\begin{equation}
\label{eq: VT}
    V_{_T}(h, s, T)= \frac{T^4}{2 \pi}\left( \sum_i n_i J_B\left(\frac{m^2_i(h, s,T)}{T^2}\right) - n_t\, J_F\left(\frac{m^2_t(h)}{T^2}\right)\right),
\end{equation}
where the thermal functions $J_{B/F}$ for the bosons and fermions are (the $+$ sign is for bosons, the $-$ for fermions)
\begin{equation}
    J_{B/F}(y) = \int^{\infty}_{0} dx \ x^2 \log{\left(1 \mp e^{- \sqrt{x^2 + y}}\right)}.
\end{equation}

\subsection{Real scalar Triplet extension of the Standard Model (RTSM)}
\label{sec:SigmaSM}
This model is a $\mathbb{Z}_2$-extension of the SM that includes a real scalar isotriplet field $\Sigma$ with zero hypercharge. The scalar sector Lagrangian reads
\begin{equation}
  \mathcal{L}= (D_{\mu} H)^{\dagger}(D^{\mu} H) + \frac{1}{2} (D_{\mu} \Sigma) (D^{\mu} \Sigma) - V_0(H, \Sigma),
\end{equation}
where the covariant derivative of $\Sigma$ is 
\begin{align}
(D_\mu \Sigma)^a = (\partial_\mu \delta^{ac} - g \epsilon^{abc} W^b_\mu) \sigma_c, 
\end{align}
and $\sigma_a$ are the three real components of the triplet $\Sigma$,
with the Higgs doublet parametrised as in \eqref{eq: Higgs doublet param}. 
The physical degrees of freedom are the two charged fields $\sigma^{\pm}= (\sigma_1 \mp i \sigma_2)/\sqrt{2}$, and the neutral $\sigma_3$, with the $\mathbb{Z}_2$ symmetry ensuring stability of the latter.

The tree-level scalar potential $V_0(H,\Sigma)$ is
\begin{equation}
   V_0(H,\Sigma) = \mu_h^2 |H|^2+\lambda_h |H|^4+\frac{\mu_\sigma^2}{2} \sigma^a\sigma^a +\frac{\lambda_\sigma}{4}\left(\sigma^a\sigma^a\right)^2 +\lambda_{h\sigma} |H|^2\sigma^a\sigma^a.
\end{equation}
In terms of the neutral CP-even scalars $h$ and $\sigma_3$, $V_0$ reduces to \eqref{eq: V0} (with the obvious substitutions $s\to\sigma_3$ and $\mu_s,\,\lambda_s,\, \lambda_{hs} \to \mu_\sigma,\,\lambda_\sigma,\, \lambda_{h\sigma}$).
The $\mathbb{Z}_2$ symmetry of the model forbids gauge-invariant cubic operators such as $H^\dagger \sigma^a H$ or $\sigma^a \sigma^b \sigma^c$. 

As for the singlet, we adopt the Parwani resummation scheme, and use the same renormalisation conditions defined in \eqref{eq: Vct}-\eqref{eq: reno condition 2}. The field-dependent mass terms that appear in the Coleman-Weinberg and thermal one-loop contributions can be found in Appendix \ref{section: thermal masses RTSM}.

\subsection{Inert two-Higgs Doublet Model (IDM)}

In the inert two-Higgs doublet model, a discrete $\mathbb Z_2$ symmetry is imposed, under which the SM fields are even and the second doublet $H_2$ is odd. This prevents $H_2$ from decaying into lighter SM particles. The lagrangian of the two doublets read
\begin{equation}
\mathcal{L}= (D_{\mu} H^{\dagger}_1)(D^{\mu} H_1) + (D_{\mu} H^{\dagger}_2)(D^{\mu} H_2) - V_0(H_1, H_2),
\end{equation}
where $H_1$ is the SM Higgs doublet and the tree-level potential $V_0$ is 
\begin{align}
V_0(H_1, H_2) = &\mu_1^2 |H_1|^2 +\mu_2^2 |H_2|^2 + \lambda_1 |H_1|^4 + \lambda_2 |H_2|^4 + \lambda_3 |H_1|^2|H_2|^2 + \\
&+ \lambda_4 | H_1^{\dagger} H_2|^2 + \frac{\lambda_5}{2}\left[ (H_1^\dagger H_2)^2 + \text{h.c}\right].  \nonumber 
\end{align}
Similar to $H_1$ in \eqref{eq: Higgs doublet param}, the second doublet can be expressed in terms of its components
\begin{equation}
H_2 = \frac{1}{\sqrt 2}\begin{pmatrix}
       \chi_4+i\chi_5 \\ h'+i\chi_6
      \end{pmatrix}.  
\end{equation}
Stability of the tree-level potential imposes the constraints
\begin{equation}
   \lambda_1>0, \qquad \lambda_2>0,\qquad \lambda_3 + \lambda_4 - \lambda_5+2\sqrt{\lambda_1\lambda_2} > 0
\end{equation}
on the quartic couplings. 

As for the other cases, we only consider the vacua landscape for the neutral and CP-even fields $h$ and $h'$, in terms of which the tree-level potential reduces to 
\begin{equation}
V_0 (h, h') = \frac{\mu_1^2}{2} h^2 + \frac{\lambda_1}{4} h^4 + \frac{\mu_2^2}{2} h'^{\,2} + \frac{\lambda_2}{4} h'^{\,4} + \frac{\lambda_{345}}{4} h^2 h'^{\,2},
\end{equation}
with $\lambda_{345}\equiv\lambda_3+\lambda_4+\lambda_5$.
In the EW vacuum $(h,h')=(v,0)$, the physical masses of the various fields are
\begin{equation}
    m^2_h=2\lambda_1 v^2,\qquad m^2_{h'}=\mu_2^2+\frac{\lambda_{345}}{2}v^2, \qquad m^2_{\chi_{4/5}}=\mu_2^2+\frac{\lambda_3}{2}v^2,\qquad m^2_{\chi_6}=\mu_2^2+\frac{\tilde\lambda_{345}}{2} v^2,  
\end{equation}
where $m^2_h$ and $m^2_{h'}$ are the masses of $h$ and $h'$, $m^2_{\chi_{4/5}}$ 
 the mass of $\chi_4$ and $\chi_5$, giving rise to the charged Higgs $H^{\pm}$, $m^2_{\chi_6}$ the mass of the CP-odd scalar $\chi_6$, and we defined $\tilde \lambda_{345}\equiv\lambda_3+\lambda_4-\lambda_5$. 

The field and temperature dependent mass matrices that enter into the one-loop contributions to the potential are given in Appendix \ref{section: thermal masses IDM}. We introduce the counterterms $\delta m^2_1,\, \delta m^2_2$ and $\delta\lambda_1$ and adopt the same renormalisation conditions introduced for the SSM to fix $v$, $m_h$ and $m_{h'}$ to their tree-level value. The only difference here stems from the Goldstone sector: the mass matrices in the $(h,h')$ background at finite temperature mix $\chi_1$ with $\chi_4$, $\chi_2$ with $\chi_5$ and $\chi_3$ with $\chi_6$.

\section{Analytic considerations}
\label{sec: analytic}

Before moving on to the numerical analysis, we make here some general considerations on the structure of the constraint equations, which serve as a guide for a better reading of the results presented in the next section.  

The four constraints in \eqref{eq: constraints}, together with the matching equations  \eqref{matching eq}, can be solved for the four unknown parameters, $v_w, \, \delta_s, \, L_h$ and $ L_s$. The wall velocity is found by first determining the temperature behind the wall $T_-$ (we recall here that $T_+$ is given in terms of $T_n$ for both detonations and deflagrations), that enters in \eqref{eq: constraints} through $h_-$, and then solving \eqref{matching eq} for $v_+$ and $v_-$, the plasma velocity in front and behind the wall.

In local thermal equilibrium, the total pressure $P_{tot}=P_h + P_s$ can be expressed as
\begin{equation}
\label{Ptot}
P_{tot}=\int_{-\infty}^{+\infty} dz \left[ \frac{\partial V}{\partial h} h' + \frac{\partial V}{\partial s} s' \right] = \Delta V - \int_{-\infty}^{+\infty} dz \frac{\partial V}{\partial T} T', 
\end{equation}
where $\Delta V \equiv V(h_-,0,T_-) -  V(0,s_+,T_+)$ is the difference between the potential of the two asymptotic states $(h_-,0, T_-)$ and $(0,s_+,T_+)$. Field-independent terms do not enter in $P_{tot}$, and the latter can equivalently be written either in terms of $V$ or in terms of $\widetilde V(h,s,T)\equiv V(h,s,T)-V(0,0,T)$.

Concerning the subtracted potential $\widetilde V$, the difference $\Delta \widetilde V$ must be negative, as the latter is the driving force of the expansion, and a solution to the equation $P_{tot}=0$ can only be found if $\int dz \,(\partial_{_T} \widetilde V)\, T'(z)$, that represents the back-reaction of the plasma, is negative too. In this case, the non-trivial temperature profile of the plasma generates a friction term that allows for a steady-state expansion of the wall. As for the potential difference $\Delta V$, its sign is not known a priori, but simple considerations can be made to estimate it. The potential $V$ is dominated by the contribution from species that are light with respect to the temperature $T$. For these, the plasma behaves as a relativistic gas. Contributions from heavier species are  Boltzmann suppressed, and the full potential $V$ can thus be approximated as 
\begin{equation}
    V(h,s,T)\simeq  -g_{*} \frac{\pi^2}{90} T^4 +\widetilde V(h,s,T),
\end{equation}
where $g_{*}$ is the effective number of relativistic degrees of freedom.
The potential difference $\Delta V$ is then given as 
\begin{equation}
\label{eq: Delta V approx}
\Delta V\simeq \frac{\pi^2}{90}\left(g_{*}^{+}T_+^4-g_{*}^{-}T_-^4\right)+\Delta \widetilde V,
\end{equation}
where the superscript $\pm$ in $g_{*}$ indicates the vacuum where it is calculated. Most of the particles are massless outside the bubble and massive inside, so that $g_{*}^+> g_{*}^-$. Unless large deviations in the temperature are found across the wall, $\Delta V$ is expected to be positive, so with reversed sign as compared to $\Delta \widetilde V$. 

The requirement for the friction term to actually hinder the wall acceleration can then be better understood considering the conservation equations \eqref{hydro eq 1}, \eqref{hydro eq 2}. For our purposes, the latter can be conveniently rewritten in terms of the potential as 
\begin{equation}
    \frac{\left(h'\right)^2 + \left(s'\right)^2}{2} - V+c_1 v_p= c_2, 
\end{equation}
and, evaluating the resulting equation far away from the wall, we get
\begin{equation}
\label{eq: Delta V}
\Delta V=c_1 (v_- -v_+), 
\end{equation}
with constant $c_1$ obtained from \eqref{hydro eq 1} at, for instance, $z\to +\infty$, 
\begin{equation}
\label{eq: c1}
    c_1=-(T\partial_{_T} V)\Big|_{(h_-,0,T_-)}\frac{v_-}{1-v_-^2}.
\end{equation}
To determine the sign of $\partial_{_T} V$, we write it as $\partial_{_T}V=\partial_{_T} V_{_T}+ \partial_{_T} V_{_{CW}}$ with  
\begin{align}
\label{eq:dTdVT_}
\partial_{_T} V_{_T} &= \frac{2 T^3}{\pi^2}\left(\sum_{i} n_{i} Y_B\left(\frac{m^2_{i}}{T^2}\right) - n_t Y_F\left(\frac{m_t^2}{T^2}\right) \right),  \\
\label{eq:dTdCW_}
\partial_{_T} V_{_{CW}}&= \sum_i \frac{n_i}{32\pi^2}m^2_i\frac{\partial m^2_i}{\partial T} \left(\log\frac{m^2_i}{\mu^2}-1\right)
\end{align}
where the functions $Y_{B/F}$ are defined as 
\begin{equation}
\label{eq: Y}
    Y_{B/F}(y) = \int_0^\infty dx \,x^2 \left( \log \left(1\mp e^{-\sqrt{x^2+y}}\right) \mp  \frac{ 2y -\partial_{_T}(T^2 y)/T }{8\left( e^{\sqrt{x^2+y}}-1\right)\sqrt{x^2+y}}\right),
\end{equation}
and the sum in (\ref{eq:dTdVT_}) extends to all bosons, while the sum in (\ref{eq:dTdCW_}) is only for degrees of freedom that acquire a thermal mass.
The terms proportional to $\partial_{_T} m^2$ in the equations above are a by-product of the Parwani resummation. For bosons, we can write $m^2_i(h,s,T)=\overline   m^2_i(h,s) + c_i \,T^2$, so that the numerator in the second term of \eqref{eq: Y} reduces to $\overline m^2_i(h,s)$.  It is then easily understood that a bosonic contribution to $\partial_{_T} V_{_T}$ is negative for $\overline m^2_i(h,s)\ge 0$, which is clearly the case at $z \to +\infty$.
Similar considerations lead to conclude that the fermionic contribution is always negative, thus implying that $\partial_{_T} V_{_T}\big|_{(h_-,0,T_-)}<0$. \\
On the other hand, the sign of $\partial_{_T} V_{_{CW}}$ is not defined, as it depends on the choice of the renormalisation scale $\mu$ (we took $\mu = v$ in this work). 
However, $\partial_{_T} V_{_{CW}}$ receives contributions only from scalars and longitudinal gauge bosons, and, in general, for reasonable choices of $\mu$, it turns out to be subdominant with respect to\footnote{In addition to that, $\partial_{_T} V_{_{CW}}$ arises only from the specific choice of resummation.} $\partial_{_T} V_{_T}$. Thus, we expect $\partial_{_T} V$ to be negative for $z \to +\infty$.

Having determined the constant $c_1$ in \eqref{eq: c1} to be positive, from \eqref{eq: Delta V} we deduce that the argument on the sign of $\Delta V$ suggests that in local thermal equilibrium a solution to $P_{tot}=0$ can only be found if $v_->v_+$. As the sign of $\partial_{_T}V$ is the same for all values of $z$ \eqref{eq: c1}, a comparison with \eqref{Ptot} also indicates that, on the solution, the temperature $T(z)$ must be a decreasing function of $z$. 
We then conclude that in local thermal equilibrium detonations are unlikely to be found\footnote{Detonations require $\Delta V < 0$ and can only arise if a cancellation takes place in \eqref{eq: Delta V approx} as to overthrow the argument on the positivity of $\Delta V$. Using $\widetilde V$, there is yet another way to see that detonations require some amount of tuning that makes them unlikely to appear in LTE. Besides the ambiguity related to the renormalisation scale, the contribution of a field $i$ to $\partial_{_T} \widetilde V$ can be shown to be positive for $m^2_i\ge0$. In the excursion from $-\infty$ to $+\infty$, only scalar masses can become negative, though this is not necessarily the case, as maxima of the potential typically are not reached in multi-field transitions. All the other fields give positive contributions to $\partial_{_T} \widetilde V$. On the other hand, detonations require $\partial_{_T} \widetilde V$ to be negative in a large enough interval for the negative part to dominate over the positive one in the integral $\int dz\,  \partial_{_T}\widetilde V\, T'(z)$. }. 
Anticipating on the results, this will also be the outcome of our numerical analysis, and it is in agreement with the conclusions of  \cite{Ai:2023see,Ai:2024btx}, where an alternative derivation based on entropy conservation was used. The LTE detonation solutions found (in a narrow region of parameter space) in \cite{Ai:2023see,Ai:2024btx} are known to be unstable, as in LTE the pressure decreases with $v_w$ for $v_w>v_{_J}$. We will come back to this point as we comment on the numerical results.

Coming back to the constraint equations \eqref{eq: constraints}, an important point both for the development of our code and for the interpretation of the numerical results, was to understand the dependence of each one of the constraints on the parameters. Starting with $P_{tot}$ in Eq.\,\eqref{Ptot}, we observe that the first term on the right hand side ($\Delta V$) is clearly independent of $L_h, L_s$ and $\delta_s$.
The second term, in contrast, has a dependence on these parameters that is governed by the profiles of the scalars. However, a few general and model-independent considerations can be made. For the sake of the present discussion, we here consider a sharp profile for the temperature, $T(z)= T_- \,\theta(z) + T_+\, \theta(-z)$. The generalisation to a smoother profile is straightforward, as we will briefly mention. 

When the potential is such that $\partial_{_T} V$ does not contain terms that mix the two fields $h$ and $s$ (take, for instance, the high-temperature potential in the quadratic truncation), changing the integration variable $z\to z/L_h$ or $z\to z/L_s$ in the various pieces shows that $P_{tot}$ only depends on $T_-$ and $\delta_s$, with the $\delta_s$ dependence arising from the non-trivial spatial profile of the temperature  $T(z)$. In this respect, we note that a smoother profile for $T(z)$ would introduce a dependence on the ratios $L_h/L_T$ and $L_s/L_T$, with $L_T$ the width of the region where the temperature has a significant variation. However, from the hydrodynamic equations \eqref{hydro eq 1}, \eqref{hydro eq 2}, we expect the field and temperature widths to be related, and so their ratio not to vary much. 
Similar considerations can be made in the more general case where $\partial_{_T} V$ contains terms that mix $h$ and $s$, and one is led to conclude that $P_{tot}$ is a function of three variables, $P_{tot}=P_{tot}(T_-, \delta_s, L_h/L_s)$. 
Combining the observations above with the fact that a solution to $P_{tot}=0$ with $\Delta V\ne 0$ necessarily requires a non-trivial profile for the temperature, we expect $P_{tot}$ to have the largest dependence on $T_-$, and in turn on $v_w$. 

Similar to $P_{tot}$, the pressure difference 
\begin{equation}
\Delta P= P_s-P_h=\int dz \left(\frac{\partial V}{\partial s} s' - \frac{\partial V}{\partial h} h'\right)
\end{equation}
is easily found to be a function of the same three variables, $\Delta P = \Delta P(T_-, \delta_s, L_h/L_s)$, and we intuitively envisage that the largest dependence is on $\delta_s$, since the pressure difference controls the displacement between the two fields\footnote{In terms of the original equations of motion, $\delta_s$ serves to realise the condition $\partial_h V=0$ at $z=0$ (centre of the $h$-bubble), while the ratio $L_h/L_s$ allows to realise the analogous condition $\partial_s V=0$ at $z=L_s \delta_s$ ($s$-bubble centre). When $\delta_s=0$, $\partial_h V$ and $\partial_s V$ both vanish for the same field configuration $h=h_-/2,\, s=s_+/2$. When $\delta_s\ne 0$, $\partial_h V\big|_{z=0}=0$ for $h=h_-/2$, $s=s_+(1+\tanh\delta_s)/2\equiv s_*/2$ and $\partial_s V\big|_{z=L_s\delta_s}=0$ for $h=h_-(1+\tanh(L_s \delta_s/L_h)/2\equiv h_*/2$ and $s=s_+/2$. The (absolute value of the) offset $\delta_s$ increases as the difference between the configurations that make $\partial_h V$ vanish and $\partial_s V$ vanish increases, with the ratio $L_s/L_h$ adjusting for the asymmetry in the deviations $h_*-h_-$ and $s_*-s_+$.}.

The considerations above are at the basis of our choice to use the equations $P_{tot}=0,\, \Delta P=0$ instead of $P_h=0,\, P_s=0$ in our numerical algorithm, as discussed in Section \ref{sec: algorithm}, and have been used to guide the root-finding procedure. 

The values of $L_h$ and $L_s$ are fixed by the pressure gradients $G_h$ and $G_s$, that can be thought of as averages of the pressure over the profiles ($\phi_i=h,s$)
\begin{equation}
G_i = \pm \int_{-\infty}^{+\infty} dz \,P_i(z) \tanh{\left(\frac
{z}{L_i}-\delta_i\right)}= \braket{P_i}_{\phi_i}
\end{equation} 
where we defined $P_i(z)\equiv E_{\phi_i} \phi_i'$. In our case, the $+$ sign applies for the Higgs field $h$, and the minus sign for $s$. Within the {\it tanh} ansatz, $G_i$ can be expressed as
{\small \begin{align}
\label{eq: Gi generic}
    G_i &= \int_{-\infty}^{+\infty} dz \left(-\phi_i''+\frac{\partial V}{\partial \phi_i}\right)\phi'_i\left(\frac{2 \phi_i}{\bar\phi_i} -1\right) = \pm \frac{2\bar \phi_i^2}{15 L_i^2} \pm \frac{2}{\bar\phi_i} \int_{-\infty}^{+\infty} dz \left(\frac{\partial V}{\partial \phi_i}\right)\phi'_i \left(\frac{2 \phi_i}{\bar\phi_i} -1\right),
\end{align}}
where $\bar \phi_i=h_-,s_+$. 
  
Following the discussions above, we can then write $G_h$ and $G_s$ as 
\begin{equation}
\label{eq: G_i}
G_h = \frac{2h_-^2}{15 L_h^2} + g_1\left(T_-,\delta_s,\frac{L_h}{L_s}\right), \qquad  G_s = -\frac{2s_+^2}{15 L_s^2} - g_2\left(T_-,\delta_s,\frac{L_h}{L_s}\right).
\end{equation}
On the solution, one has 
\begin{equation}
   \frac{h_-^2}{s_+^2} = \frac{g_1 (T_-,\delta_s,L_h/L_s)}{g_2(T_-,\delta_s,L_h/L_s)}\frac{L_h^2}{L_s^2}.
   \label{eq: ratio width}
\end{equation}
As one can infer from \eqref{eq: Gi generic}, the ratio $g_1/g_2$ gives a measure (weighted by $(\phi_i^2)'$) of the differences between the gradient of the potential encountered in the $h$ and $s$ directions. If no large hierarchy in the shape and height of $V$ along $h$ and $s$ is present, it is reasonable to to expect $g_1/g_2\sim  \mathcal O(1)$, and the ratio of the widths can be approximated to $L_h/L_s\sim h_-/s_+$. Such a relation can be taken as a rough estimate, to be used, for example, as an initial guess in a numerical analysis. As for the individual widths, according to the explicit calculation of \cite{Moore:1995ua,Moore:1995si}, the ratio between the square of the jump in the field $\bar \phi_i^2$ and the square of the width $L_i^2$ should be proportional to the potential difference $\Delta V$. They are thus expected to be related to $\Delta V$ as $L^2_i\propto \bar\phi_i^2/\Delta V$.

\section{Numerical results}
\label{sec: numerical}
In this section, we present the numerical results for the bubble wall velocity and, more generally, for the system of equations that describe the wall dynamics, obtained using our code that implements the algorithm described in Section \ref{sec: algorithm}. 
The analysis was performed for the three models introduced in Section \ref{sec: Models}. As already stressed, our main interest is in regions of the parameter space where the transition proceeds with a two-step pattern, schematically $(0,0)\to(0,s)\to(h,0)$, with the last one being first order. In this case the potential barrier between the two states involved in the EW transition is generated at tree-level in most of the parameter space. Our algorithm needs as input some specifics of the transition such as the nucleation temperature $T_n$ and the values of the fields in the false and true vacuum at $T_n$. Therefore, for each model considered, we performed a preliminary analysis of their parameter space with the \texttt{Cosmotransitions} package \cite{Wainwright:2011kj} and with BSMPT \cite{Basler:2018cwe,Basler:2020nrq,Basler:2024aaf} for crosschecking, to determine the transition pattern and individuate the region of interest.

\subsection{SSM}

\begin{figure}[t!]

\centering
\includegraphics[scale=0.35]{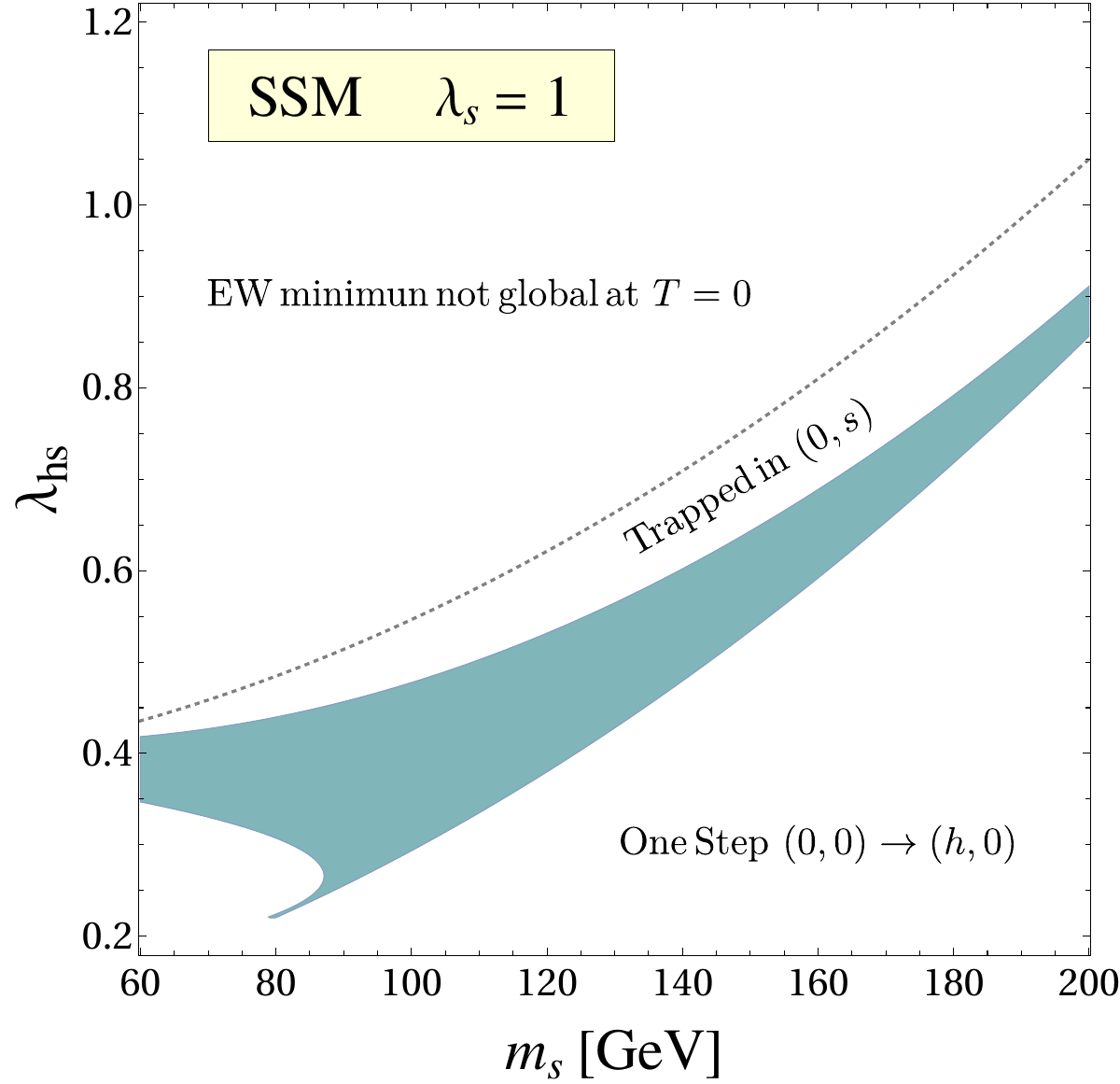}\,\,\,\,
\includegraphics[scale=0.35]{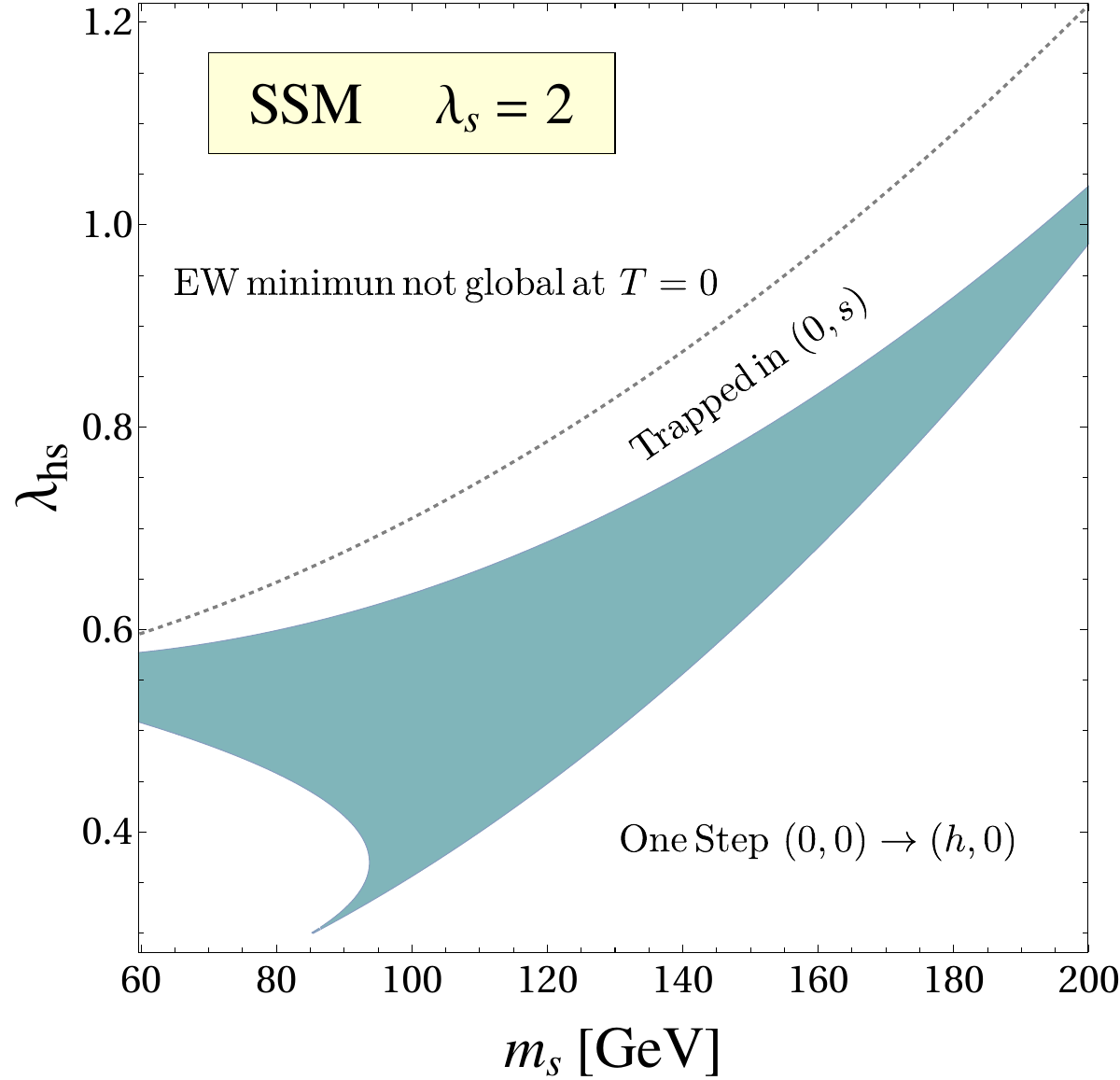}
\caption{Regions on the plane $(m_s,\lambda_{hs})$ with different patterns for the EW phase transition for the SSM with  $\lambda_s=1$ ({\it left panel}) and $\lambda_s=2$ ({\it right panel}). In the coloured region the EWPhT is two-step.}
\label{SSMParSpace}
\end{figure}

For the singlet extension of the Standard Model we present the results for two different choices of the singlet self-coupling, $\lambda_s=1$ and $\lambda_s=2$. 

The region where the EW vacuum at $T=0$ is reached through a two-step phase transition is reported in Fig.\,\ref{SSMParSpace}. As can be seen, for larger values of the singlet self-coupling $\lambda_s$, the two-step region is shifted to larger values of $\lambda_{hs}$. This can be understood taking a high temperature approximation for the effective potential (tree-level plus quadratic corrections)
\begin{equation}
\label{pot quadratic}
    V(h,s,T)=\frac12 \mu_h^2h^2+ \frac12 \mu_s^2 s^2 +\frac{\lambda_h}{4}h^4+\frac{\lambda_s}{4}s^4+\frac{\lambda_{hs}}{2}h^2 s^2 +\frac{c_h T^2}{2} h^2 +\frac{c_s T^2}{2} s^2,
\end{equation}
where all the model-dependence is encoded in the coefficients $c_h$ and $c_s$ of the thermal masses, and $m^2_s=\mu_s^2+\lambda_{hs} v^2$. For $\mu_s^2+c_s T^2<0$, a non-trivial minimum is found in the $s$ direction, $w^2(T)=-(\mu_s^2+c_s T^2)/\lambda_s$. The Hessian calculated on $(h,s)=(0,w(T))$ is diagonal,
\begin{equation}
    \mathcal H\left(0,w(T)\right)=\begin{pmatrix}
        \mu_h^2+c_hT^2+\frac{\lambda_{hs}}{2} w^2(T) && 0 \\
        0 && \mu_s^2+c_sT^2+3\lambda_s w^2(T)
    \end{pmatrix},
\end{equation}
and stability around it requires that its eigenvalues are positive. As $w^2(T)$ is inversely proportional to $\lambda_s$, it is immediately apparent that an increase in $\lambda_s$ must be compensated by a simultaneous increase of the portal $\lambda_{hs}$ for the state $(0,w(T))$ to be stable along the Higgs direction. 

A comparison of the region in Fig.\,\ref{SSMParSpace} with the one obtained through its analytic determination with the quadratic potential \eqref{pot quadratic} (see, for instance, \cite{DeCurtis:2019rxl})
shows a substantial overlap between the two, and confirms the 
crucial role played by the tree-level barrier in strong two-step PhTs. Below the coloured region of Fig.\,\ref{SSMParSpace}, the transition towards the EW vacuum consists of only one step, either first or second order, schematically $(0,0)\to (h,0)$, with the SM result recovered in the limit $\lambda_{hs}\to 0$. For large $\lambda_{hs}$ above the coloured region, the EW vacuum is not the global minimum of the potential at $T=0$, as an increase in $\lambda_{hs}$ with fixed value of $m^2_s$ determines an increase in the absolute value of $\mu^2_s$. Before this happens, for intermediate values of $\lambda_{hs}$ above the upper boundary of the two-step region, one finds that, though the EW vacuum is the global minimum of the potential at zero temperature, there is no transition towards it. This is due to the absence of a sufficiently large difference in potential values between the two vacua at $T=0$. In fact, for fixed $m_s$ the critical temperature $T_c$ and the nucleation temperature $T_n$ decrease with $\lambda_{hs}$. Above a certain maximal value of $\lambda_{hs}$, $T_c$ drops to such low values that the conditions for nucleation cannot be satisfied, so that the system remains trapped in the false vacuum all the way down to $T=0$.  

Given the parameters describing the EWPhT obtained with these preliminary scans, the bubble wall velocity $v_w$, the wall thicknesses $L_h, L_s$, and the displacement $\delta_s$ are obtained with our algorithm. Stationary solutions were found in the whole two-step parameter space but a strip on the upper part of it. This is related to the non-appearance of detonation solutions in local thermal equilibrium, as discussed in Section \ref{sec: analytic}.  

\subsubsection*{SSM, $\lambda_s=1$}

\begin{figure}[t!]
\centering
\includegraphics[width=0.6\linewidth]{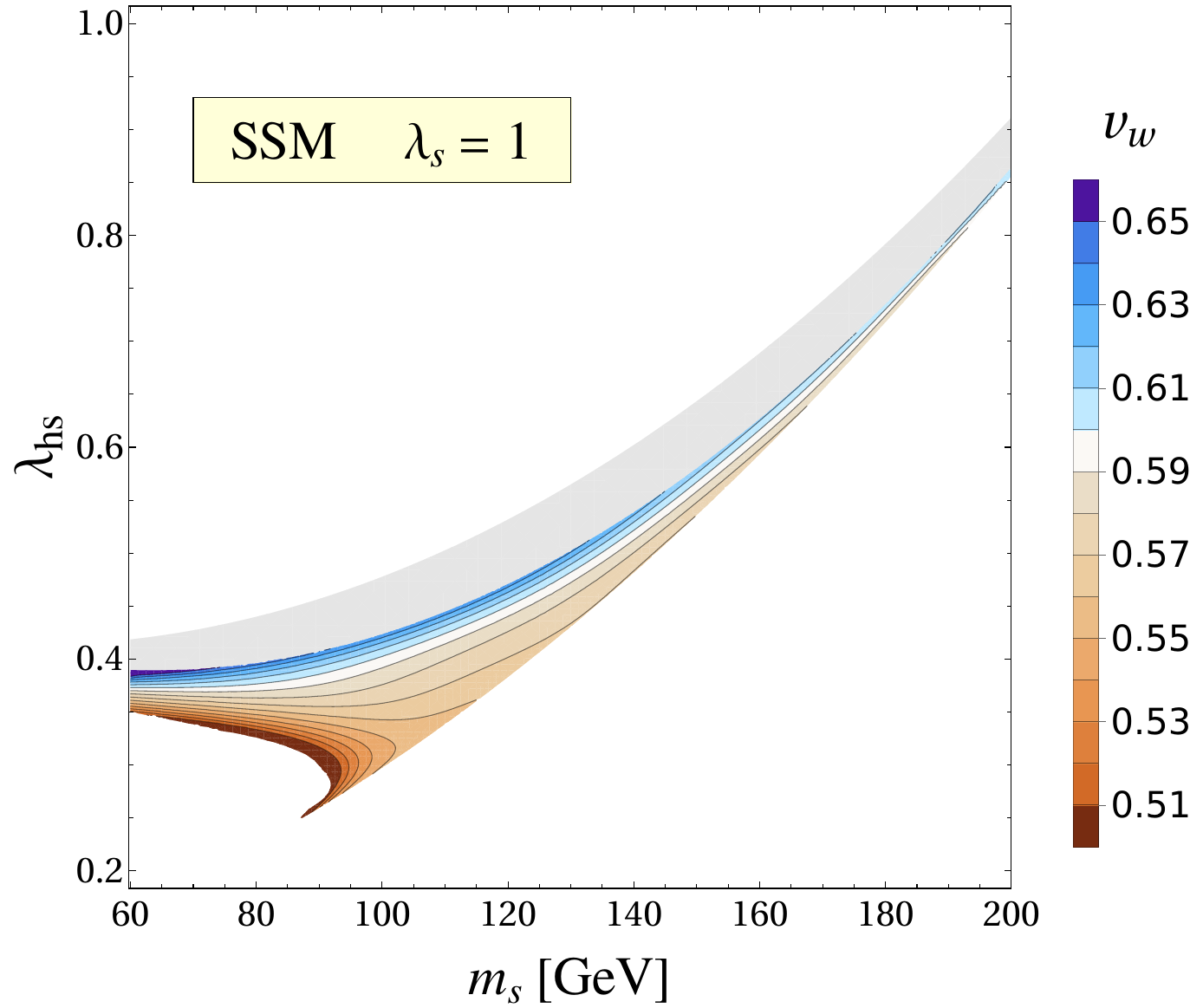}
\caption{Contour plot of the wall velocity $v_w$ as a function of the
parameters $m_s$ and $\lambda_{{hs}}$  for the SSM with $\lambda_s=1$. The bubble wall velocity is constant on the black lines, with the colour gradient describing its variation and the colour bar serving as a legend. In the grey region no steady-state solutions are found in LTE.}
\label{ContourV}
\end{figure}

\begin{figure}
    \centering
\includegraphics[width=0.45\linewidth]{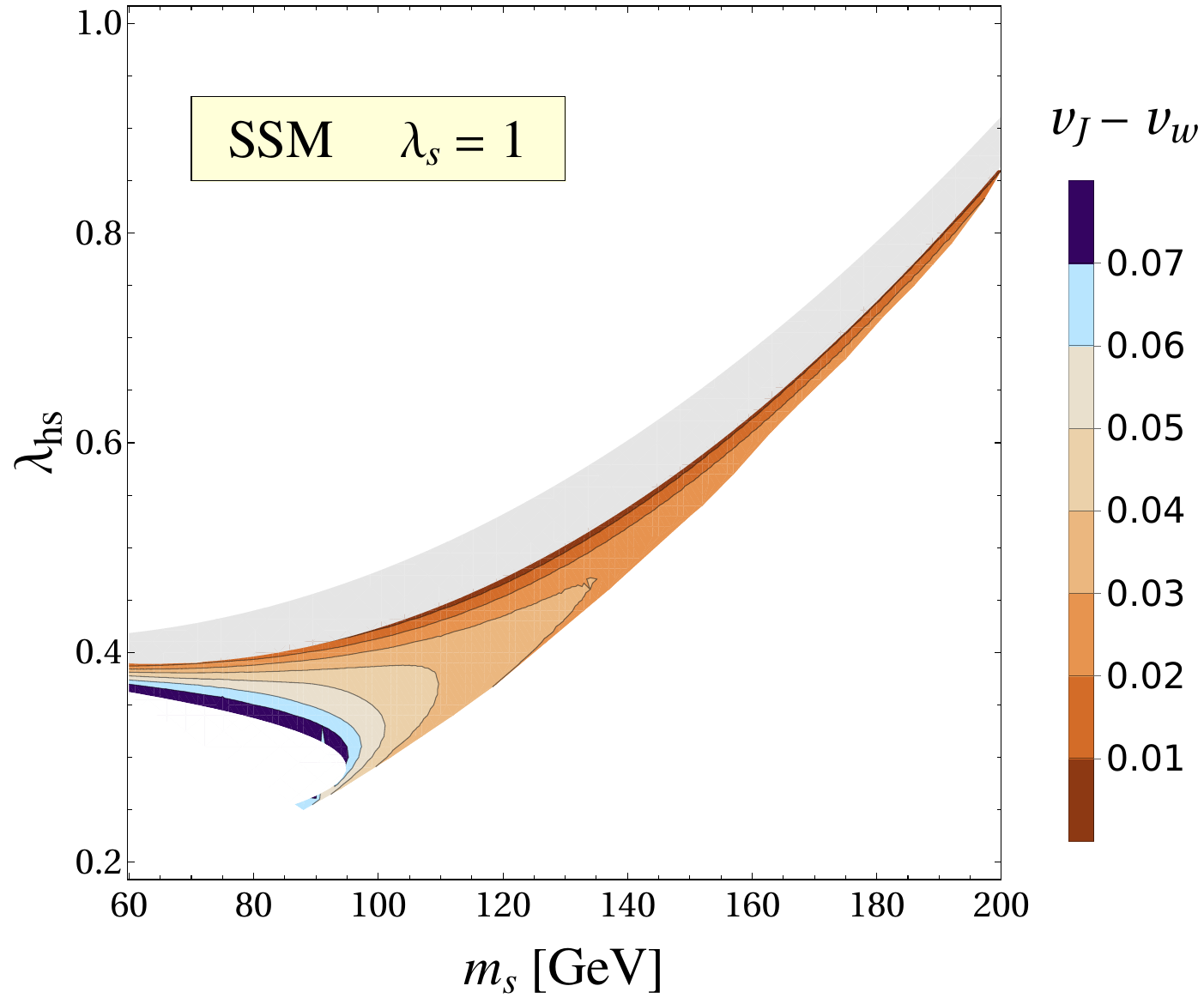}
\hspace{1mm}
    \includegraphics[width=0.5\linewidth]{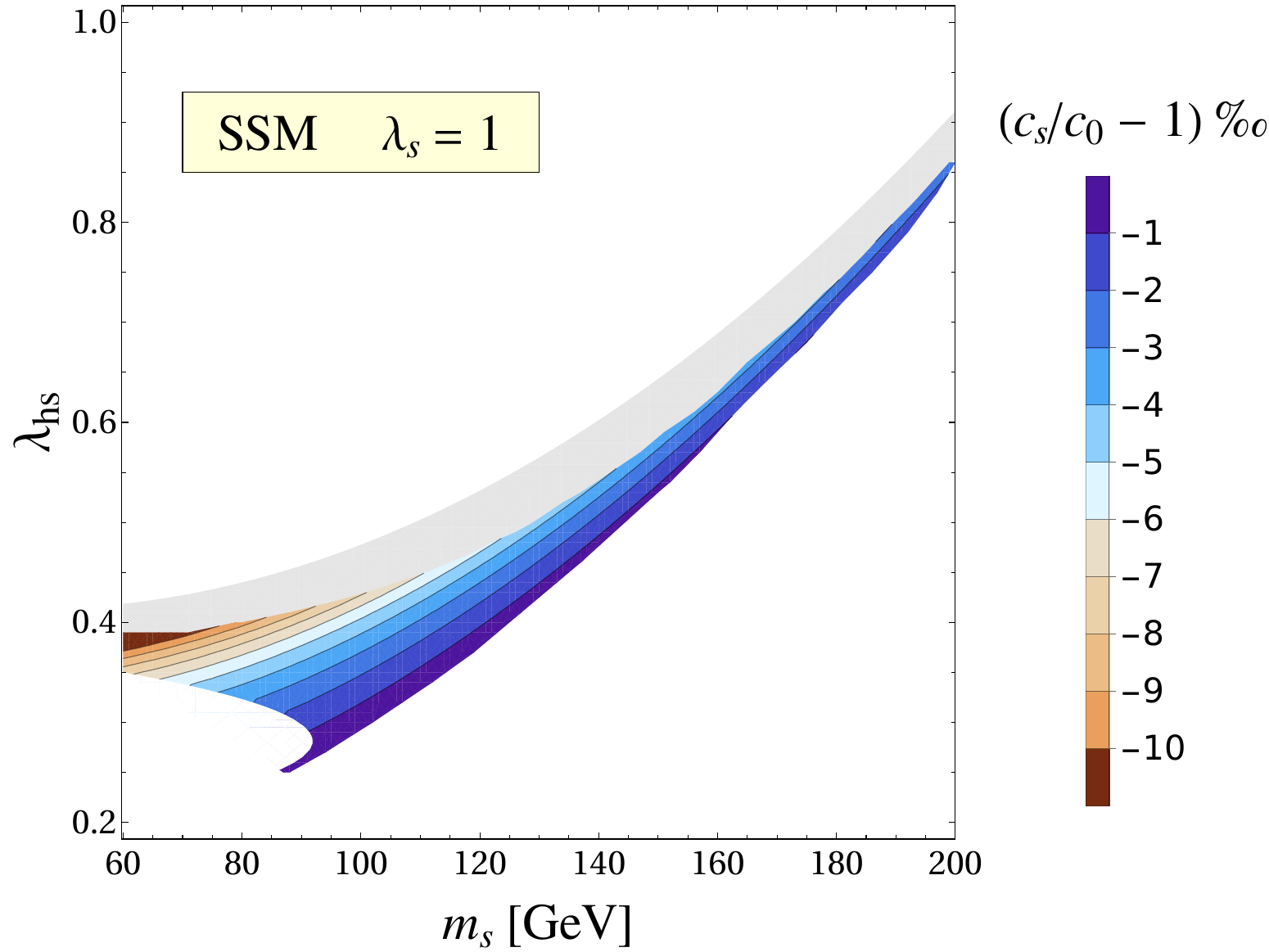}
    \caption{Contour plots of the difference between the Jouguet and the bubble wall velocities $v_{_J} - v_{_w}$ ({\it left panel}),  and   of the deviation of the speed of sound $c_s$ from $c_0 = 1/\sqrt{3}$ ({\it right panel}), in the plane ($m_s$, $\lambda_{hs}$) for the SSM with $\lambda_s=1$. In the grey band no steady-state solution is found in LTE.}
    \label{fig: vJ minus v & cs}
\end{figure}

The results for the wall velocity in the ($m_s,\lambda_{hs}$) plane are shown in Fig.\,\ref{ContourV}.  For constant $m_s$ the velocity tends to increase as $\lambda_{hs}$ increases until $v_w$ gets too close to the Jouguet velocity $v_{_J}$ and the framework described above does not provide an accurate description of the wall dynamics any more. 
As already stressed, the limit $v_w\to v_{_J}$ from below, in fact, corresponds to the case of an hybrid solution where the wall gets infinitely close to the shock front, and the description of the bubble as a planar domain wall breaks down before the limit is reached. 
The left panel in Fig.\,\ref{fig: vJ minus v & cs} confirms  that the upper boundary of $v_w$ is given by the condition $v_w\sim v_{_J}$.

As seen in Fig.\,\ref{fig: vJ minus v & cs} (left), our analysis does not find any detonation solution, in agreement with the considerations of Section \ref{sec: analytic}. Instead, the grey region above the upper boundary of the coloured one should be interpreted as an ultra-relativistic detonation regime, if the appropriate friction was included. Conversely, smaller values of $v_w$ are found in the lower part of the plot (especially in the left corner), where iso-lines tend to become denser and the transition is weaker, with the potential barrier between the vacua becoming flatter as one approaches it. For completeness, the results for the speed of sound behind the wall $c_s$ are presented in the right panel of Fig.\,\ref{fig: vJ minus v & cs}, showing an opposite behaviour to that of $v_w$, i.e.\,\,it decreases while going from the lower to the upper boundary of the parameter space. It is worth noting that $c_s$ turns out to be smaller (up to the percent level) than the speed of sound in the vacuum $c_0=1/\sqrt 3$, as it should, with larger deviations for stronger transitions.

\begin{figure}[t]
\centering
{\includegraphics[width=4.8 cm]{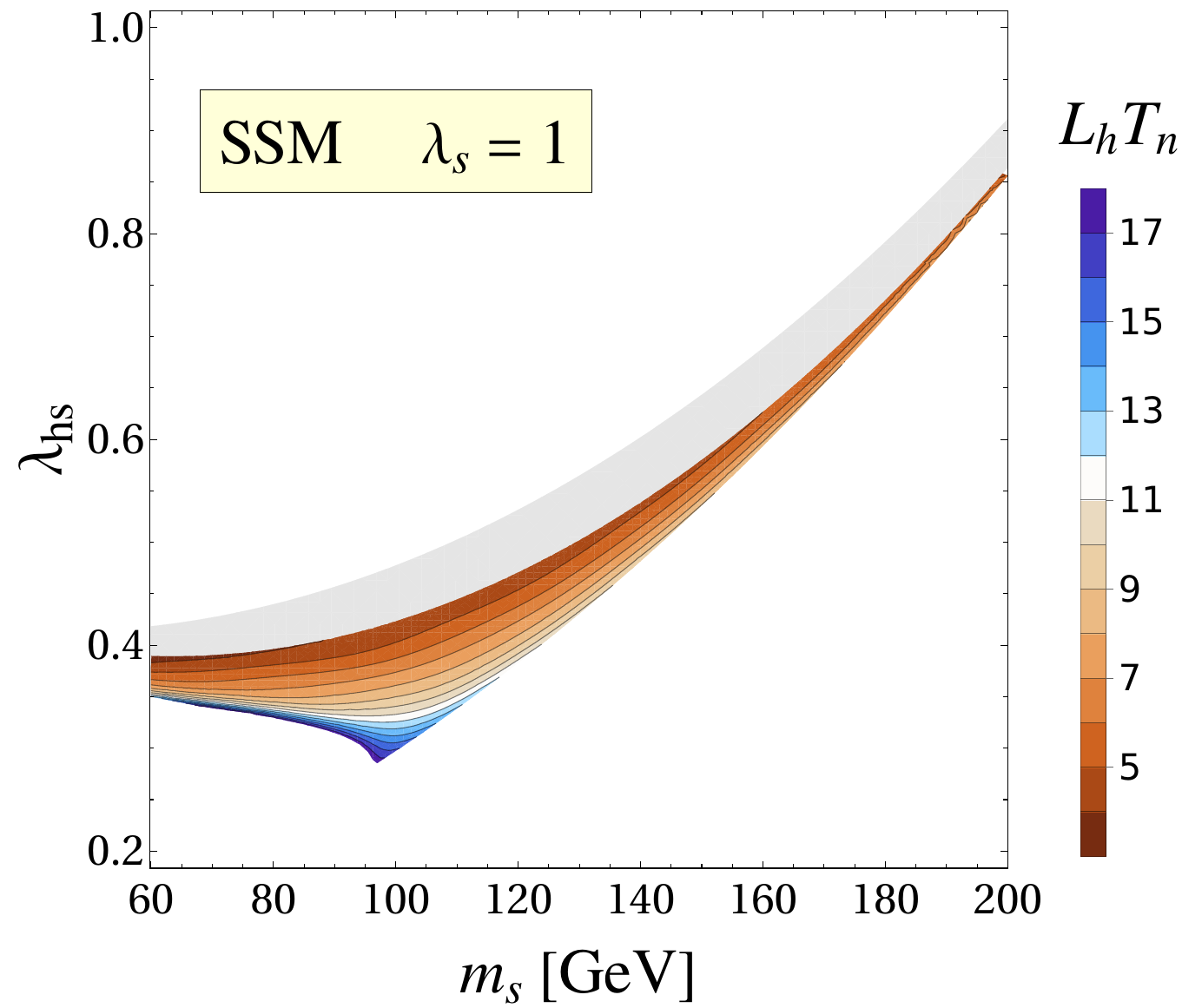}}
\hspace{0.5mm}
{\includegraphics[width=4.8cm]{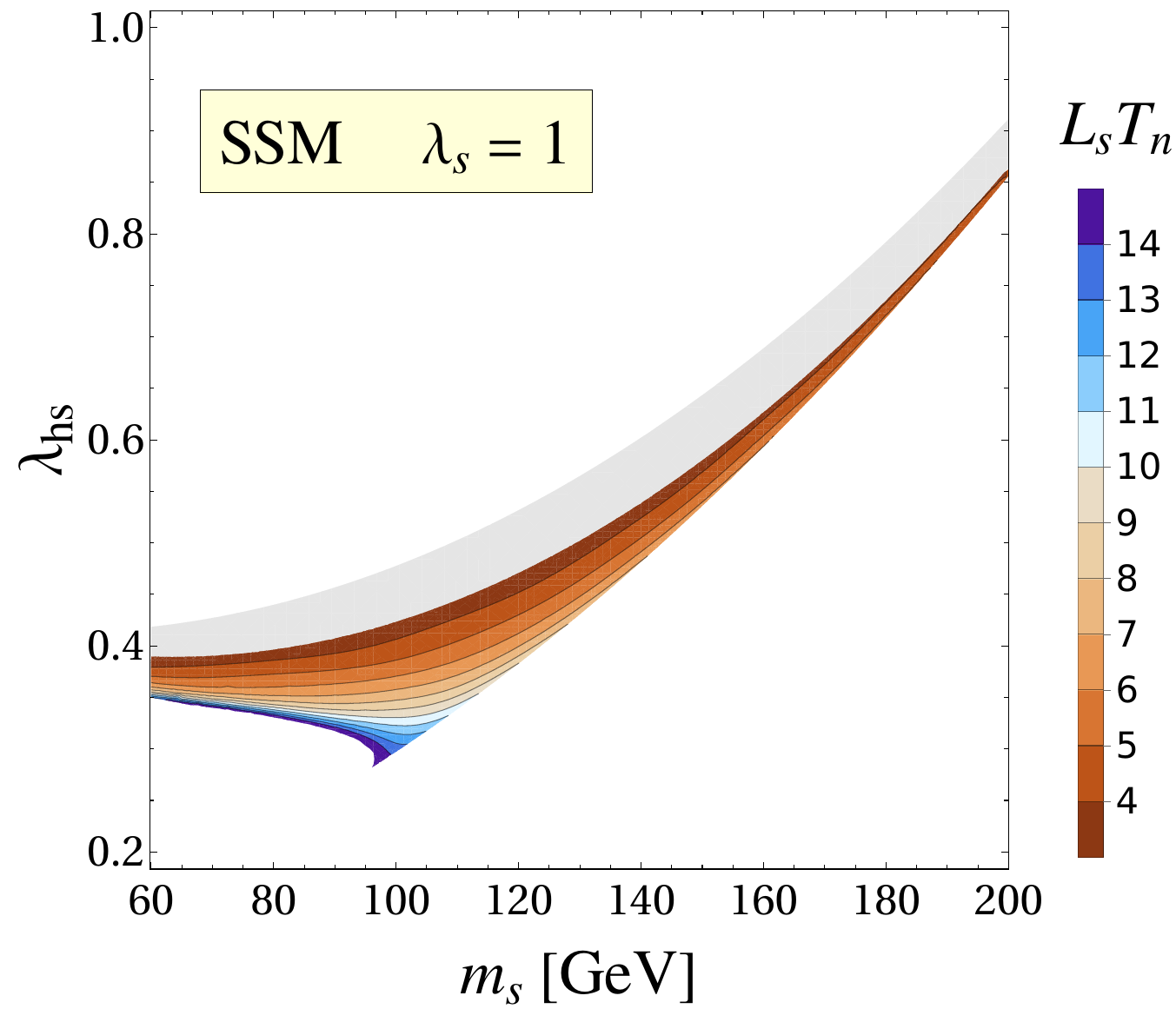}}
\hspace{0.5mm}
\includegraphics[width=4.9cm]{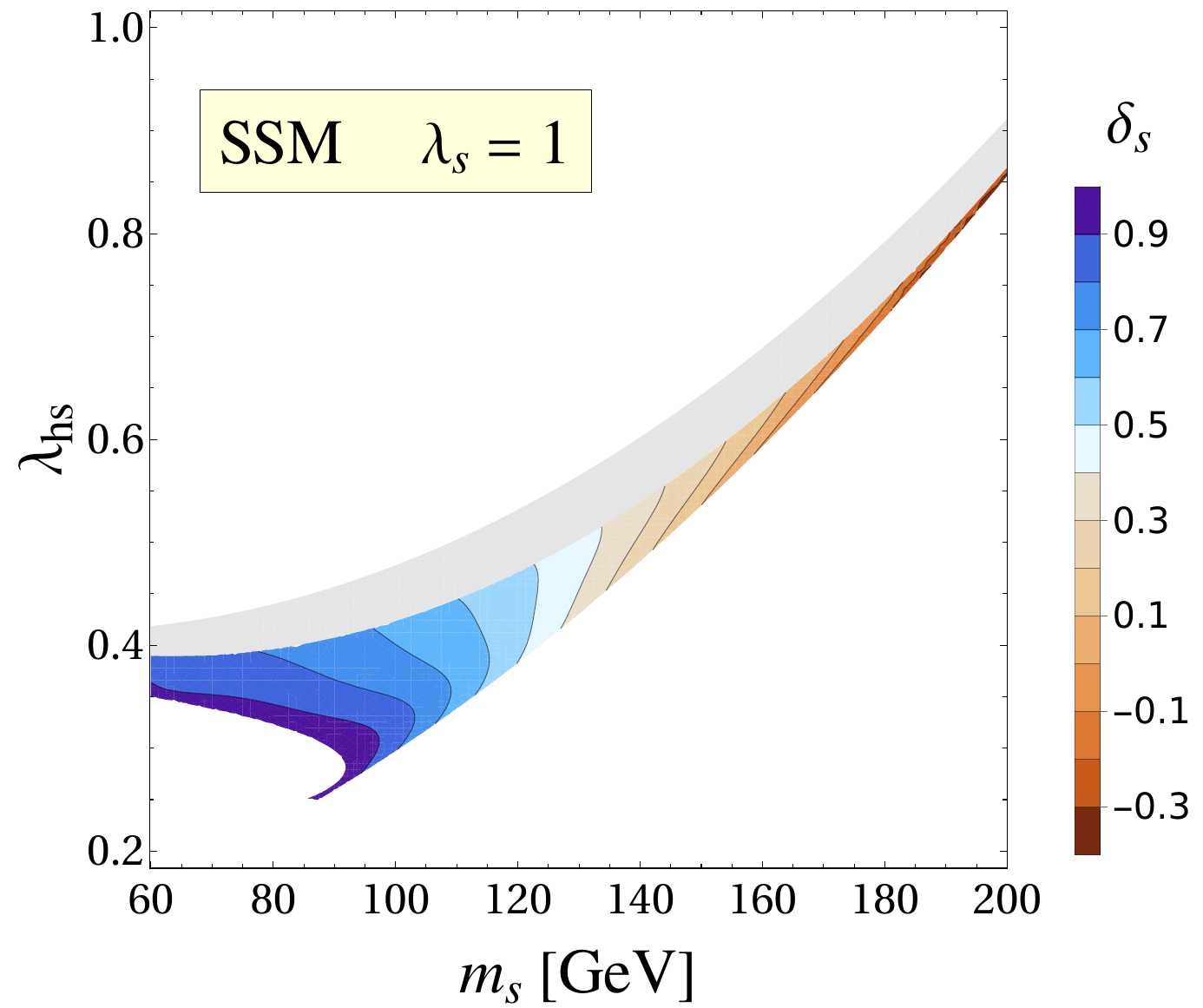}
\caption{Contour plots of the wall widths $L_h$, $L_s$, and of the displacement $\delta_s$ ({\it left}, {\it central}, and {\it right} panel, respectively) in the plane ($m_s$, $\lambda_{hs}$) for the SSM with $\lambda_s=1$. In the grey band no steady-state solution is found in LTE.}
\label{contourLhLsds}
\end{figure}

A similar analysis can be performed for the other parameters, $L_h$, $L_s$ and $\delta_s$, whose contour plots are shown in Fig.\,\ref{contourLhLsds}. The wall widths have a behaviour opposite to the velocity, i.e.\, they decrease as the strength of the phase transition increases, as one would intuitively guess. This reflects the physical relation between energy dissipation and wall dynamics, with lower velocities allowing for greater spreading of the profiles. As mentioned in Section \ref{sec: scalar EOMs}, it is worth to recall here that, when $L_{h} T_n\to 1$ ($L_sT_n\to 1)$, the approximations leading to the scalar equations of motion \eqref{eq: scalar EOMs} break down, as they are obtained taking a WKB approximation to evaluate thermal averages \cite{Moore:1995ua,Moore:1995si}. We also observe that, for fixed $m_s$, the widths $L_h$ and $L_s$ decrease with $\lambda_{hs}$, that is with the PhT strength. As the latter is proportional to the potential difference in the two vacua, this also suggests that the widths tend to decrease with $\Delta V$. This is in good agreement with expectations. 
Moreover, throughout the parameter space, the ratio $L_h/L_s \simeq h_-/s_+ \gtrsim 1$, and it decreases for decreasing $\lambda_{hs}$, as per the arguments given in Section \ref{sec: analytic}.

\begin{figure}
    \centering
    \includegraphics[width=0.44\linewidth]{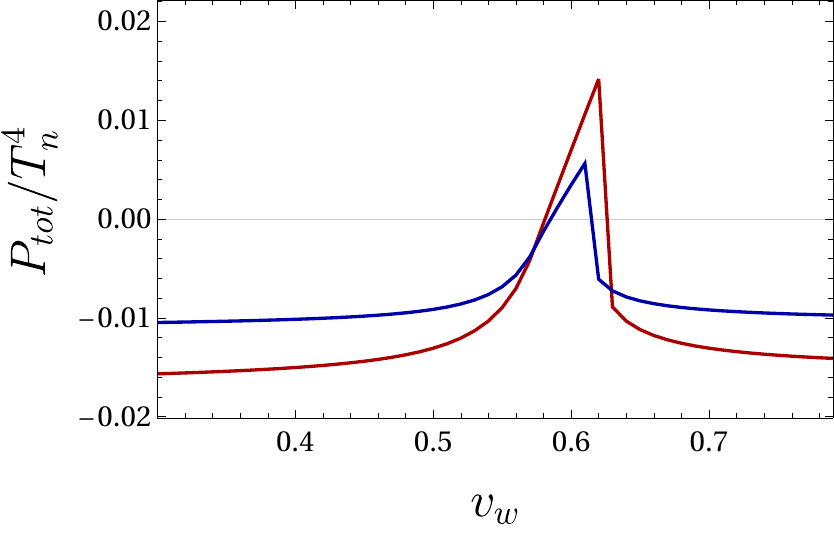}
    \hspace{3mm}
       \includegraphics[width=0.45\linewidth]{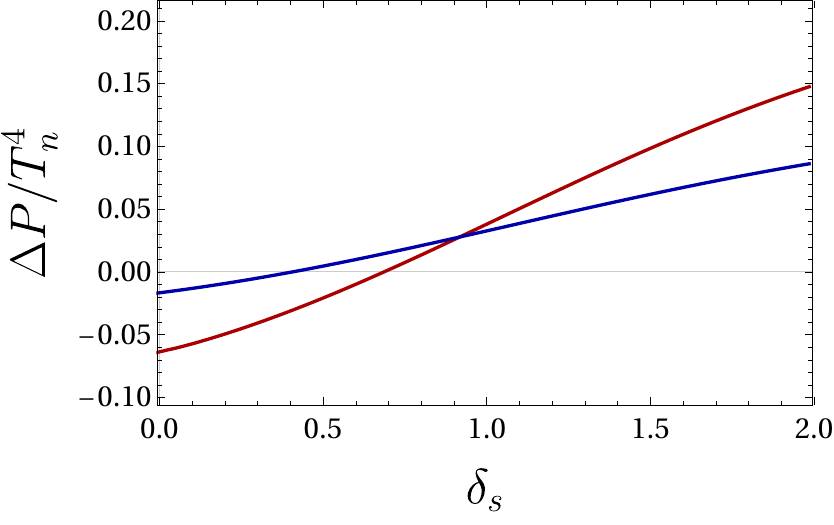} \\
       \vspace{5mm}
    \includegraphics[width=0.445\linewidth]{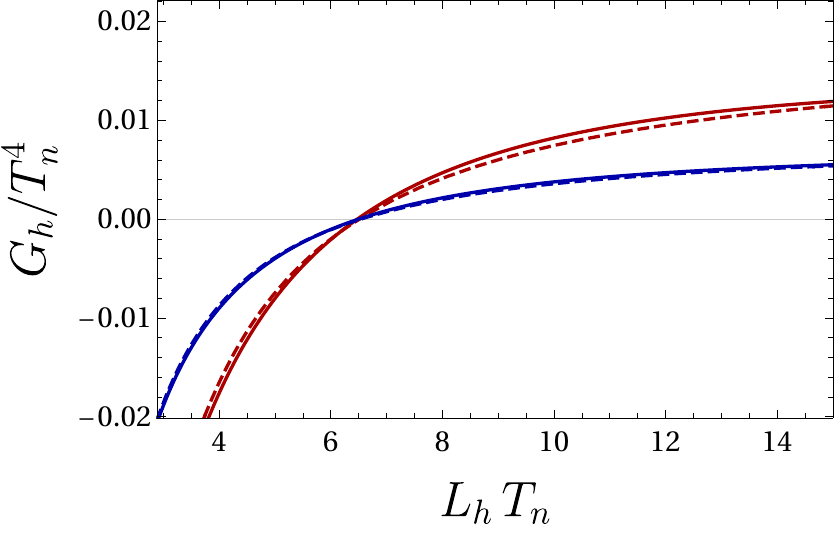}
        \hspace{3mm}
    \includegraphics[width=0.45\linewidth]{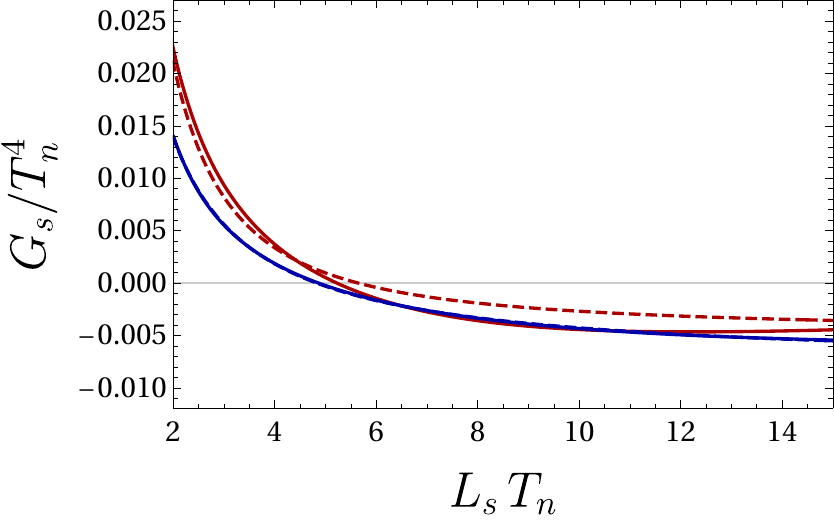}
    \caption{{\it Upper panel}: Total pressure $P_{tot}=P_h + P_s$ as a function of the wall velocity $v_w$ for two benchmark points in the SSM with $\lambda_s=1$: BP1 ($m_s=105$ GeV, $\lambda_{hs}=0.39$, red curve) and BP2 ($m_s=131$ GeV, $\lambda_{hs}=0.47$, blue curve)
    ({\it left}).  Pressure difference $\Delta P =P_s-P_h$ as a function of $\delta_s$ ({\it right}). {\it Lower panel}: Pressure gradient $G_h$ as a function of the wall width $L_h$. The dashed lines are for the fitting function $G_h(L_h)$, whose form is given in the text ({\it left}).  Pressure gradient $G_s$ as a function of the wall width $L_s$. The dashed lines are for the fitting function $G_s(L_s)$ ({\it right}). }
    \label{fig: plasma constraints}
\end{figure}

As discussed in Section \ref{sec: analytic}, analytic considerations suggest that the bubble wall velocity $v_w$ mainly depends on the total pressure $P_{tot}=P_h+P_s$, the displacement $\delta_s$ on the pressure difference $\Delta P= P_s-P_h$, and the wall widths $L_h$, $L_s$ on $G_h$ and $G_s$, respectively, so that in the following we will typically refer to $P_{tot}$ and $\Delta P$ rather than $P_h$ and $P_s$ when discussing constraints. Fixing the other three quantities to the solution, in Fig.\,\ref{fig: plasma constraints} the dependence of each one of the four constraints above on their ``primary" variable is shown for two different benchmark points.

In the upper left panel, we see that, starting from negative values, $P_{tot}$ increases with $v_w$ until it eventually crosses the axis. A peak is then reached at the Jouguet velocity, beyond which $P_{tot}$ decreases and relaxes to negative values again. Actually, 
as stressed several times, the hydrodynamic equations cannot be used for $v_w$ too close to the Jouguet velocity. In particular, while the first zero of $P_{tot}(v_w)$ represents the deflagration solution found in our analysis, the second zero, located right beyond the Jouguet velocity, should not be taken for a detonation solution. The fact that $P_{tot}(v_w)$ is monotonically decreasing for $v_w>v_{_J}$ means that an increase in the wall velocity does not induce an increase in the friction acting on the wall, and thus the second zero does not represent a stable stationary expansion. This is a further manifestation of the non-appearance of detonation solutions in local equilibrium. In this respect, it is important to remind that, as shown for instance in \cite{DeCurtis:2023hil, DeCurtis:2024hvh,Laurent:2022jrs}, out-of-equilibrium effects give a positive and increasing contribution to $P_{tot}$ for $v_w>v_{_J}$, opening the door to the possibility of finding detonations. Concerning $\Delta P$ (upper right panel) we see that, as expected, the offset $\delta_s$ adjusts to equate the two pressures $P_h$ and $P_s$. When $P_h>P_s \, (\Delta P<0)$, an increase in $\delta_s$, corresponding to a shift of the $s$-profile in the direction opposite to propagation, is required. The converse is true for $P_h<P_s$.

As for the other constraints, the pressure gradients are shown in the second row of Fig.~\ref{fig: plasma constraints}. The dashed lines represent a fit to the data of the form $G^{fit}(x)= c_1 + c_2 \,x^{-1} + c_3\, x^{-2}$, that can be appreciated to reproduce the results to a good approximation, especially around the solution $G=0$. It can be verified that the coefficient $c_3$ is $c_{3,\,h} \sim -2h_-^2/15$ for $G_h^{fit}$ and $c_{3,\,s}\sim 2 s_+^2/15$ for $G_s^{fit}$, in agreement with expectations from Section \ref{sec: analytic}. As both the constraints are shown in terms of only one width, with the other one constant, a net dependence of $G_h$ from $L_h$ and of $G_s$ from $L_s$ comes from the functions $g_1(T_-,\delta_s,L_h/L_s)$ and $g_2(T_-,\delta_s,L_h/L_s)$, respectively. The latter are responsible for the terms $x^0$ and $x^{-1}$ in $G^{fit}(x)$. Since the fitting function above nicely reproduces the numerical results around $G=0$ for both the pressure gradients $G_h$ and $G_s$, the widths can be approximately determined as the roots of the functions  $G_h^{fit}(x)$ and $G_s^{fit}(x)$. Noting that $c_2^2/4c_1^2$ is always subdominant with respect to $c_3/c_1$ (at least for our benchmark points), it is then understood that the widths determined in this way are approximately given as\footnote{For $G_h^{fit}$, $c_{1,\,h}>0$ and $c_{3,\,h}<0$, while for $G_s^{fit}$ one has $c_{1,\,s}<0$ and $c_{3,\,s}>0$.} $L_h^2\sim -c_{3,\,h}/c_{1\,h} $ and $L_s^2\sim -c_{3,\,s}/c_{1,\,s}$, so that $L_h^2/L_s^2\sim (c_{3,\,h}/c_{3,\,s})(c_{1,\,h}/c_{1,\,s})\sim (h_-^2/s_+^2)(c_{1,\,h}/c_{1,\,s})$. For the two benchmark points in the figure, $\sqrt{c_{1,h}/c_{1,s}}\sim 1.6$ (red curve) and $\sqrt{c_{1,h}/c_{1,s}}\sim 1.9$ (blue curve).
 
\begin{figure}[t]
	\centering
	\includegraphics[width=0.48\linewidth]{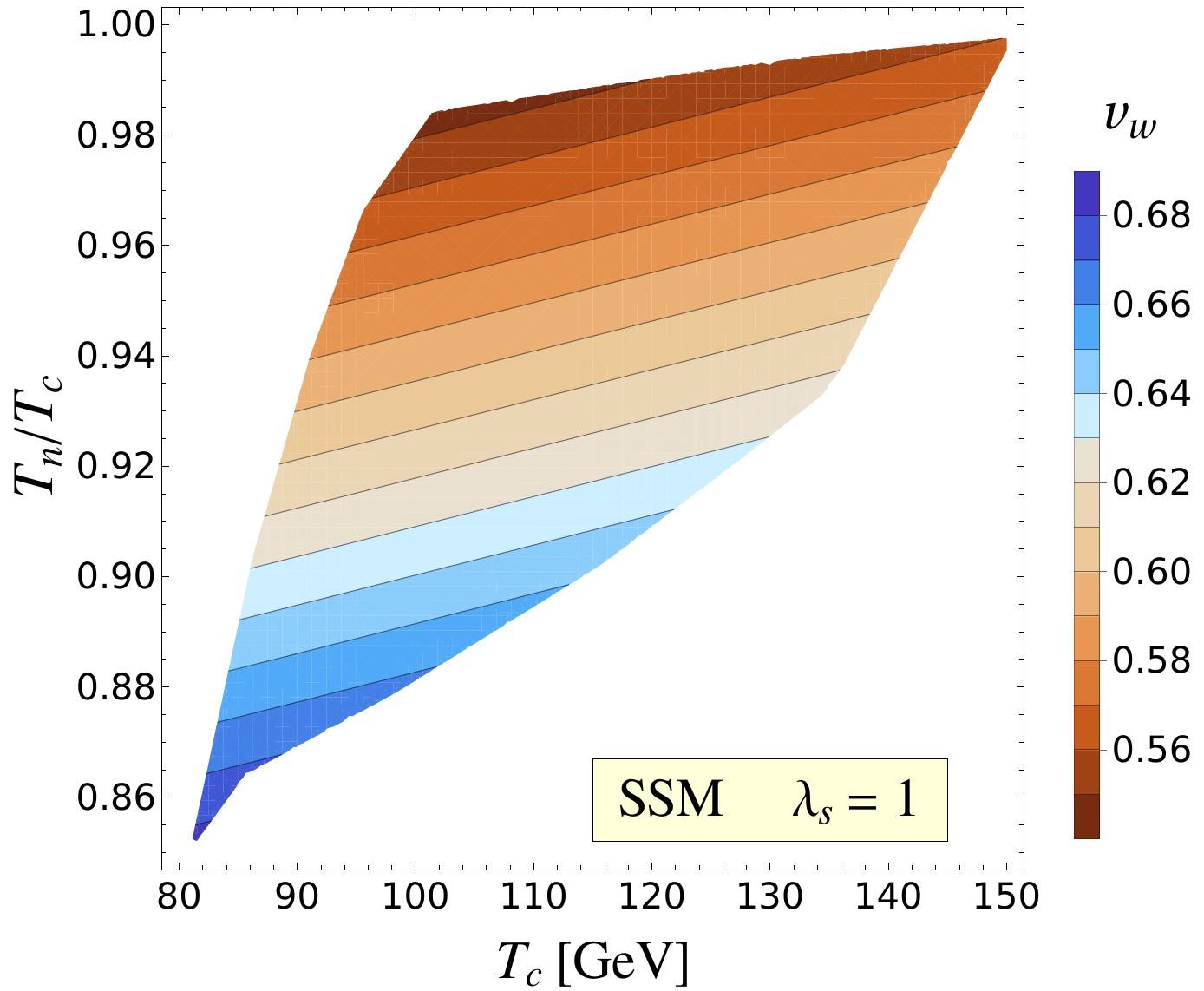} \,\, 
	\includegraphics[width=0.495\linewidth]{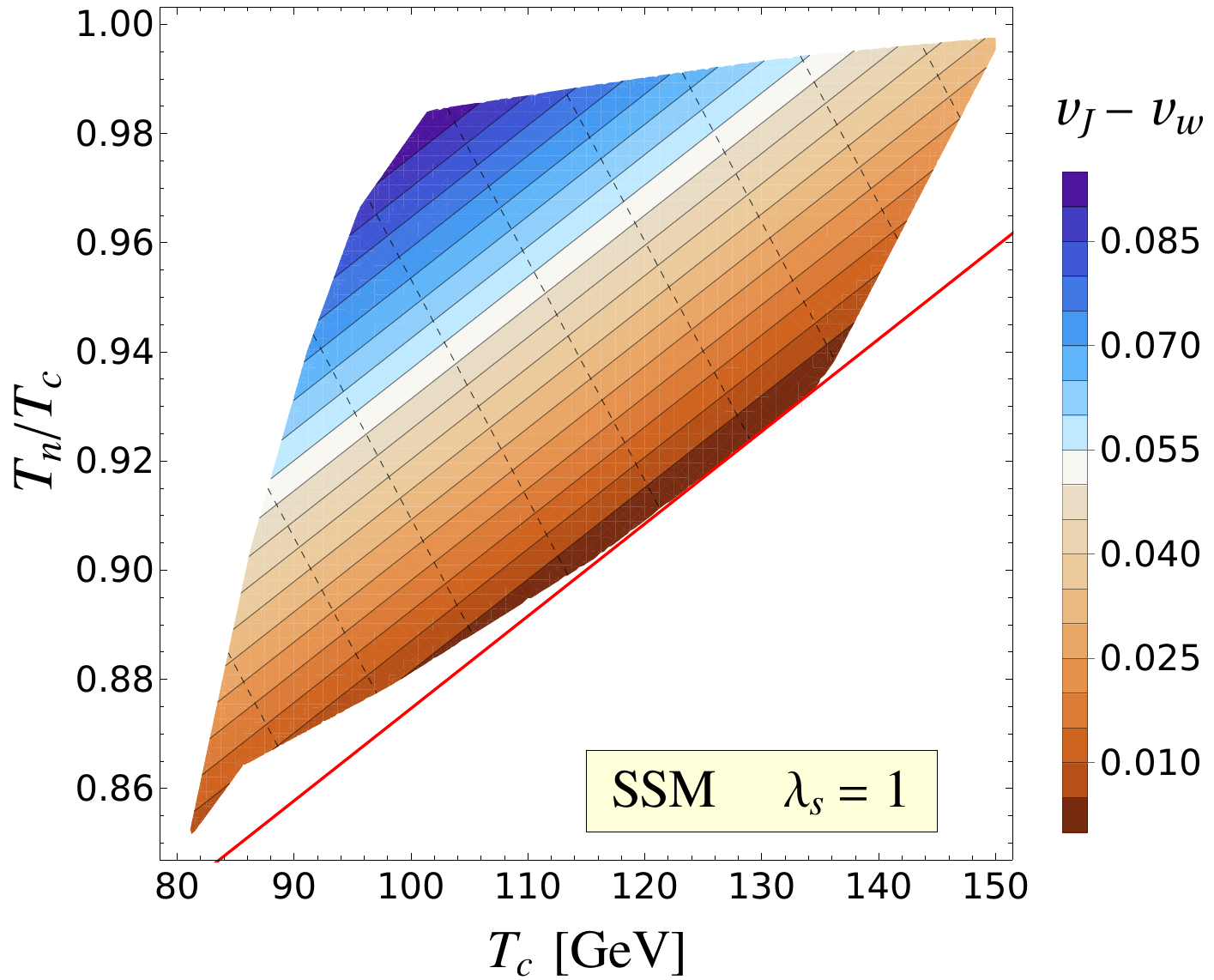}
	\caption{ {\it Left panel}. Contour plot of the wall velocity $v_w$ in  the plane ($T_c$,  $T_n/T_c$)  for the SSM with $\lambda_s=1$.  The straight lines represent the curves of constant $v_w$ for the fitting function in Table \ref{table}. {\it Right panel}. Contour plot of ($v_{_J}-v_w$).  The straight  lines represent the curves of constant $v_{_J}-v_w$ obtained using linear fitting  functions for both $v_w$ and $v_{_J}$ (see Table \ref{table}).   
    The dashed lines represent curves of constant $v_{_J}$ ($v_{_J}$ decreases going from the left to the right part of the coloured region). On the red line, $v_w=v_{_J}$, as explained in the text.}
	\label{SSMCorr}
\end{figure}

Moving forward, we now study the dependence of the bubble wall velocity on the PhT parameters. For instance, in the left panel of Fig.\,\ref{SSMCorr} we show the wall velocity as a function of the critical temperature $T_c$ and of the ratio $T_n/T_c$, that provides a measure of the amount of supercooling related to the transition. To better appreciate the outcome, we have excluded from the plot a small portion of space with weak transitions ($v_w\lesssim 0.54$), that, due to the near-flatness of the potential, contains larger numerical errors and introduce small fluctuations in the figure. The points excluded all collapse on the upper left part of the plot. 

The figure displays a particularly clear dependence of $v_w$ on $T_c$ and $T_n/T_c$, that is very well approximated by a linear function, which is given in Table \ref{table}. 
The velocity shows a tendency to decrease as the supercooling decreases, that is for larger $T_n/T_c$, and to increase with the critical temperature. The first feature amounts to the wall being faster for stronger transitions, which is quite intuitive. The second one can be understood on the same ground as follows (using the quadratic approximation for the potential, Eq.~\eqref{pot quadratic}, as a guide). On a line of constant $T_n/T_c$, an increase in the critical temperature corresponds to an increase in the difference $T_c-T_n$. At temperatures such that non-trivial minima exist in both field directions, the cancellation between $\mu_h^2$ and $c_h T^2$ sets the scale of the minimum $\bar h^2(T)$ (similarly for $\bar s^2(T)$), with the potential in the vacuum being quadratically sensitive to this cancellation, $V(\bar h(T),0,T)\sim - (\mu_h^2+c_h T^2)^2/\lambda_h$, $V(0,\bar s(T),T)\sim -(\mu_s^2+c_s T^2)^2/\lambda_s$. At $T_c$, both $\mu_h^2+c_h T_c^2$ and $\mu_s^2+c_sT_c^2$ are negative, so $-\mu_h^2>c_h T_c^2$, $-\mu_s^2> c_s T_c^2$. The larger the excursion from $T_c$ to $T_n$, the more $\mu_h^2+c_h T_n^2$ and $\mu_s^2+c_s T_n^2$ tend to $\mu_h^2$ and $\mu_s^2$, respectively, and the potential difference between the two vacua $\Delta V$ at $T=T_n$ grows and approaches its zero-temperature value. We then understand that an increase in $T_c$ with fixed $T_n/T_c$ corresponds to an increase in $\Delta V$ at $T=T_n$, and thus an increase in the strength of the transition. 

In the right panel of Fig.\,\ref{SSMCorr}, we show the difference $v_{_J}-v_w$ in the ($T_c,T_n/T_c)$ plane, together with the lines of constant $v_{_J}$ (dashed lines). The isolines of $v_w$ in the left panel of the figure and of $v_{_J}$ (right panel, dashed lines) are nearly orthogonal. The limit $v_w\to v_{_J}$ from below is realised when two lines with same value (of $v_w$ and $v_{_J}$, respectively) cross each other. This gives a boundary for deflagrations, and more generally for steady-state solutions in LTE, and is represented by the red line in the figure. With our fitting functions, the condition $v_w \lesssim v_{_J}$ in terms of $T_c$ and $T_n/T_c$ is given in Table \ref{table}, 
that is an upper boundary $T_n/T_c|_{min}$ on the amount of supercooling. 

\begin{figure}
    \centering
    \includegraphics[width=0.45\linewidth]{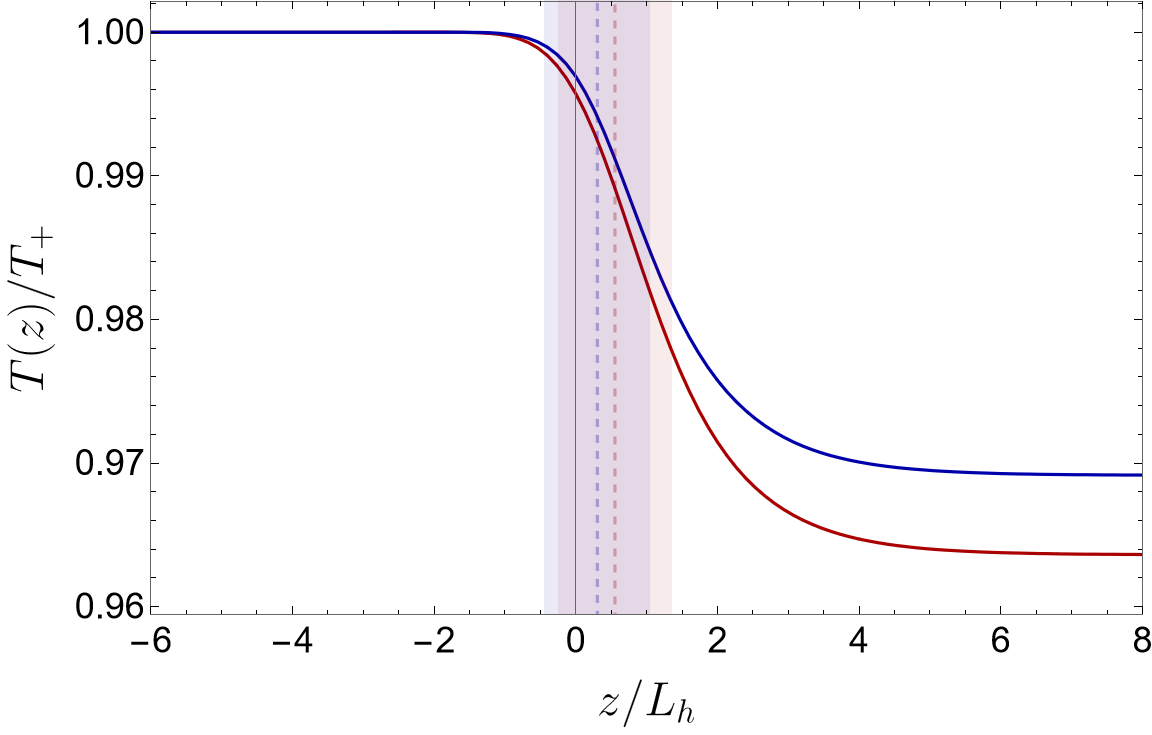} \,\,\,\,\,\,   \includegraphics[width=0.45\linewidth]{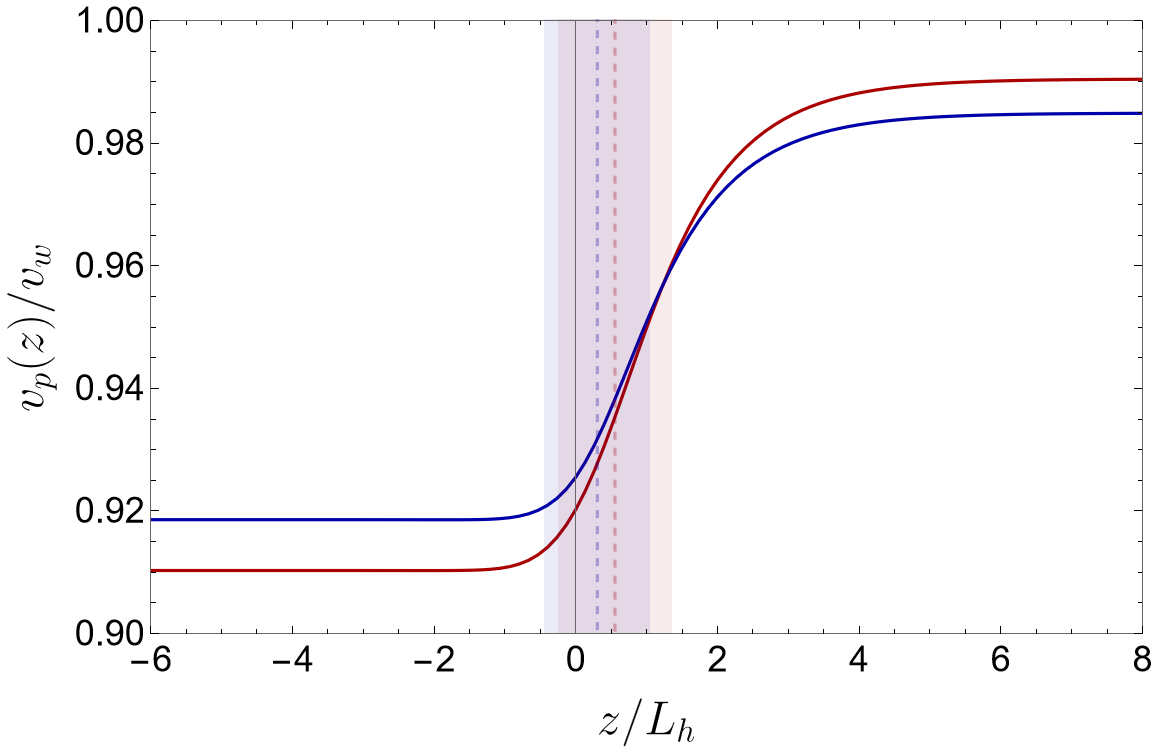}
    \caption{{\it Left panel}. Plasma temperature profile normalised by $T_+$, $T(z)/T_+$ as a function of $z/L_h$ in the SSM with $\lambda_s=1$ for two different benchmark points: BP1 ($m_s=105$ GeV, $\lambda_{hs}=0.39$, red curve) and BP2 ($m_s=131$ GeV, $\lambda_{hs}=0.47$, blue curve). The coordinate $z=0$ indicate the $h$ bubble wall, with the horizontal axis pointing towards the inside of the bubble. Dashed lines locate the centre of the $s$-profile, with the band indicating the width of the latter. {\it Right panel}. Plot of the plasma velocity normalized by the wall velocity, $v_p(z)/v_w.$ The colours are for the same benchmark points as for the left panel. }
\label{fig: plasma profiles}
\end{figure}

Before closing this section, let us turn our attention to the plasma. Its temperature and velocity profiles $T(z)$ and $v_p(z)$, normalised by $T_+$ and $v_w$ respectively, are shown in Fig.\,\ref{fig: plasma profiles} for two benchmark points, namely $(m_s / \text{GeV},\lambda_{hs})=(105, 0.39)$ (red curves) and $(m_s / \text{GeV},\lambda_{hs})=(131, 0.47)$ (blue curves).  In the plots, the axis $z=0$ represents the location of the $h$ planar domain wall, with positive (negative) values of $z$ indicating the region inside (outside) the $h$-bubble. The coordinate is normalised by $L_h$, so that on the axis the $h$ width is equal to one. The centre of the $s$-bubble is indicated by dashed lines, with the coloured bands showing the corresponding widths. Besides the single profiles $v_p(z)$ and $T(z)$, our analysis finds that the product $\gamma(z) T(z)$ is constant. It was recently argued in \cite{Ai:2021kak} that, in local thermal equilibrium, entropy conservation enforces this condition. The latter can be used as an additional hydrodynamic equation alongside \eqref{matching eq}.

Coming back to the profiles themselves, Fig.\,\ref{fig: plasma profiles} shows a decreasing temperature, and an increasing velocity, as one moves towards the domain wall from the symmetric phase, as expected in the  deflagration/hybrid regime. In particular, the profiles are nearly-flat far from the wall, and the plasma is approximately unperturbed up until the bubble front, with a significant variation only inside the bubble and in the vicinity of the wall. It is worth to remind here that the gradient of the temperature is fully responsible for the friction force exerted by the plasma on the expanding domain wall in local equilibrium, as per Eq.\,\eqref{Ptot}, and the latter can be easily determined from the knowledge of the profile $T(z)$. This also means that the contribution to the friction comes, to a large extent, from the region right behind the bubble wall.

\subsubsection*{SSM, $\lambda_s=2$}

To check the dependence on the singlet self coupling, we repeated the analysis for $\lambda_s=2$. To keep the discussion more concise, here and in the following we refrain from presenting the results for all the quantities, as we did in the previous section, but concentrate only on the main ones. 

\begin{figure}[t]
\centering
\includegraphics[width=0.49\linewidth]{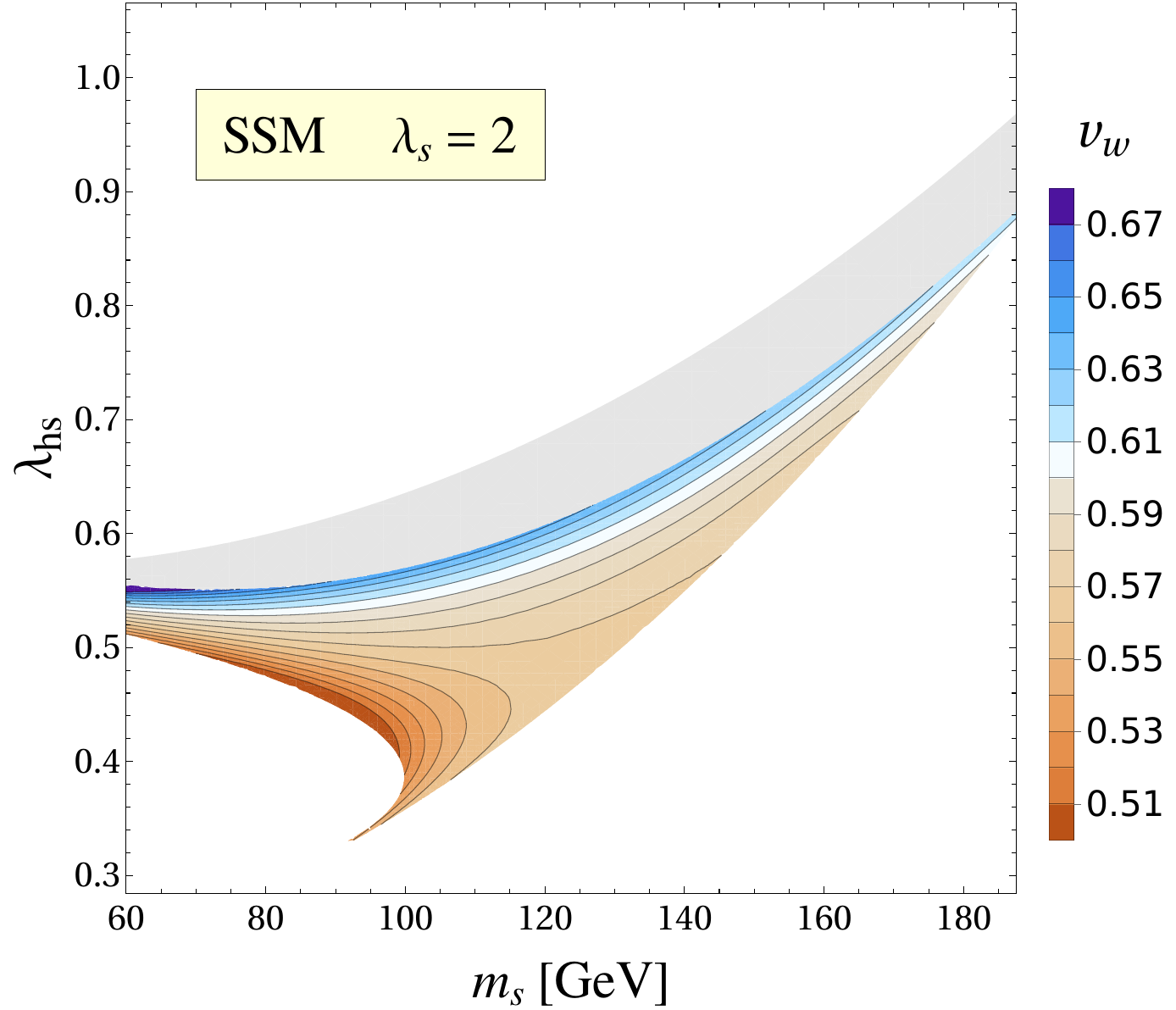}\,
\includegraphics[width=0.49\linewidth]{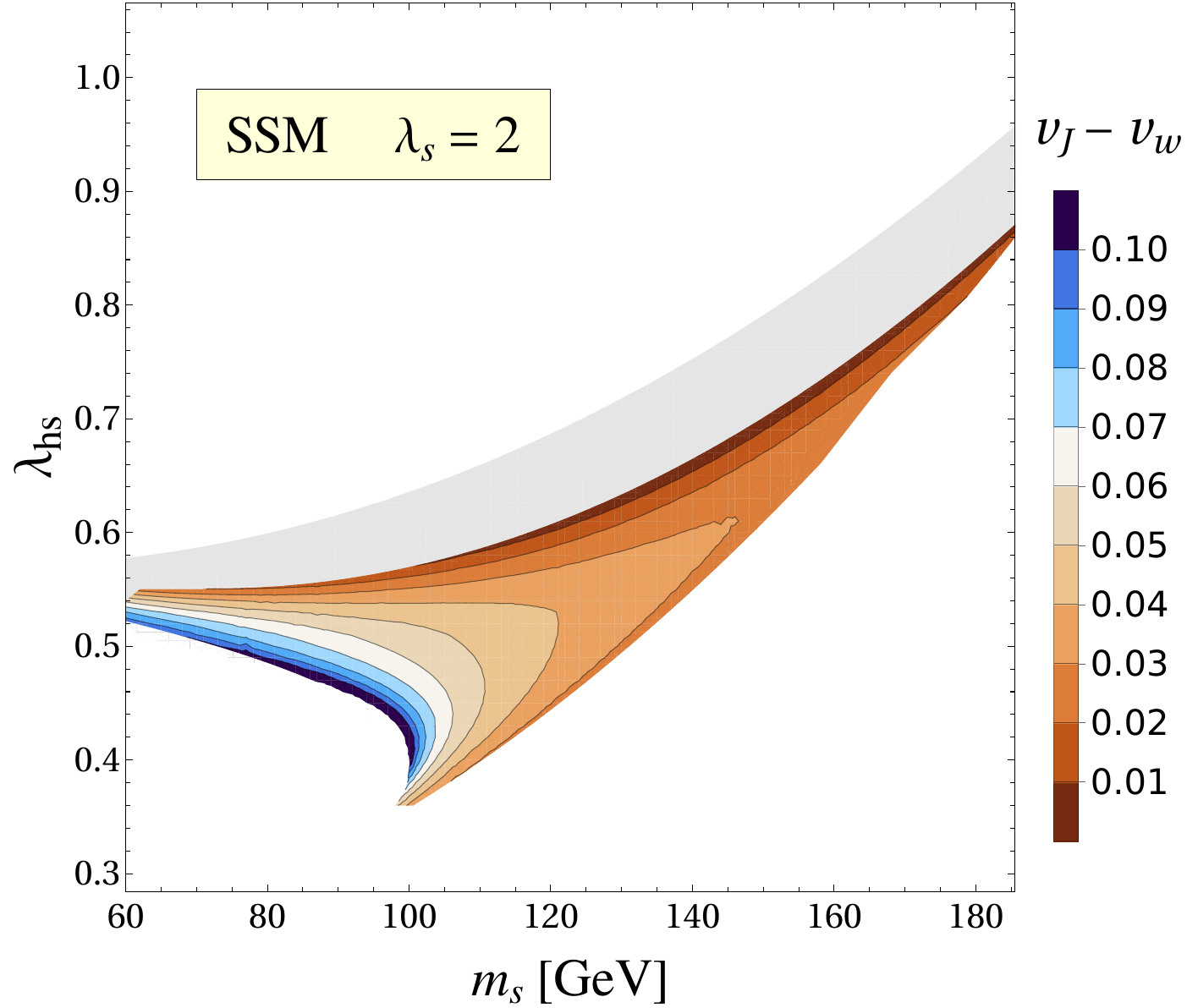}
\caption{ Contour plots of the bubble wall velocity ({\it left panel}) and of the difference between the Jouguet and bubble wall velocities ({\it right panel}), in the $(m_s,\lambda_{hs})$ plane for the SSM with $\lambda_s=2$.  }
\label{fig: vw SSM param space ls=2}
\end{figure}

The results for the wall velocity and the difference $v_{_J}-v_w$ in the $m_s-\lambda_{hs}$ plane are shown in Fig.\,\ref{fig: vw SSM param space ls=2}. It can be immediately appreciated that the dependence of both $v_w$ and $v_{_J}-v_w$ on $m_s$ and $\lambda_{hs}$ is quite similar to the one in Fig.\,\ref{ContourV} and \ref{fig: vJ minus v & cs}, respectively. The overall range of variation is also very similar, with the weakest transitions displaying a velocity $v_w\sim 0.5$, and the strongest ones, in the upper left corner of the plot, having $v_w\sim 0.67$. In agreement with the arguments of Section \ref{sec: analytic}, only deflagration solutions were found in this case too, and only in a subset of the entire two-step parameter space. No LTE solution was found in the grey region of Fig.\,\ref{fig: vw SSM param space ls=2}. The right panel of Fig.\,\ref{fig: vw SSM param space ls=2} shows that the upper boundary of the region with a steady-state solution is reached as $v_w$ approaches $v_{_J}$ from below.       

\begin{figure}[t!]
    \centering
    \includegraphics[width=0.48\linewidth]{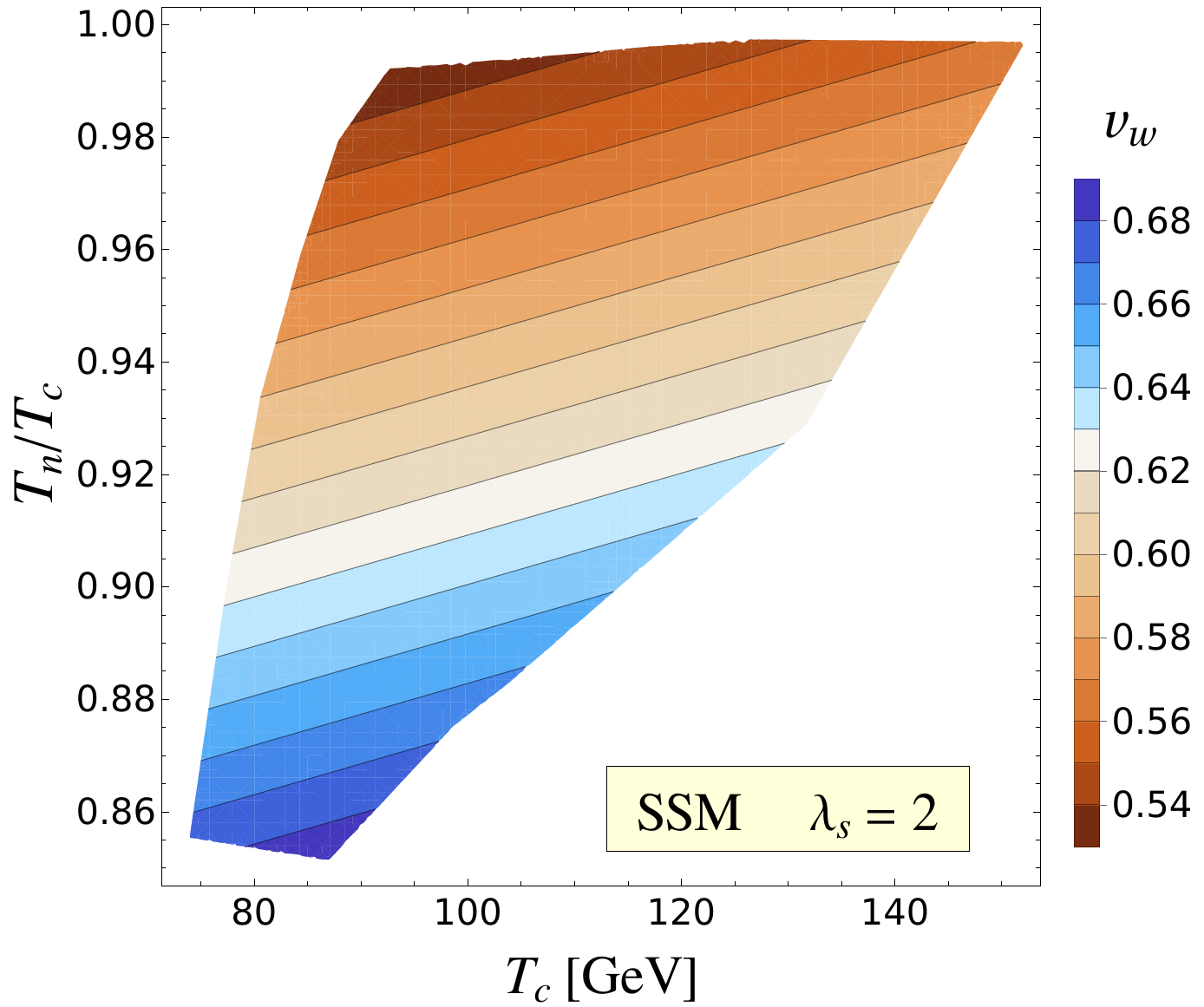} 
    \,\,\includegraphics[width=0.48\linewidth]{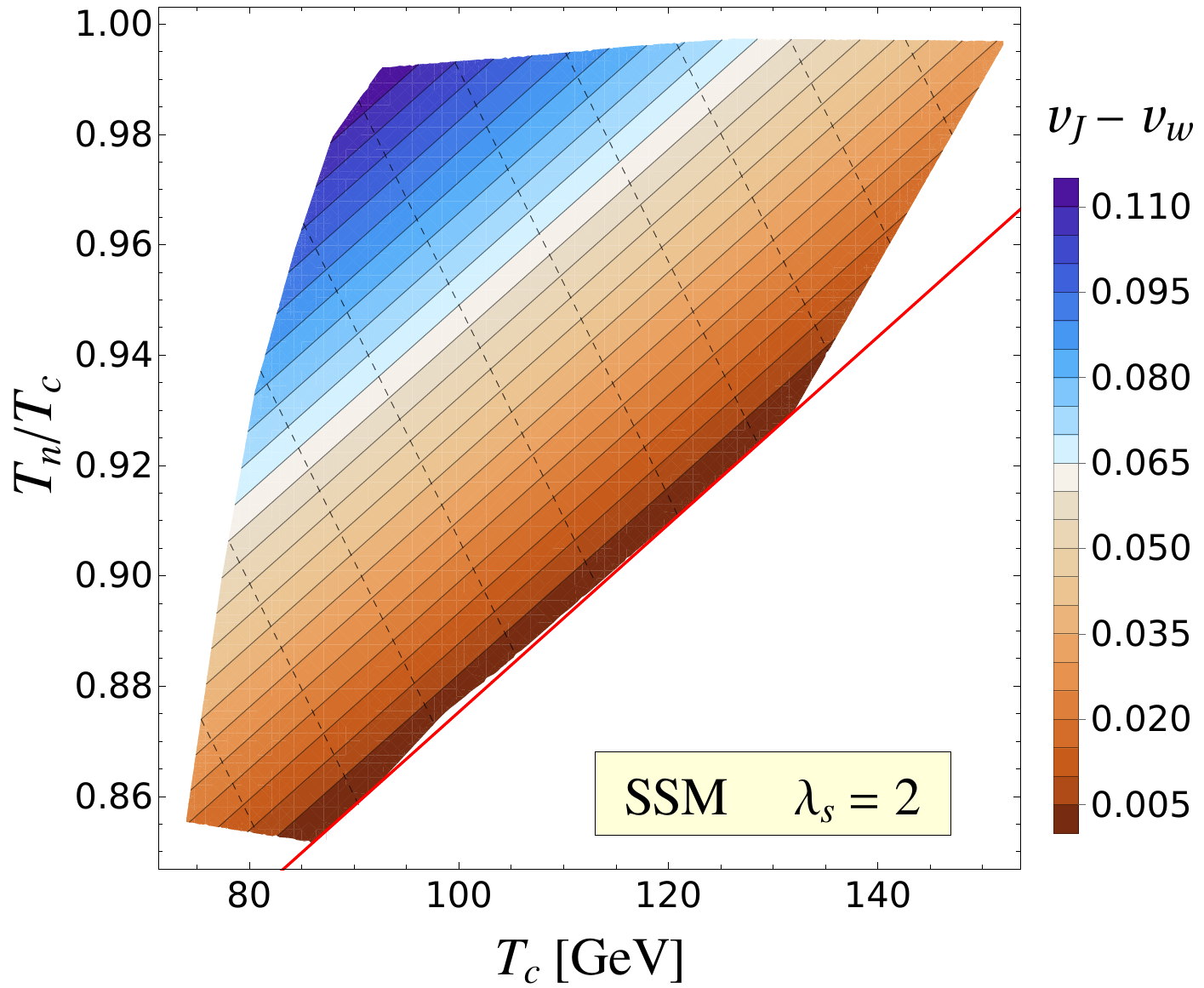}
    \caption{{\it Left panel}. Contour plot of the wall velocity $v_w$ in the $(T_c,T_n/T_c)$ plane for the SSM with $\lambda_s = 2$.   
{\it Right panel}. Contour plot of $v_{_J}-v_w$, with dashed lines representing curves of constant $v_{_J}$. On the red line, $v_w=v_{_J}$.}
    \label{fig: vw T-space ls2}
\end{figure}

The dependence of the velocity on the parameters of the transition $T_c$ and $T_n/T_c$ is shown in  the left panel of Fig.\,\ref{fig: vw T-space ls2}. Again the results are reproduced to a very good approximation by a linear fit, the coefficients of which are reported in Table \ref{table}. The agreement of the numerical fit to the one obtained for $\lambda_s=1$ can hardly be missed. The largest difference between the coefficients is below the percent level. This can be taken as indicating a very mild dependence of the bubble wall velocity, and more generally of the dynamics of the phase transition, on the self-coupling of the additional scalar.

The behaviour of the difference $v_{_J}-v_w$ in the $(T_c,T_n/T_c)$ plane is shown in the right panel of Fig.~\ref{fig: vw T-space ls2}. The similarity to the $\lambda_s=1$ case is again evident: the iso-lines of the Jouguet velocity (dashed lines) are nearly orthogonal to those of $v_w$, and this again leads to an ``upper" boundary for deflagrations, with the condition $v_w\lesssim v_{_J}$ obtained with our linear fits reported in Table~\ref{table}. 
The agreement between the two realisations with different $\lambda_s$ is again below the the percent level.

The results for the other parameters not shown in this section are all very similar to those presented for $\lambda_s=1$.

\subsection{RTSM}

We now proceed with our analysis and consider the Real Triplet extension of the Standard Model. Similar to the previous case, we performed the computation for two benchmark values of the extra scalar self-coupling, $\lambda_\sigma=1$ and $\lambda_\sigma=2$. 

\begin{figure}[t]
    \centering
\includegraphics[scale=0.365]{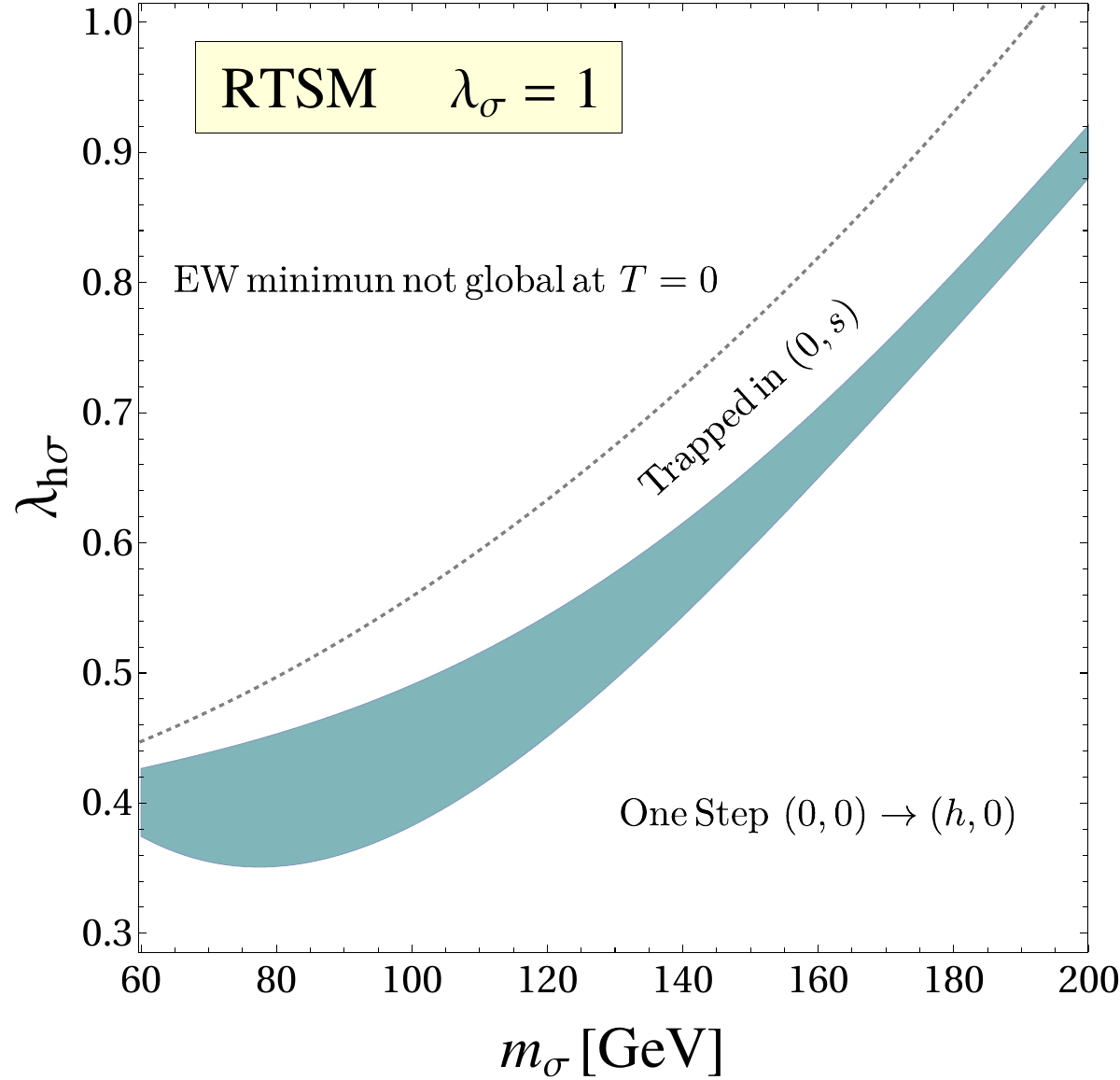}\,
\includegraphics[scale=0.365]{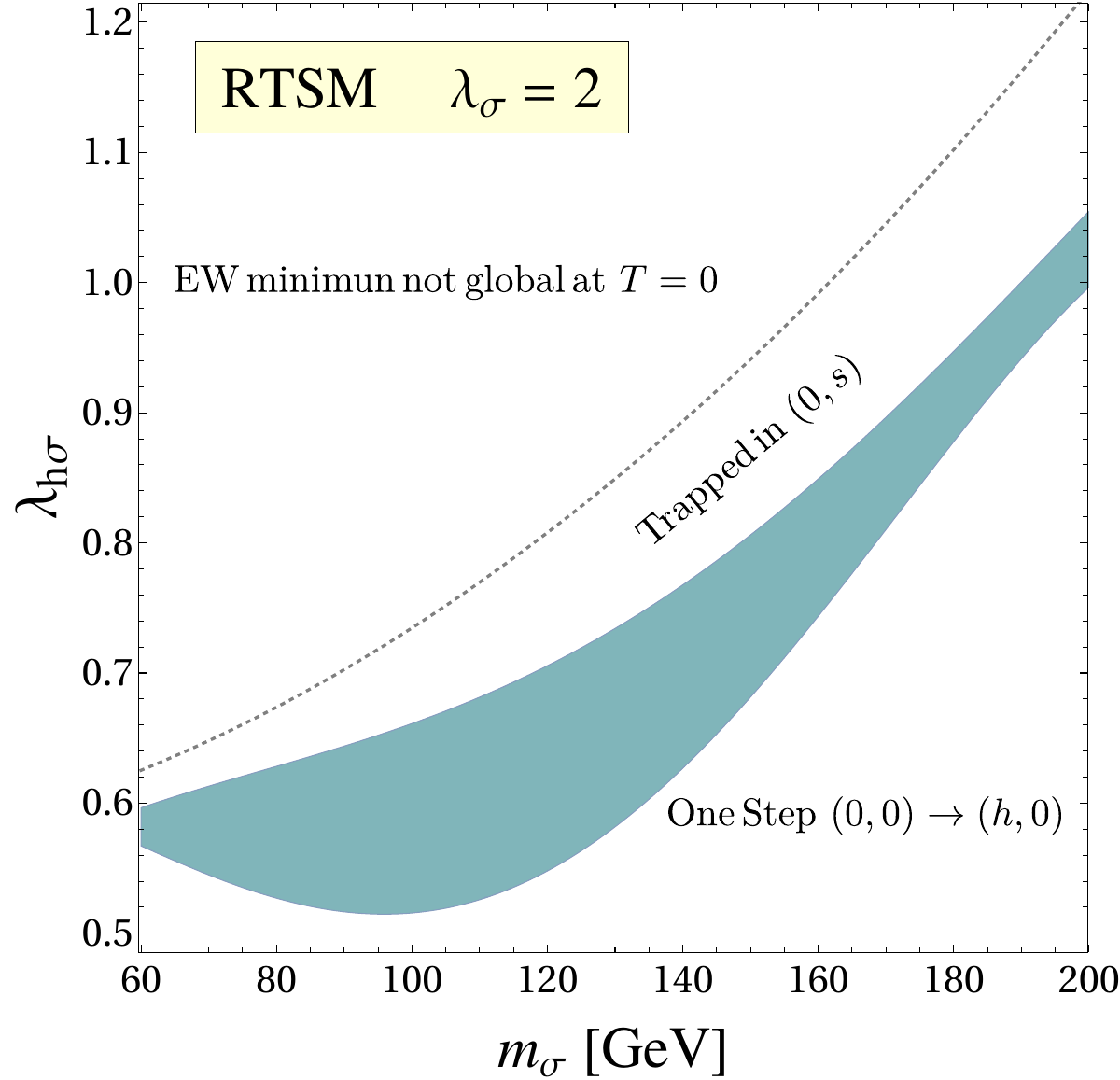}
    \caption{Regions in the plane $(m_\sigma,\lambda_{ h\sigma})$ with corresponding patterns for the EW phase transition in the RTSM 
    for $\lambda_\sigma=1$ ({\it left panel}) and $\lambda_\sigma=2$ ({\it right panel}). In the coloured region the EW phase transition is two-step.}
    \label{fig: param space RTSM}
\end{figure}

The parameter space with a two-step PhT found with \texttt{CosmoTransitions} is given by the coloured regions in Fig.~\ref{fig: param space RTSM} for  $\lambda_\sigma=1$ (left panel) and $\lambda_\sigma=2$ (right panel). As for the SSM, the two-step region gets shifted to larger values of $\lambda_{h\sigma}$ for larger values of the self-coupling $\lambda_\sigma$, and only in a portion of these regions is it possible to find a deflagration solution in local thermal equilibrium. 

The results turn out to be qualitatively model-independent, so that, for both $\lambda_\sigma=1$ and $\lambda_\sigma=2$, we will only show the plots for the wall velocity and the difference between the Jouguet and the bubble wall velocities ($v_{_J}-v_w$).

\begin{figure}
    \centering
    \includegraphics[width=0.49\linewidth]{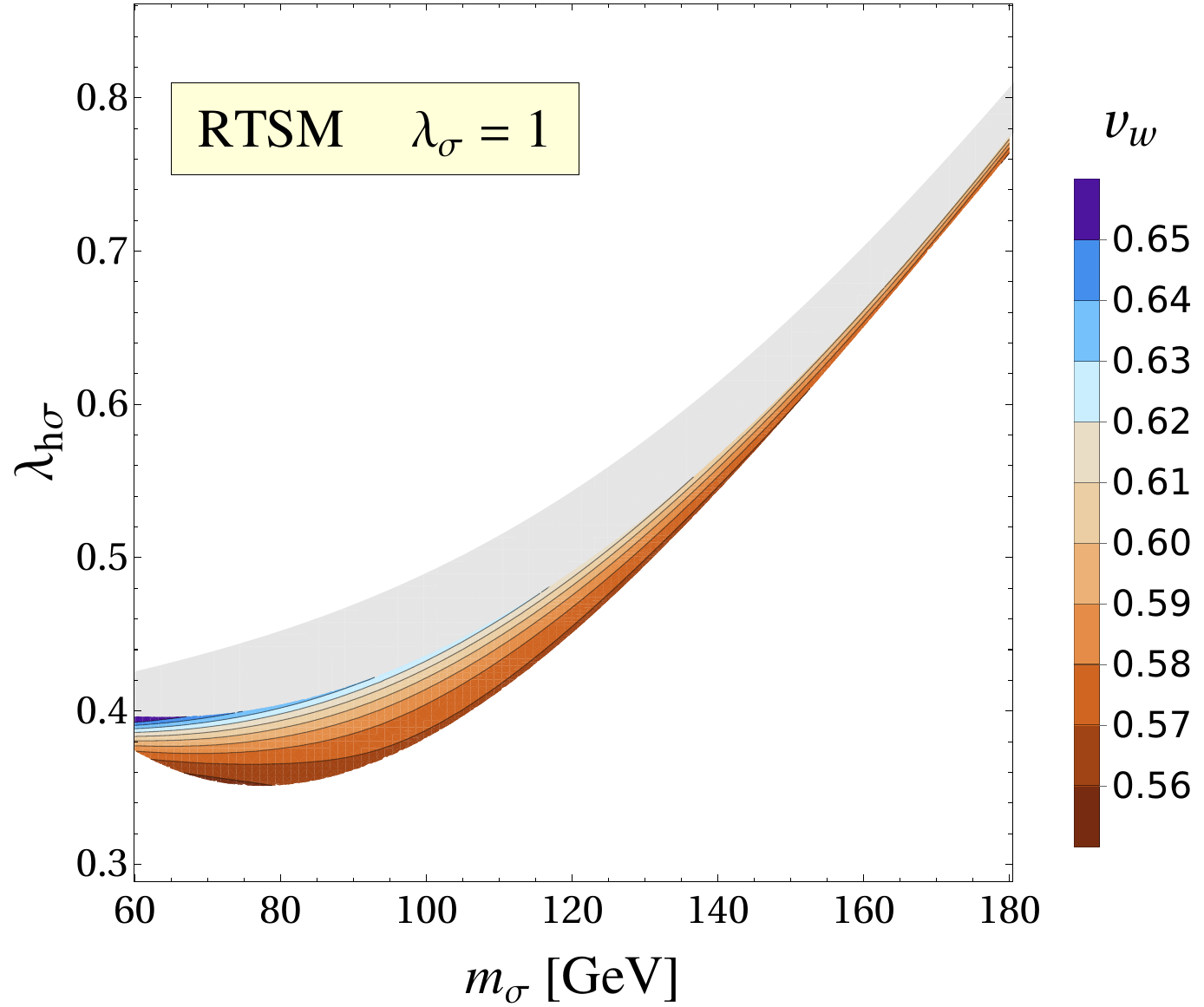}
    \includegraphics[width=0.49\linewidth]{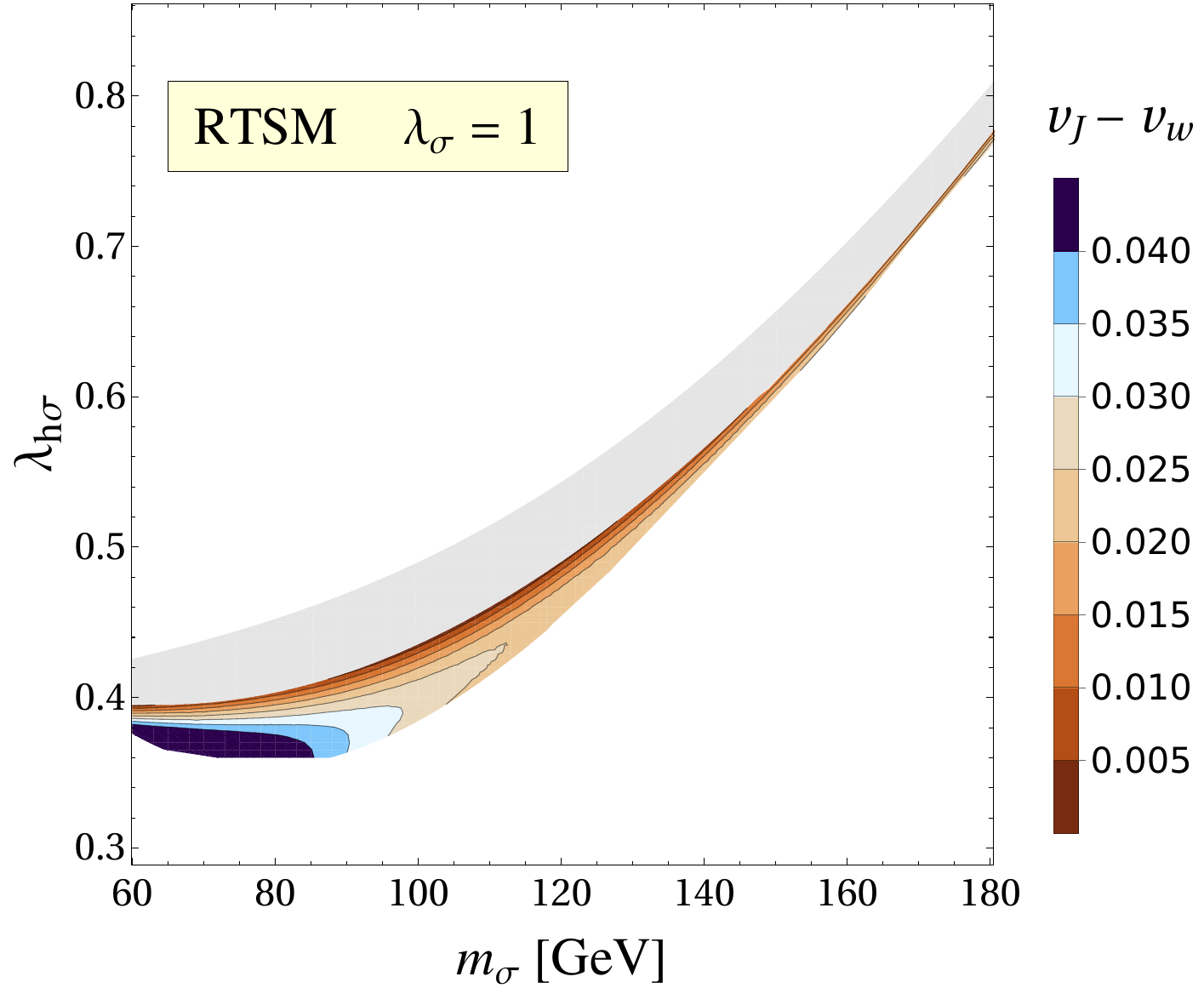}
    \caption{{\it Left panel}. Contour plot of $v_w$ in the ($m_\sigma$, $\lambda_{h\sigma}$) plane for the RTSM with $\lambda_\sigma=1$. {\it Right panel}.  Contour plot of the difference between the Jouguet and the bubble wall velocities ($v_{_J}-v_w$).}
    \label{fig: vw RTSM ls1 param space}
\end{figure}

\begin{figure}
    \centering
\includegraphics[width=0.49\linewidth]{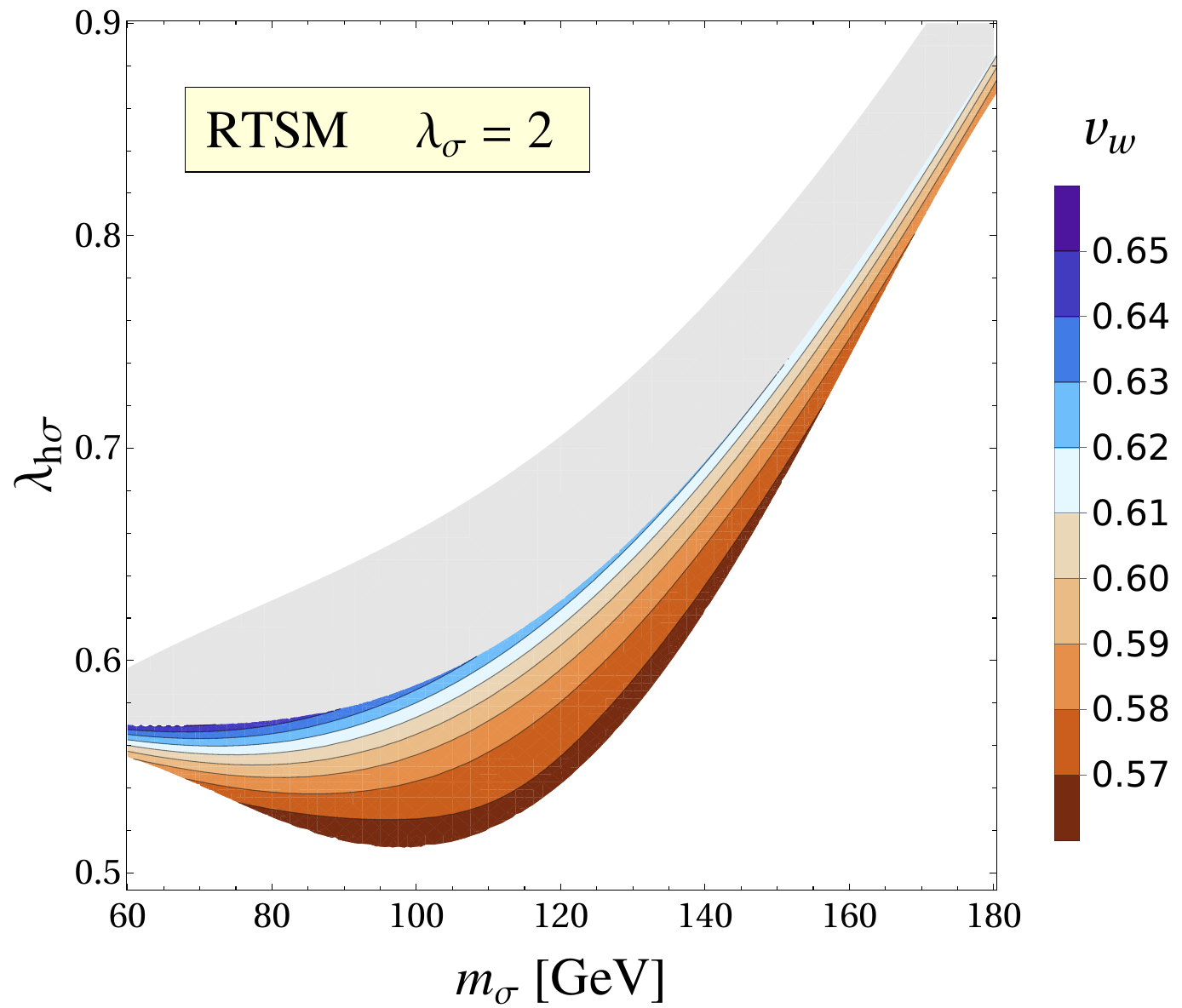}
\includegraphics[width=0.5\linewidth]{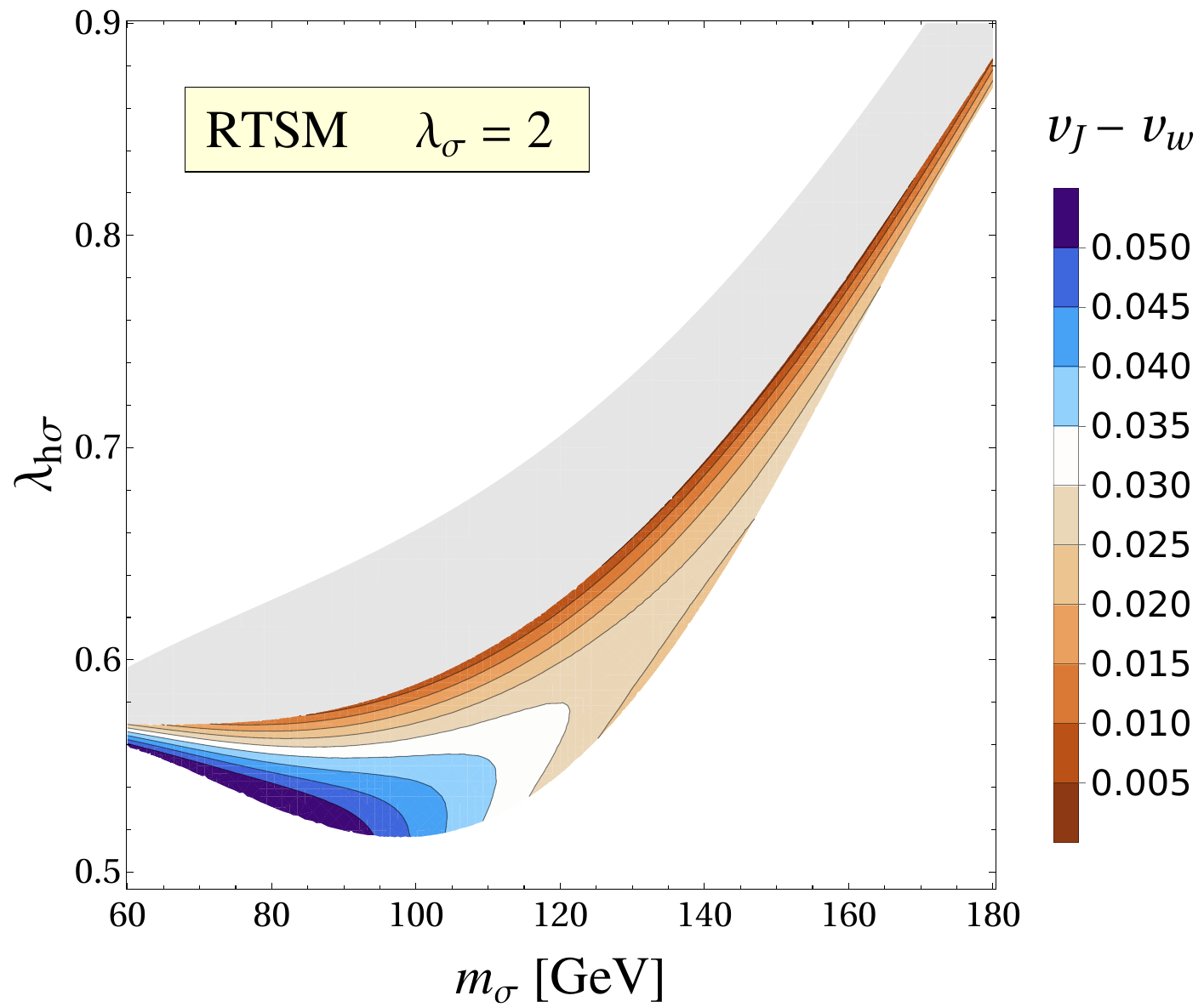}
\caption{{\it Left panel}. Contour plot of $v_w$ in the ($m_\sigma$, $\lambda_{h\sigma}$) plane  for the RTSM with $\lambda_\sigma=2$.  {\it Right panel}.  Contour plot of the difference between the Jouguet and the bubble wall velocities ($v_{_J}-v_w$).}
\label{fig: vw param space RTSM ls2}
\end{figure}

The bubble wall velocity in the ($m_\sigma,\lambda_{h\sigma}$) plane is shown in the left panel of Fig.~\ref{fig: vw RTSM ls1 param space} for $\lambda_\sigma=1$, with the difference between the Jouguet velocity and $v_w$ in the right panel of the same figure. The analogous results for $\lambda_\sigma=2$ are shown in Fig.\,\ref{fig: vw param space RTSM ls2}. As can be seen, they have the same features as those found in the analysis of the SSM, with $v_w$ and $v_{_J}$  increasing from the lower to the upper boundary of the deflagration parameter space, where $v_w$ approaches $v_{_J}$. The largest values of $v_w$ and $v_{_J}$ are found in the upper left corner of the plot, with the smallest ones in the lower left corner. Again, no detonation solution with $v_w>v_{_J}$ was found in the grey regions of Fig.~\ref{fig: vw RTSM ls1 param space} and \ref{fig: vw param space RTSM ls2}.

\begin{figure}
    \centering    \includegraphics[width=0.48\linewidth]{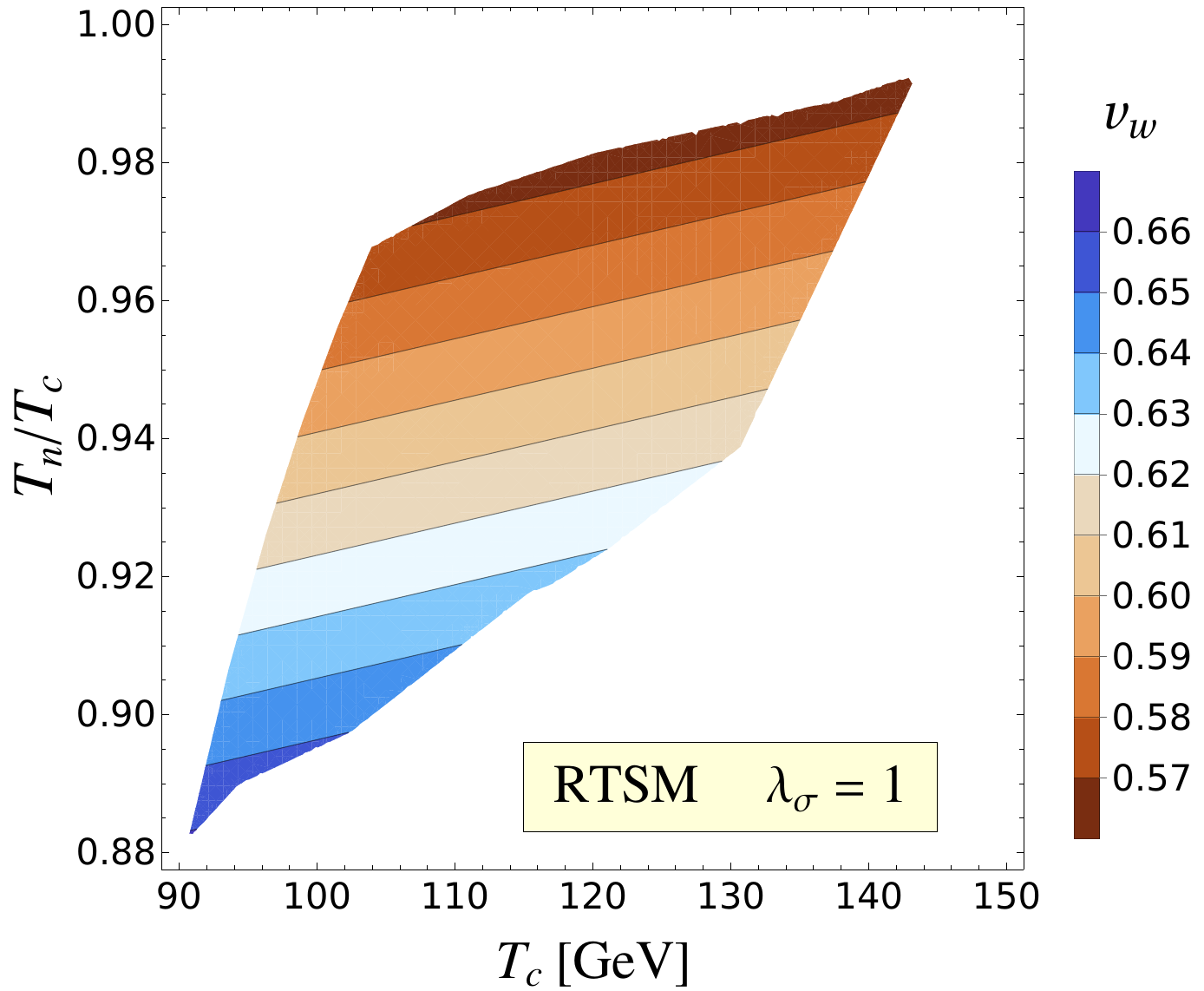} \,\, 
\includegraphics[width=0.48\linewidth]{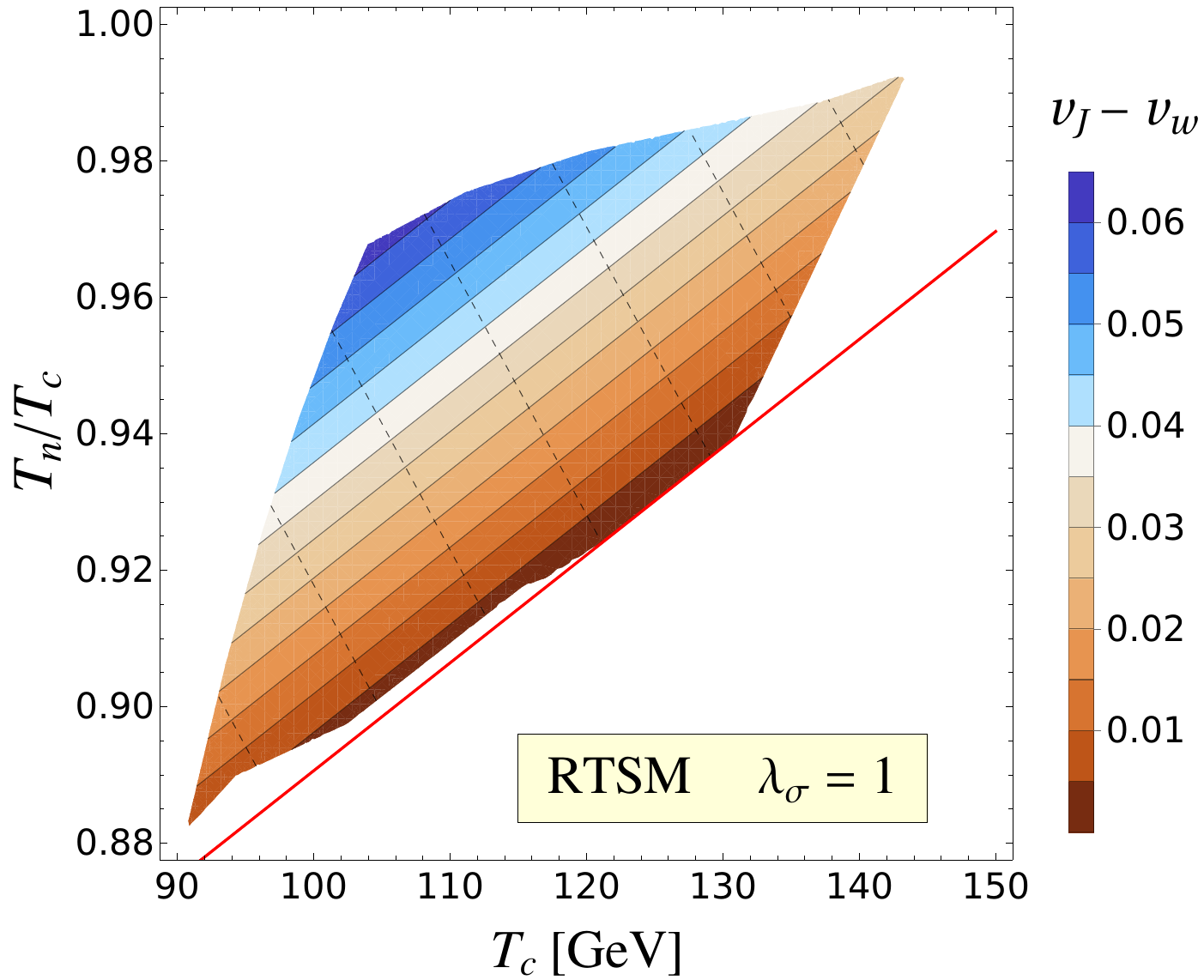}
    \caption{{\it Left panel}. Contour plot of the wall velocity $v_w$ in the $T_c-T_n/T_c$ space for the RTSM with $\lambda_\sigma = 1$.  
{\it Right panel}. Contour plot of $v_{_J}-v_w$. Dashed lines are for constant $v_{_J}$. On the red line, $v_w=v_{_J}$.}
    \label{fig: vw RTSM ls1 Tspace}
\end{figure}

\begin{figure}
    \centering
    \includegraphics[width=0.48\linewidth]{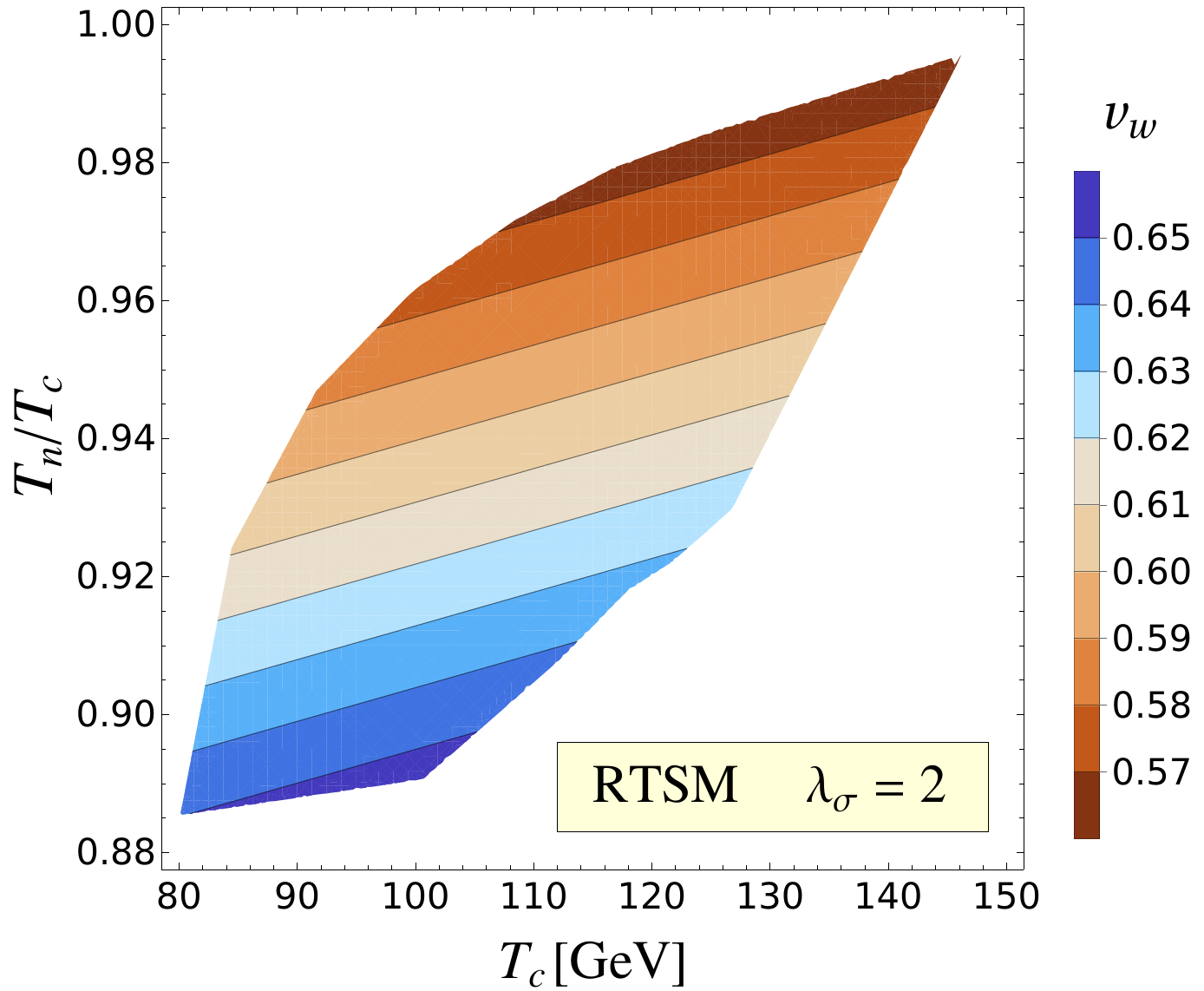} \,\, \includegraphics[width=0.48\linewidth]{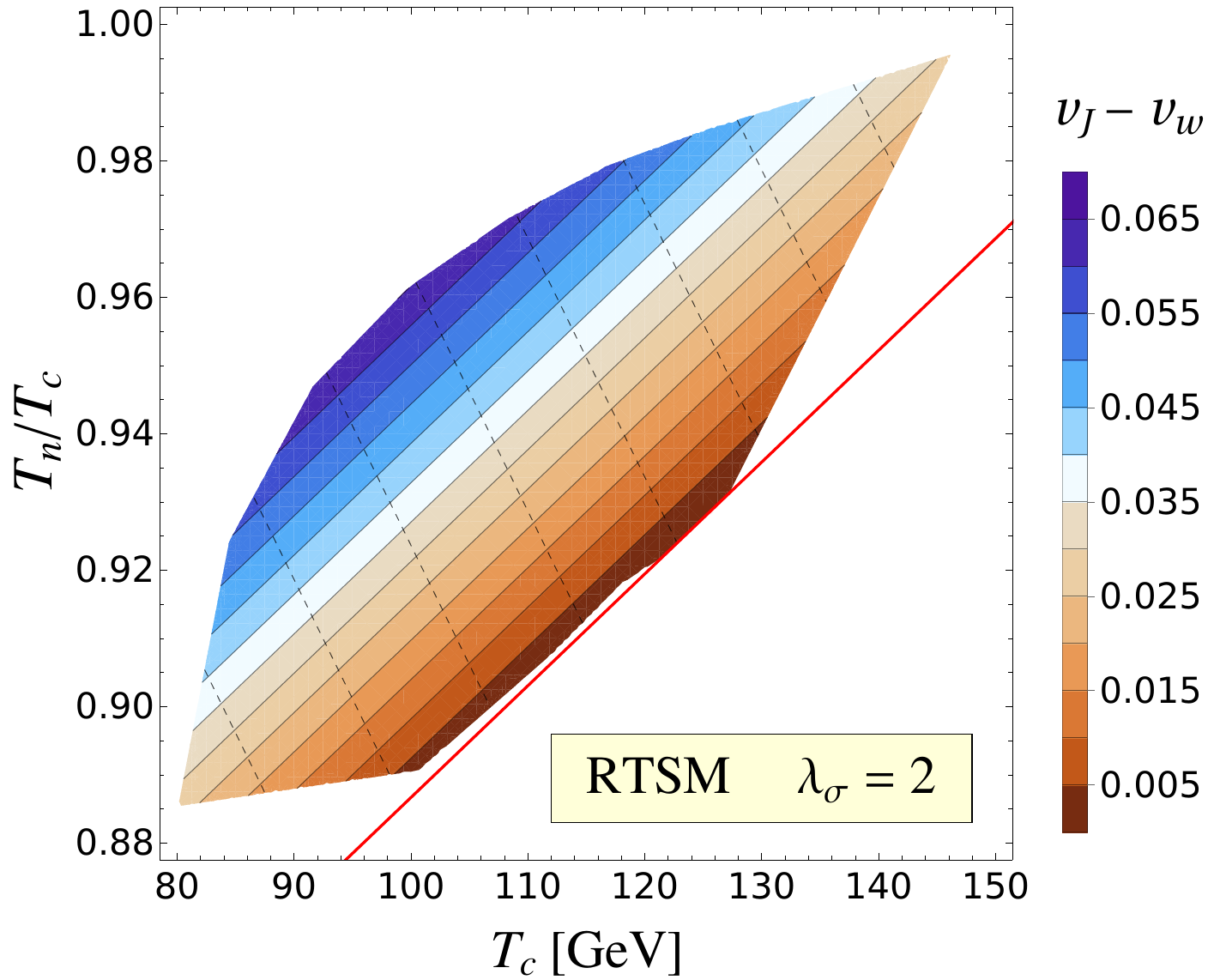}
    \caption{{\it Left panel. }Contour plot of the wall velocity $v_w$ in the ($T_c,\,T_n/T_c$) plane for the RTSM with $\lambda_\sigma = 2$. {\it Right panel}. Contour plot of $v_{_J}-v_w$. Dashed lines are for constant $v_{_J}$. On the red line, $v_w=v_{_J}$.}
    \label{fig: RTSM ls2 vw T projection}
\end{figure}

A plot of $v_w$ in the $T_c$ - $T_n/T_c$ plane is given in Fig.~\ref{fig: vw RTSM ls1 Tspace} for $\lambda_\sigma = 1$ and in Fig.~\ref{fig: RTSM ls2 vw T projection} for $\lambda_\sigma = 2$. To a good approximation, the velocity shows a linear dependence on the parameters of the transition, in all similar to the one found in the SSM for both $\lambda_s=1$ and $\lambda_s=2$. The fitting functions are given in Table \ref{table}. The contour plots of the difference $v_{_J}-v_w$ are in the right panel of Fig.~\ref{fig: vw RTSM ls1 Tspace} and \ref{fig: RTSM ls2 vw T projection}, together with the lines of constant Jouguet velocity (dashed lines). The features discussed for the SSM are again found here. As in the other case, the lines of constant $v_w$ and $v_{_J}$ lead to an upper bound on the amount of supercooling.
The expression for the fitting functions are given in Table~\ref{table}.

\subsection{IDM}

\begin{figure}
    \centering
    \includegraphics[width=0.485\linewidth]{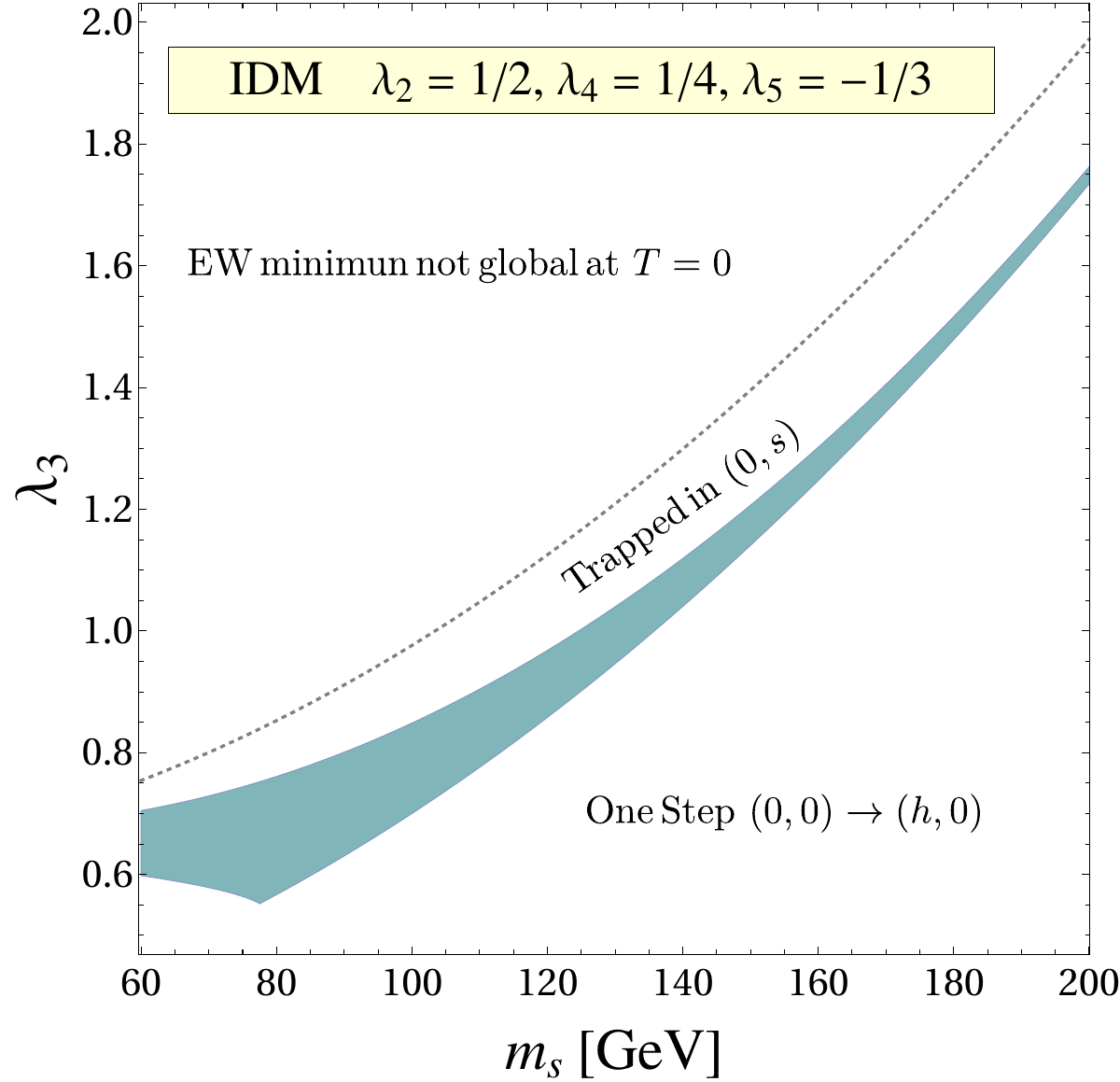} \,\,
    \includegraphics[width=0.485\linewidth]{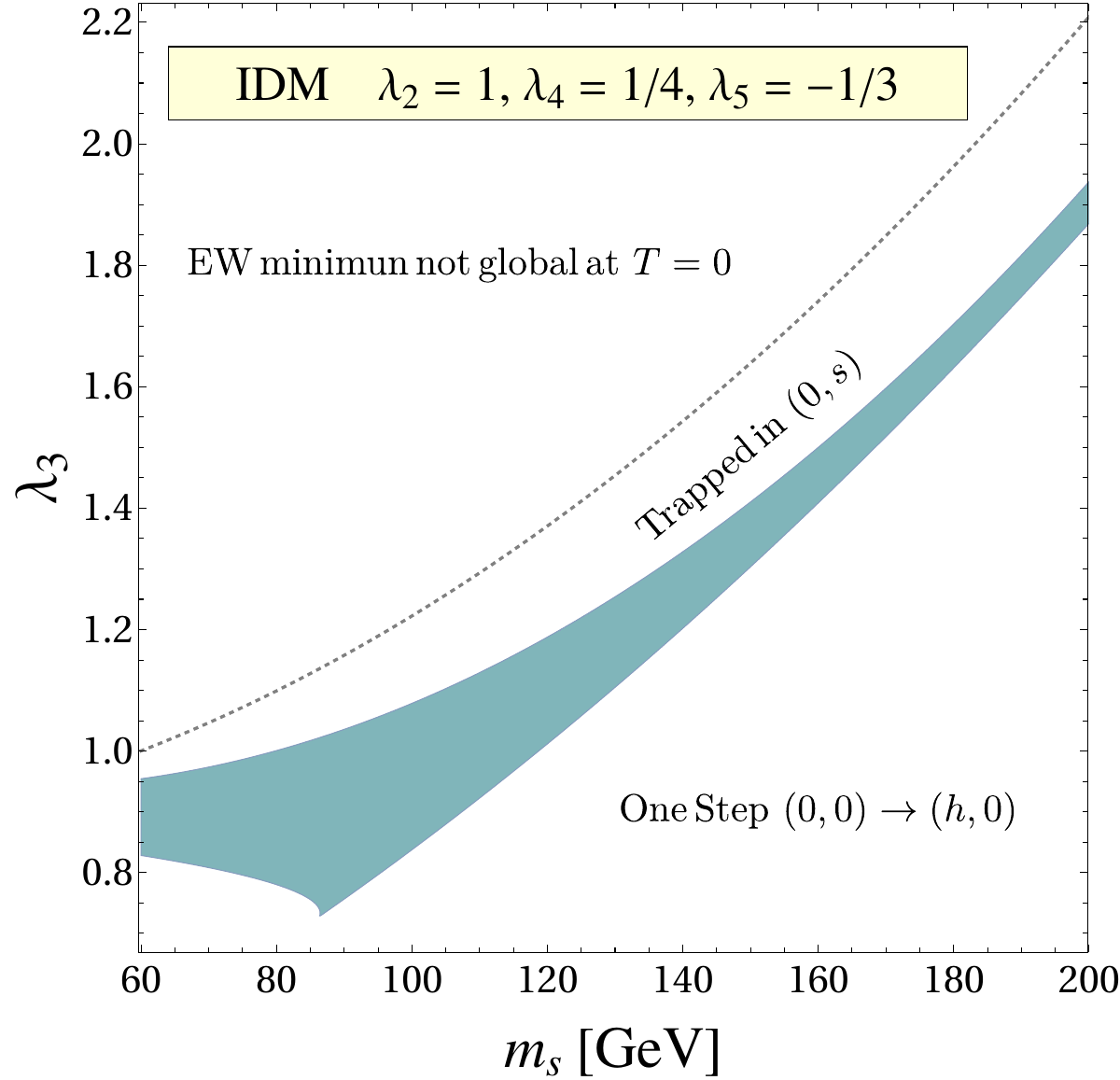}
    \caption{Regions in the plane $(m_s,\lambda_3$) with corresponding patterns for the for the EWPhT 
    in the IDM  
    for $\lambda_4=1/4$, $\lambda_5=-1/3$, $\lambda_2=1/2$ ({\it left panel}) and $\lambda_2=1$ 
    ({\it right panel}). In the coloured region the EWPhT is two-step.}
    \label{fig: IDM param space}
\end{figure}

Finally, we show in this section the results for the scan of the IDM. Similar to the cases above, we fixed the value of the $H_2$ self-coupling. Moreover, since the relevant coupling for the potential of the two neutral CP-even fields is $\lambda_{345}$, we chose to perform the analysis fixing the values of $\lambda_4$ and $\lambda_5$ to $\lambda_4=1/4$ and $\lambda_5=-1/3$, and varying $\lambda_3$. 

The region of parameter space with a two-step phase transition is depicted in Fig.~\ref{fig: IDM param space} for the two choices of the $H_2$ self-coupling $\lambda_2=1/2$ (left panel) and $\lambda_2=1$ (right panel), with the second one involving larger values of $\lambda_3$, as expected.

\begin{figure}
    \centering
    \includegraphics[width=0.485\linewidth]{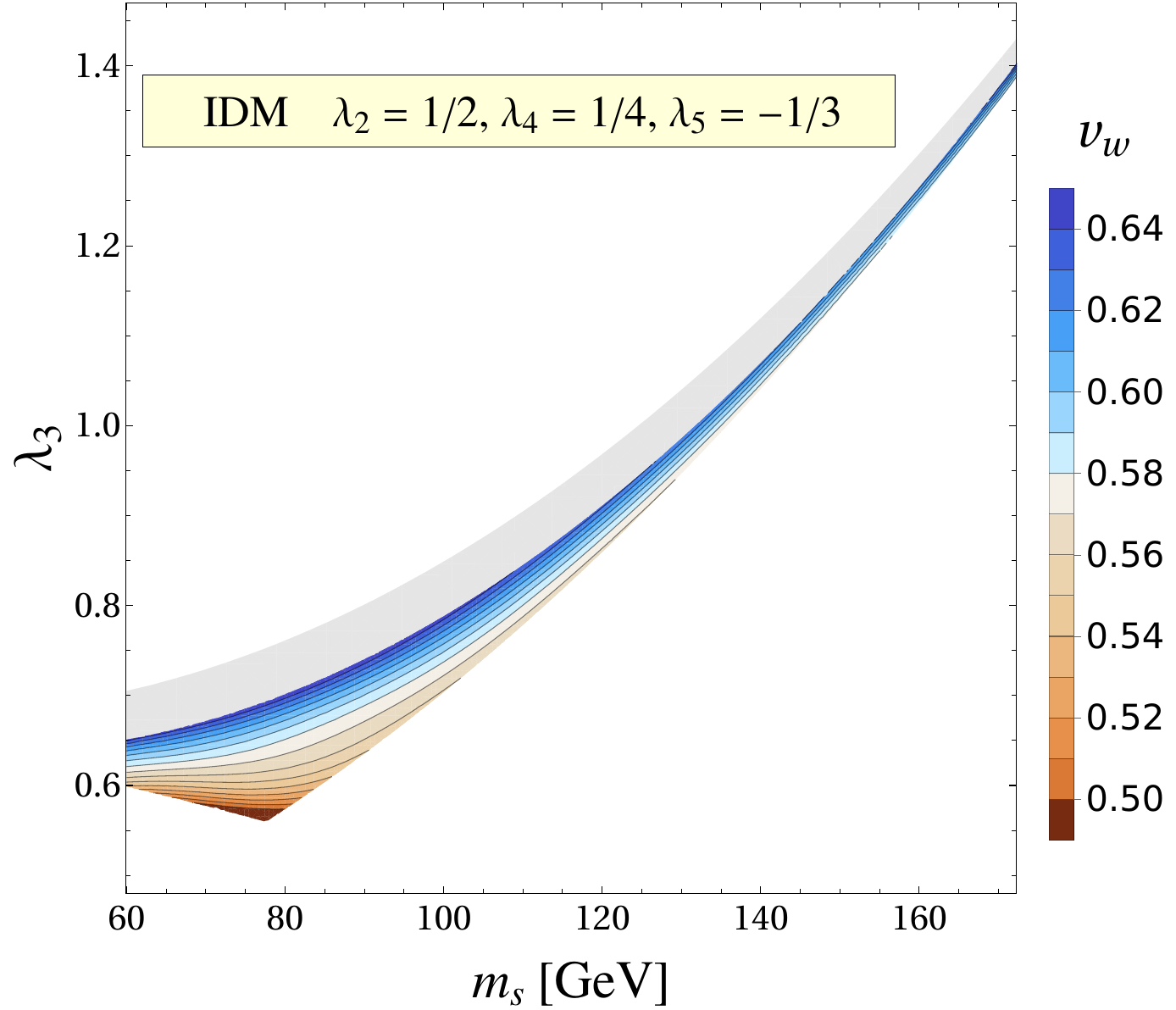}\,\,\includegraphics[width=0.49\linewidth]{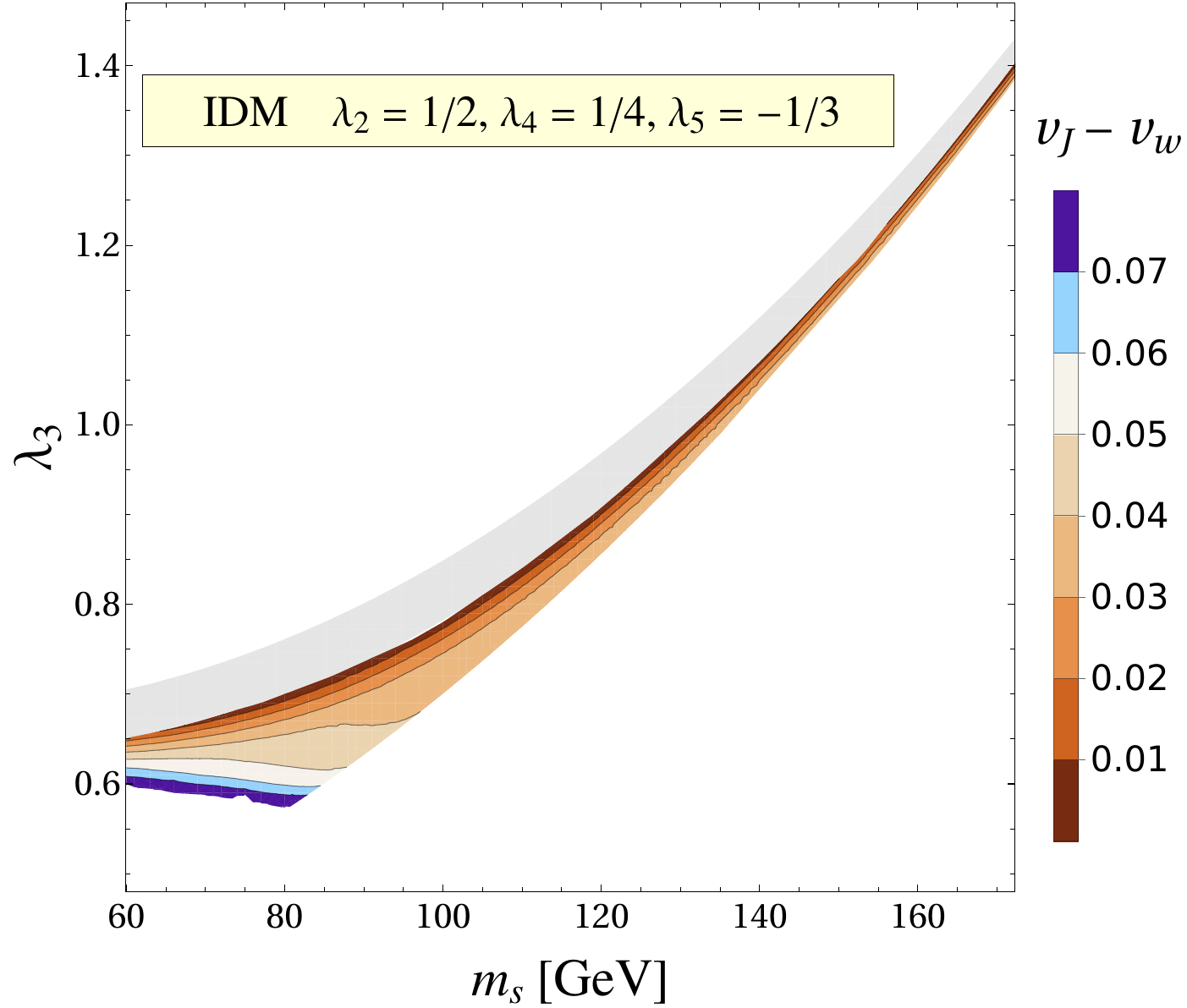}
    \caption{{\it Left panel}. Contour plot of $v_w$ in the ($m_s$, $\lambda_3$) plane for the IDM with $\lambda_2=1/2$, $\lambda_4=1/4$, $\lambda_5=-1/3$. {\it Right panel}.  Contour plot of the the difference $v_{_J} - v_w$.}
    \label{fig: IDM vw paramspace}
\end{figure}

\begin{figure}
    \centering
    \includegraphics[width=0.48\linewidth]{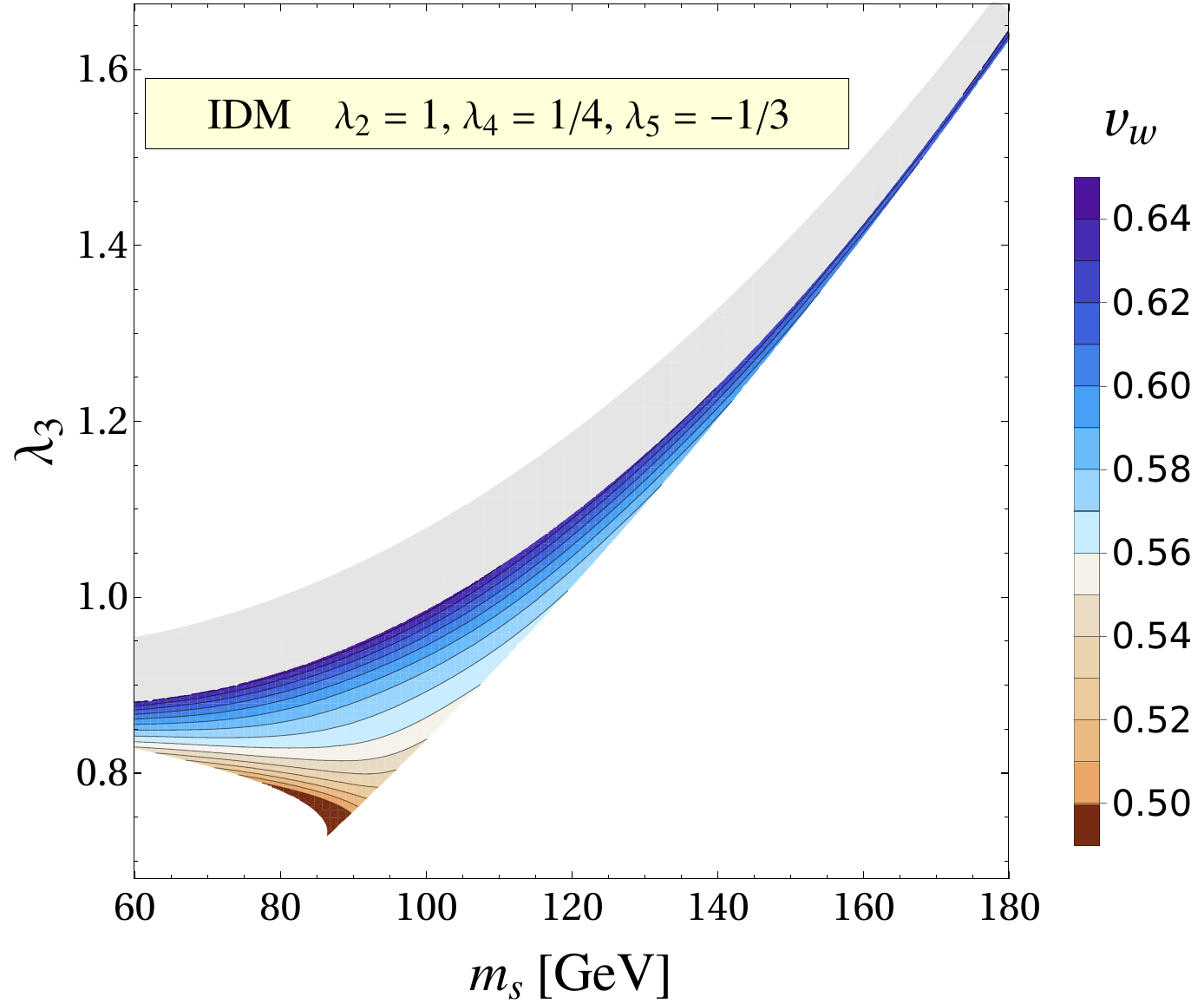}\,\,\includegraphics[width=0.48\linewidth]{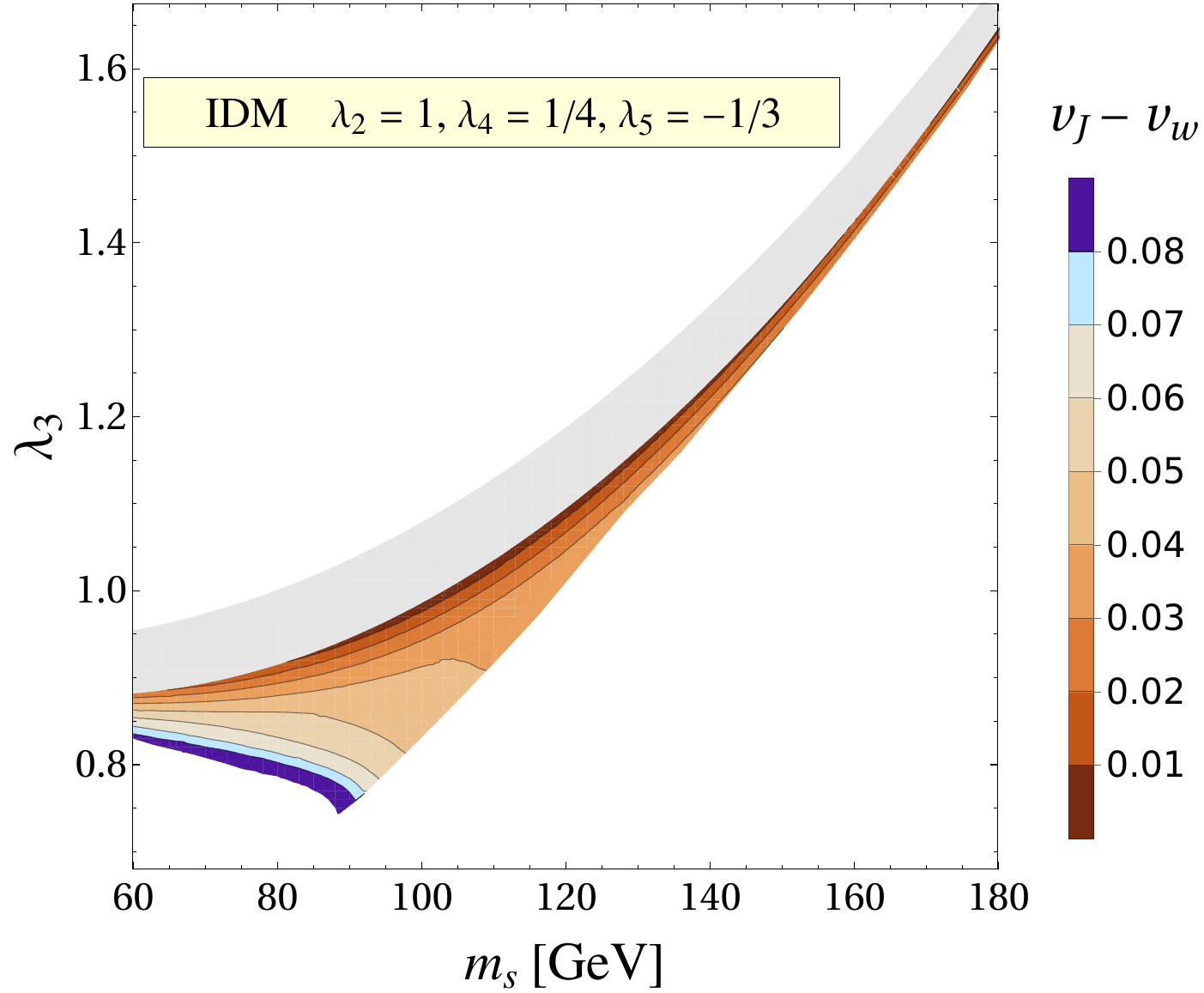}
    \caption{{\it Left panel}. Contour plot of $v_w$ in the ($m_s$, $\lambda_3$) plane for the IDM with $\lambda_2=1$, $\lambda_4=1/4$ and $\lambda_5=-1/3$. {\it Right panel}. Contour plot of  $(v_{_J}-v_w)$.}
    \label{fig: IDM vw paramspace l2 = 1}
\end{figure}

We show once more the results for $v_w$ and for $v_{_J}-v_w$ in the parameter space and in the $T_c$ - $T_n/T_c$ plane. Concerning the first ones, the plots of the wall velocity (left panel) and of the difference (right panel) are in Fig.~\ref{fig: IDM vw paramspace} for $\lambda_2=1/2$ and in Fig.\ref{fig: IDM vw paramspace l2 = 1} for $\lambda_2=1$. They do not show any significant difference with respect to the corresponding plots in the SSM and in the RTSM, with $v_w$ getting closer to $v_{_J}$ as the upper boundary of the (coloured) parameter space is approached. The bubble runs away in the grey region above it.

\begin{figure}
    \centering
    \includegraphics[width=0.48\linewidth]{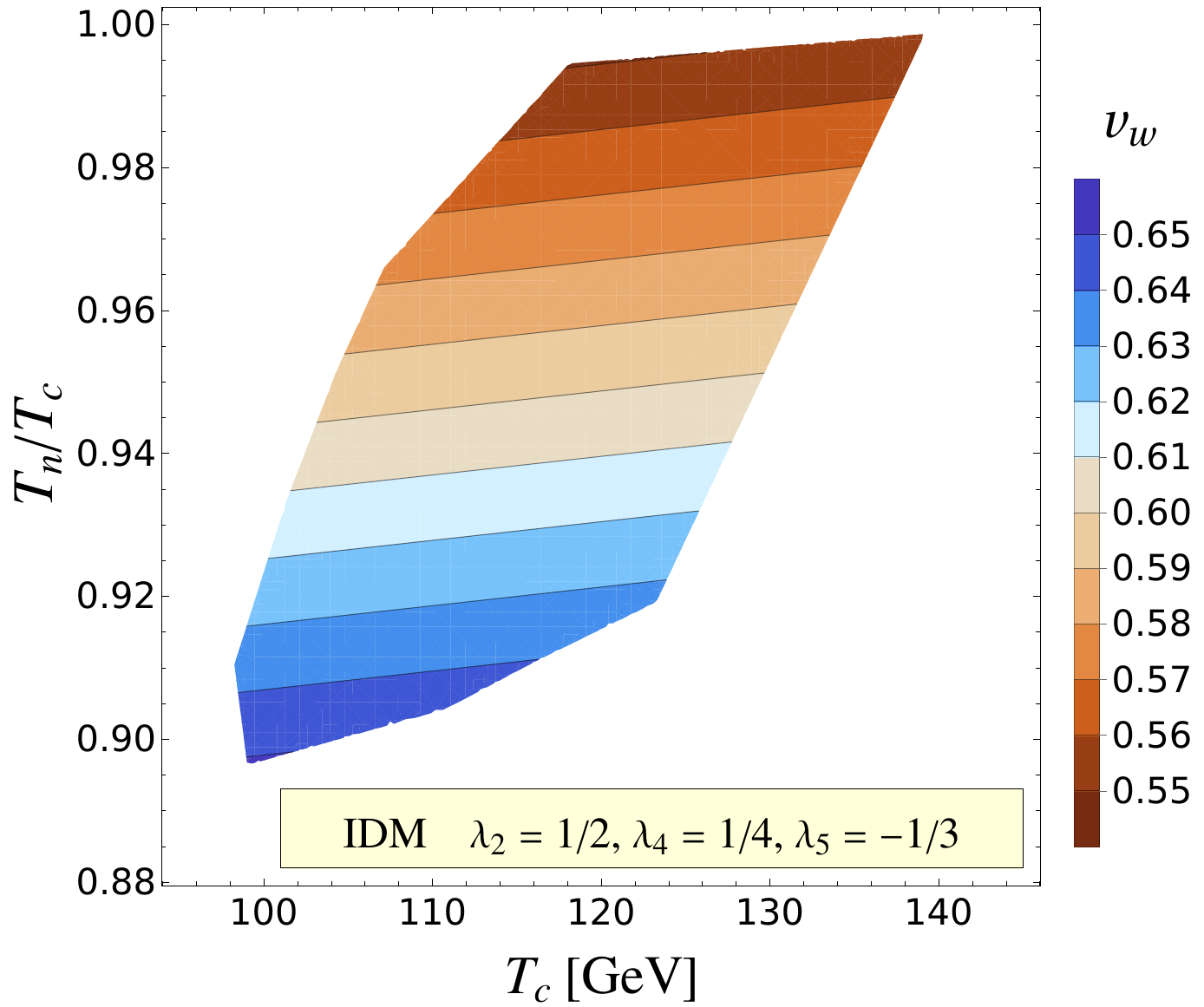}\,\,\includegraphics[width=0.48\linewidth]{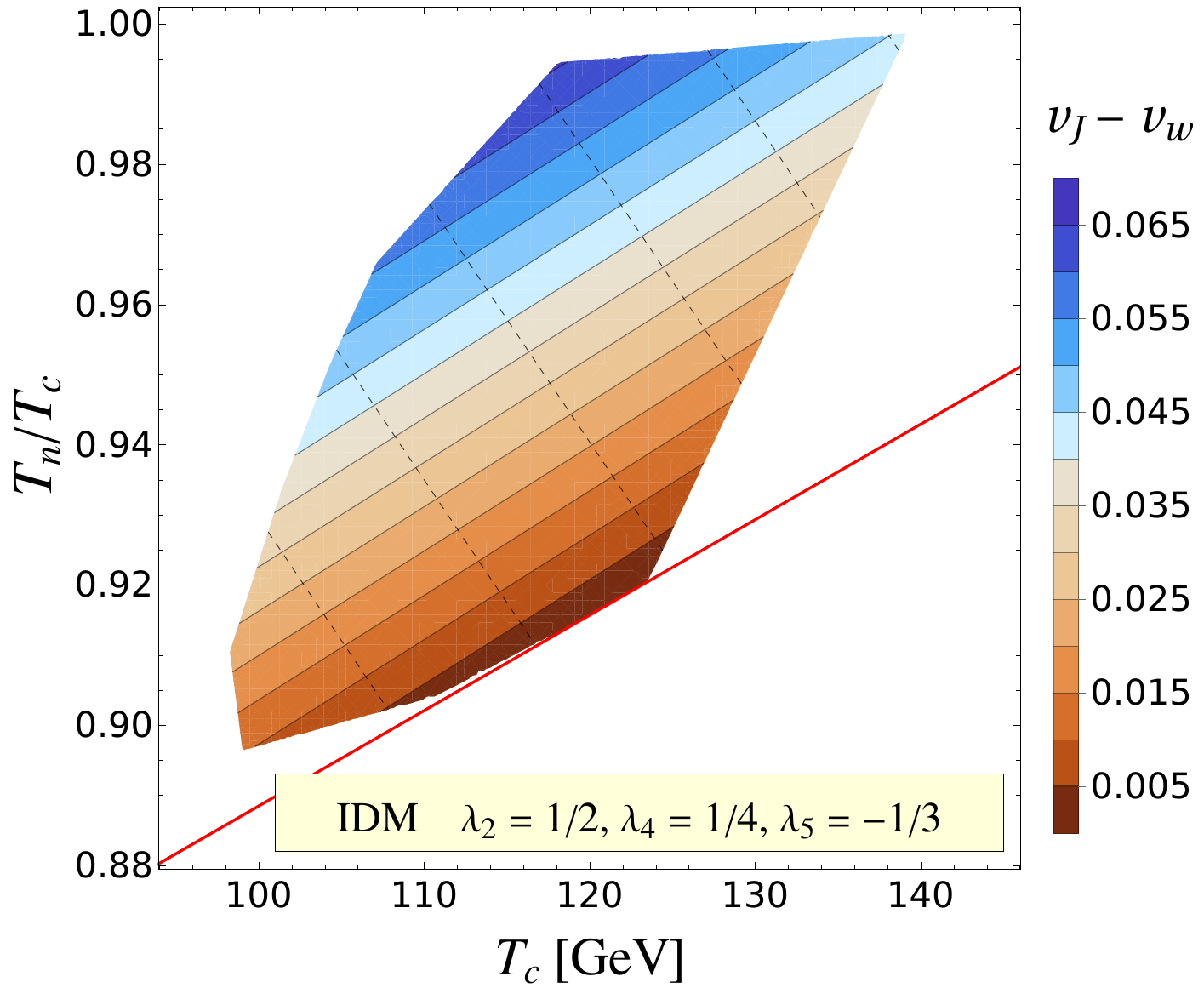}
    \caption{{\it Left panel}. Contour plot of the wall velocity $v_w$ in the ($T_c,T_n/T_c$) plane for the IDM with $\lambda_2 = 1/2$, $\lambda_4=1/4$ and $\lambda_5=-1/3$. {\it Right panel}. Contour plot of $v_{_J}-v_w$. Dashed lines are for constant $v_{_J}$. On the red line, $v_w=v_{_J}$.}
    \label{fig: IDM vw Tspace}
\end{figure}

\begin{figure}
    \centering
    \includegraphics[width=0.48\linewidth]{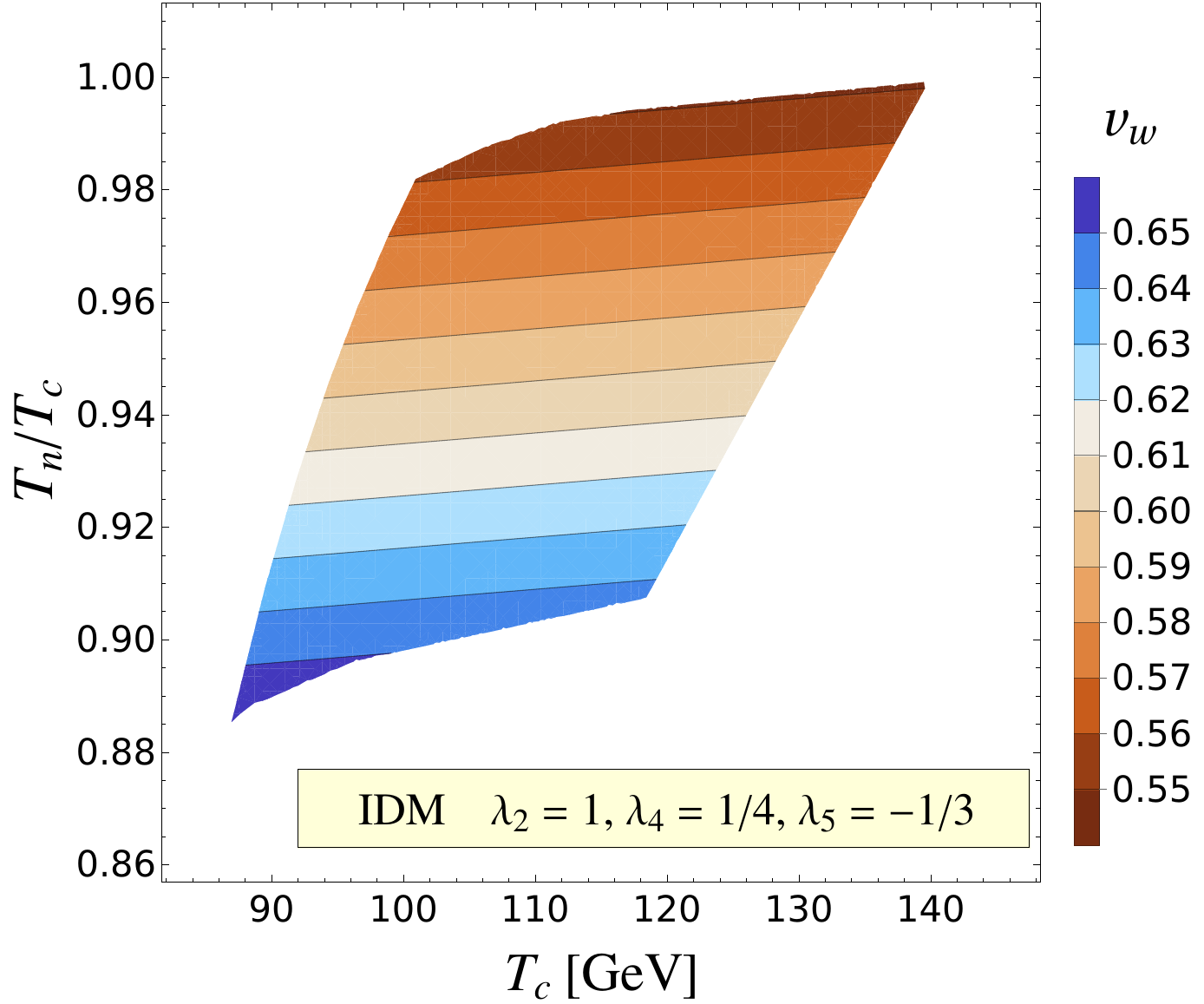}\,\,\includegraphics[width=0.48\linewidth]{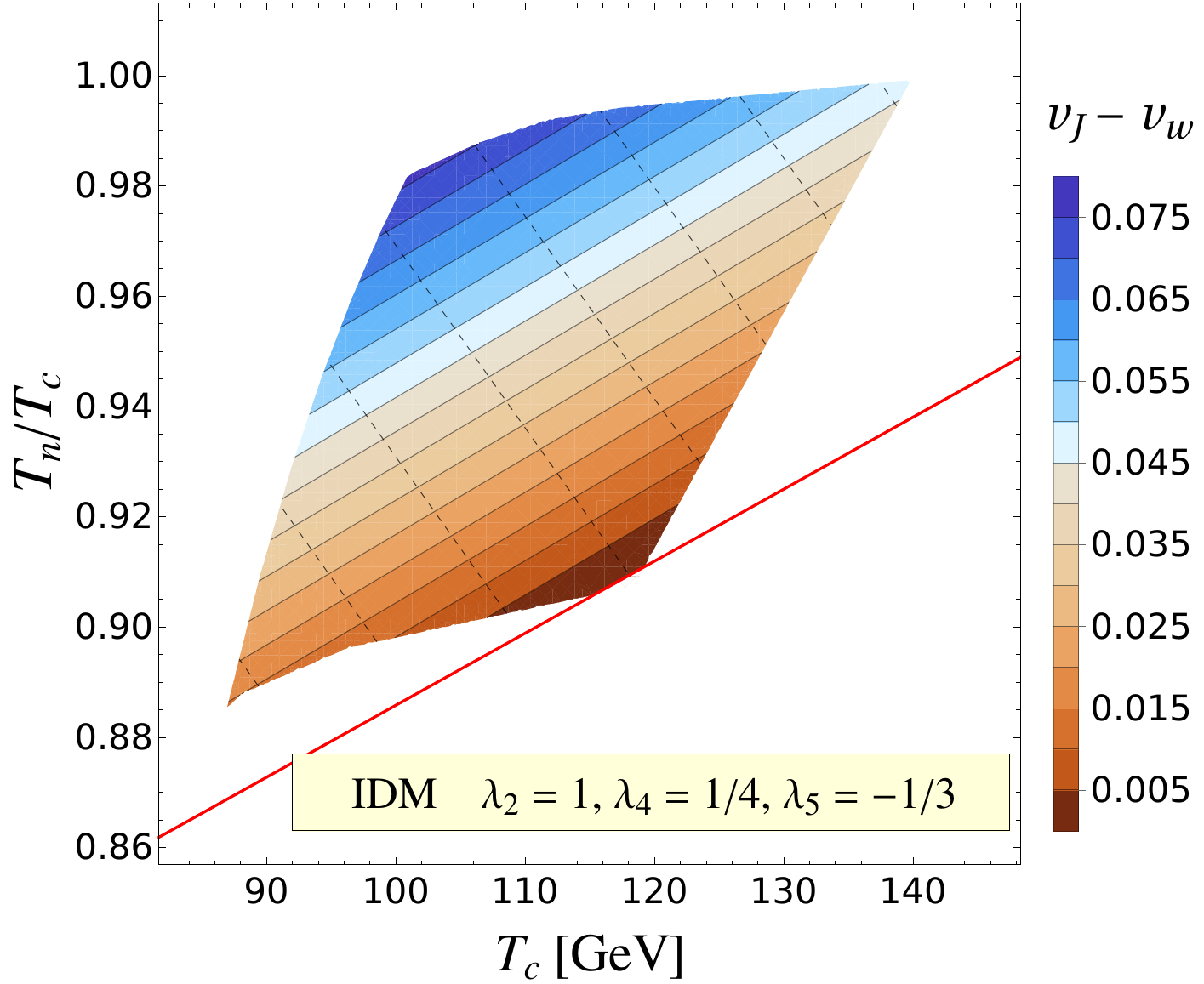}
    \caption{{\it Left panel}. Contour plot of the wall velocity $v_w$ in the ($T_c,T_n/T_c$) plane for the IDM with $\lambda_2 = 1$, $\lambda_4=1/4$ and $\lambda_5=-1/3$.  {\it Right panel}. Contour plot of ($v_{_J}-v_w$). Dashed lines are for constant $v_{_J}$.  On the red line, $v_w=v_{_J}$. }
    \label{fig: IDM vw Tspace l2 =1}
\end{figure}

The wall velocity in the $T_c$ - $T_n/T_c$ plane is shown in the left panel of Fig.\,\ref{fig: IDM vw Tspace} for $\lambda_2=1$ and in the left panel of Fig.\,\ref{fig: IDM vw Tspace l2 =1} for $\lambda_2=1$. The qualitative behaviour is as observed in the previous cases, with the lines of constant $v_w$ that appear somehow flatter. Numerical data are again fitted to a good approximation by a linear function, that can be found together with the others in Table~\ref{table}. 
The lines of constant Jouguet velocity still form a large angle with those of constant $v_w$, as can be seen in the right panel of the same figures. The portion of $T_c$ - $T_n/T_c$ plane available for deflagrations is then restricted to $T_n/T_c\ge (T_n/T_c)_{min}$, with $(T_n/T_c)_{min}$ given in Table~\ref{table}.

\subsection{Comments and comparison between the models}

\begin{table}[htbp!]
	\centering
	\begin{tabular}{p{2.6cm}|p{3.7cm}|p{3.7cm}|p{2.6cm}}

		Model &  $v_w(T_c/v, T_n/T_c)$ & $v_{_J}(T_c/v, T_n/T_c)$ & $T_n/T_c\big|_{min}(T_c/v)$ \\  
					\hline\hline		
		SSM $\lambda_s=1$ &  $ 1.60 + 0.15\, x -1.14 \,y$ & $0.96-0.23\,x-0.23\,y$ & $0.71 + 0.42 \, x$\\
		SSM $\lambda_s=2 $ & $ 1.60 + 0.15 \, x - 1.14\,  y$ & $0.96-0.22\,x-0.23\,y$ & $0.71 + 0.42 \, x$\\
		\hline 
		RTSM $\lambda_\sigma=1$ & $1.60+0.13 \,x-1.12\, y$ & $0.96-0.22\,x-0.24\,y$  & $0.73 + 0.39 \, x$ \\
		RTSM $\lambda_\sigma=2$ & $1.59+0.13\, x-1.12 \,y$ & $0.97-0.22\,x-0.25\,y$ & $0.72 + 0.40 \, x$\\
		\hline 
		IDM $\lambda_2=1/2$ &  $1.60 + 0.07 \,x-1.10\, y$ & $0.97-0.21\,x-0.25\,y$ & $0.75 + 0.34\,  x$\\
		IDM $\lambda_2=1$ & $1.60+ 0.05\, x-1.08 \,y$ & $0.98-0.21\,x-0.26\,y$ & $0.76 + 0.32\, x$ \\
		\hline\hline 
	\end{tabular}
    \caption{Fitting functions for the wall velocity ({\it second column}) and the Jouguet velocity ({\it third column}) in terms of $T_c/v\, (\equiv x)$ and $T_n/T_c\,(\equiv y)$, and of the line of minimal $T_n/T_c$ ({\it fourth column}) for the different models studied.}
    \label{table}
\end{table}

We show the fitting functions for the wall velocity, the Jouguet velocity and the equation determining the upper bound on the supercooling found in the $T_c$ - $T_n/T_c$
plane for the different models in Table~\ref{table}. Approximate independence of all of them from the self-coupling of the additional scalar is clearly suggested by the results, as its variation induces differences at the percent level at most. 

A slight dependence emerges when comparing different models. In particular, the coefficient of $T_c/v$ in the fitting function of $v_w$ appears to have the largest variation, though the velocity mostly depends on $T_n/T_c$. The coefficient of $T_n/T_c$ shows a slight tendency to decrease as the additional scalar becomes more strongly coupled. As for the Jouguet velocity $v_{_J}$, the coefficient of $T_c/v$ slightly decreases the larger the $SU(2)$ representation of the scalar, while the coefficient of $T_n/T_c$ slightly increases. The fitting functions for $T_n/T_c\big|_{min}$ also show a mild model dependence. We observe a tendency to be shifted upwards for scalars in larger representations of $SU(2)$, as well as a tendency to become flatter.   

\section{Applications}
\label{sec: applications}

As mentioned in the introduction, the determination of the bubble wall dynamics is crucial to establish the cosmological signals related to a putative first order EWPhT. In this section we present results for the emitted spectrum of gravitational waves and for the matter asymmetry obtained with the parameters calculated in the previous section. 
A similar analysis was recently presented in \cite{Carena:2025flp}.

\subsection{Gravitational waves}

Gravitational waves are one of the most compelling cosmological relics of first order phase transitions. There are three main mechanisms of production: bubble collisions, sound waves, that are generated in the plasma after the collisions and before the expansion dissipates the kinetic energy, and turbulence effects that take longer to dissipate. 
For the thermal transitions of interest to our analysis, 
numerical simulations have shown that sound waves give the dominant contribution to the gravitational wave spectrum \cite{Hindmarsh:2013xza}. 
The frequency $f^{\rm peak}$ and the amplitude $\Omega^{\rm peak}_{\rm SW}$ at the peak of the spectrum for the sound wave contribution can be found in \cite{Espinosa:2010hh, Hindmarsh:2013xza, Caprini:2015zlo,Hindmarsh:2015qta,Hindmarsh:2016lnk,Hindmarsh:2017gnf,Weir:2017wfa, Ellis:2018mja,Caprini:2019egz,Ellis:2020awk,Caprini:2024hue}.

\begin{figure}
\centering
\includegraphics[scale=0.28]{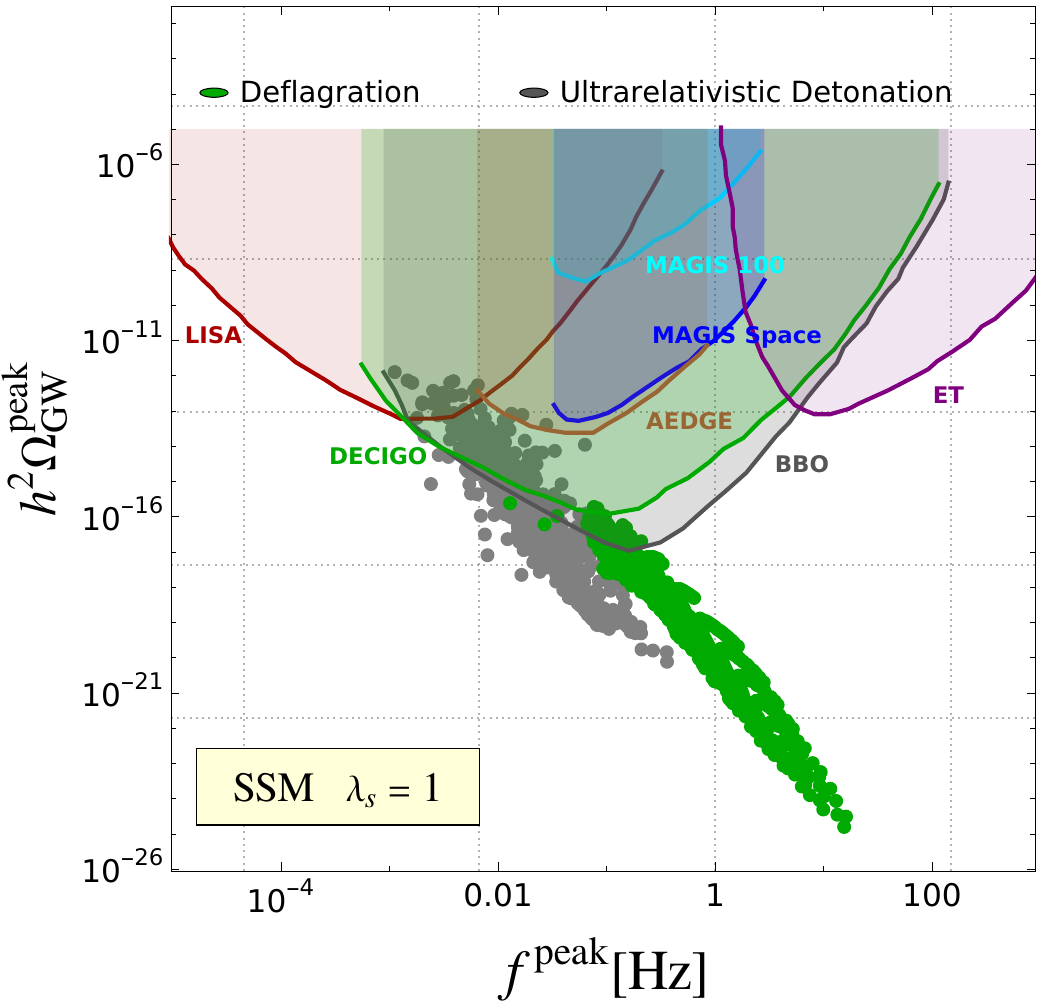}
\includegraphics[scale=0.28]{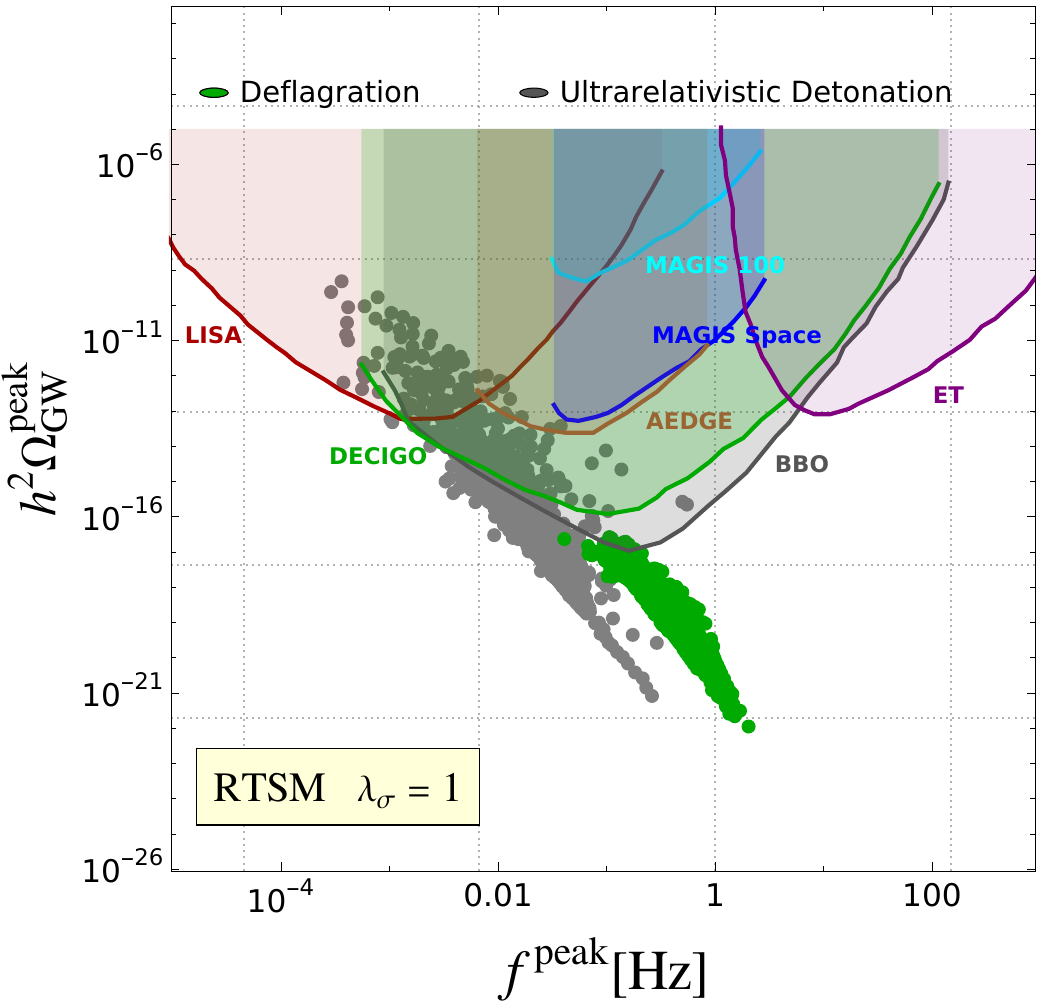}
\includegraphics[scale=0.28]{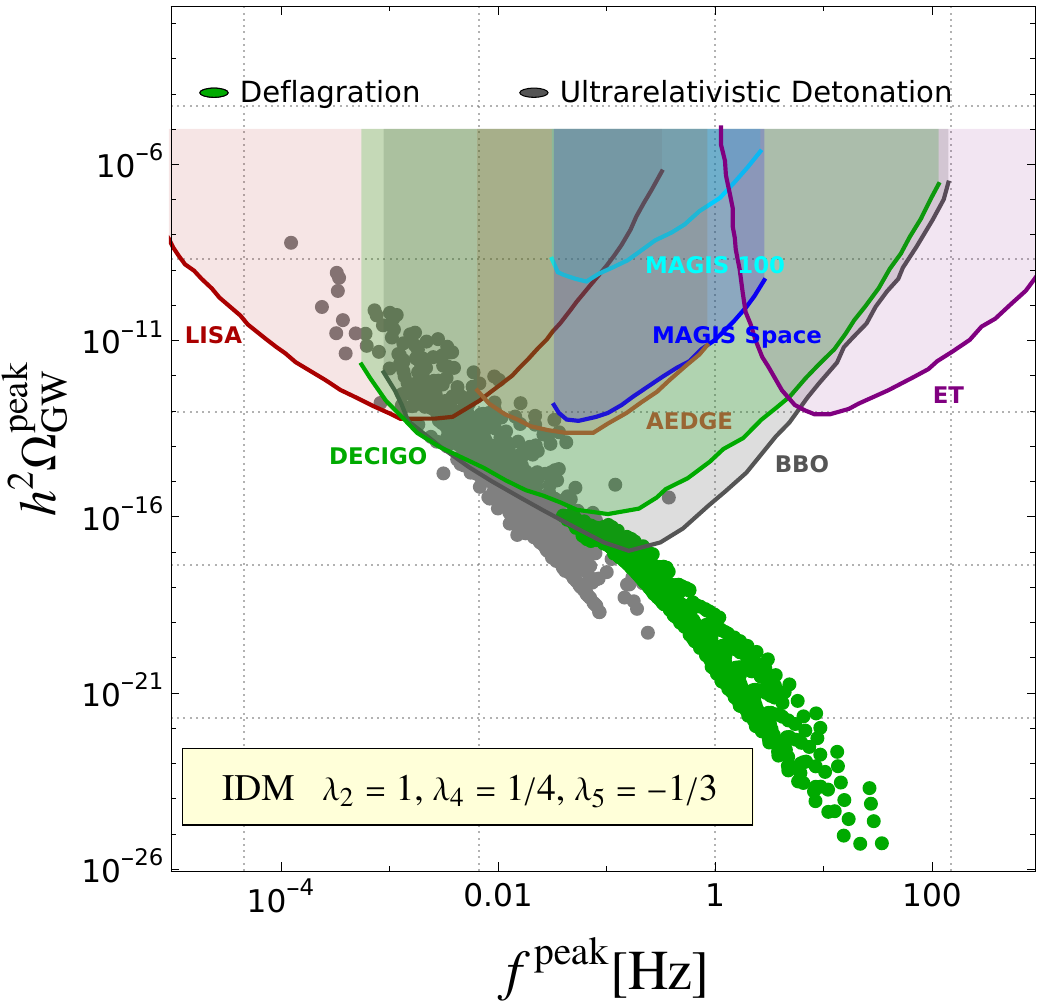}
\caption{Peak of the gravitational wave signal $h^2 \Omega_{\rm SW}^{\rm peak}$ as function of the peak frequency $f_{\rm peak}$ for the SSM with $\lambda_s = 1$ (left), for the RTSM with $\lambda_\sigma= 1$ (center), and for the IDM with $\lambda_2 = 1$, $\lambda_4 = 1/4$, $\lambda_5 = -1/3$ (right). Green (gray) points correspond to deflagration (ultra-relativistic detonation) profiles. Also shown are the sensitivity curves of some future interferometers. }
\label{gwspectra}
\end{figure}

We show here the results for the peak amplitude and frequency for the three BSM scenarios considered. As showcases we limit ourselves to   $\lambda_s=1$ for the SSM,  $\lambda_\sigma=1$ for the RTSM and  $\lambda_2=1$ (with $\lambda_4=1/4$ and $\lambda_5=-1/3$) for the IDM. As can be seen in Fig.\,\ref{gwspectra} the three models show common predictions. Most of the deflagrations are below the sensitivity of the experimental proposals. Within the LTE approximation, we find that a small fraction of the two-step parameter space with a deflagration solution can lead to gravitational wave signals observable at BBO and very marginally at DECIGO. On the other hand, ultra-relativistic detonations tend to be more in reach of the proposed experiments, with LISA probing a fraction of the parameter space, which, as we verified, is larger for values of the self-coupling. This is expected since the detonation region opens up with as the latter grows, allowing larger values of the portal coupling and, correspondingly, stronger transitions.  \\
We expect these conclusions to get somehow worse with the inclusion of out-of-equilibrium contributions, as the latter decrease the value of the wall velocity with respect to the corresponding LTE solution for deflagrations and possibly convert some LTE detonations into deflagrations. 

\subsection{Electroweak baryogenesis}

A crucial ingredient for EW baryogenesis is the presence of CP violation in the interactions between the wall and matter. 
This effect can be described, in the simplest scenarios, through an effective dimension-5 operator, such as $s \, \bar Q_L H t_R$, with complex coefficient, where $Q_L$ and $t_R$ are the left- and right-handed top-quark fields, $H$ is the Higgs doublet and $s$ an additional scalar with a non-trivial spatial profile. The CP violation becomes efficient during the EWPhT, when both scalar fields acquire a space dependent VEV. As such, the phase in the top mass cannot be reabsorbed by a redefinition of the fermionic field.
This operator emerges, for instance, from the non-linear dynamics of composite Higgs models~\cite{Espinosa:2011eu,DeCurtis:2019rxl}. \\
The relevant term in the Lagrangian is 
\begin{equation}
\mathcal L \supset \frac{y_t}{\sqrt 2} h(z) \bar t_L\left(1+ic_5\frac{s(z)}{\Lambda}\right)t_R + \, \text{h.c.}, 
\label{eq: dim5 operator}
\end{equation}
where $\Lambda$ is the scale of the effective interaction and $c_5$ its Wilson coefficient. The latter gives rise to a field-dependent top-quark mass and phase
\begin{equation}
   \widetilde m_t(z) = \frac{y_t h(z)}{\sqrt 2} \sqrt{1+c_5^2\frac{s^2(z)}{\Lambda^2}}\, ; \qquad \theta(z) = \arctan \left(c_5\frac{s(z)}{\Lambda}\right). 
   \label{eq: mass EWBG}
\end{equation}

In the following, we will focus on the SSM scenario. 

We have verified that the addition of this dimension-5 operator has a negligible impact on the PhT dynamics, so we can safely  use all the results found in the previous section. With the mass term \eqref{eq: mass EWBG}, the Boltzmann equation for the perturbations $\delta f_i$ can be projected into its CP-even and its CP-odd parts. The former determines the out-of equilibrium dynamics of the bubble wall, more precisely the (out-of-equilibrium) friction acting on the expanding front, while the CP-odd equation is the relevant one for the calculation of the baryon asymmetry. The latter is eventually determined by solving the equations for the CP-odd component of the chemical potential of the species in the plasma.
For details we refer to Ref.\,\cite{Cline:2020jre}, whose method we closely follow here for the calculation. The CP-odd equations are solved by exploiting a moment expansion, truncated to second order. This strategy has been recently generalised to an arbitrary number $n$ of moments in \cite{Kainulainen:2024qpm}.

Compared to the LTE calculation of the wall velocity through the entropy conservation method, the upshot of our procedure is that we fully determine the plasma and field profiles (through the parameters $\delta_s, \,L_h$ and $L_s$). These enter in both the CP-even and the CP-odd components of the Boltzmann equation, and the final result for the asymmetry crucially depends on them. Our results thus provide a complete determination of the baryon asymmetry within the LTE approximation.

\begin{figure}
    \centering \includegraphics[scale=0.39]{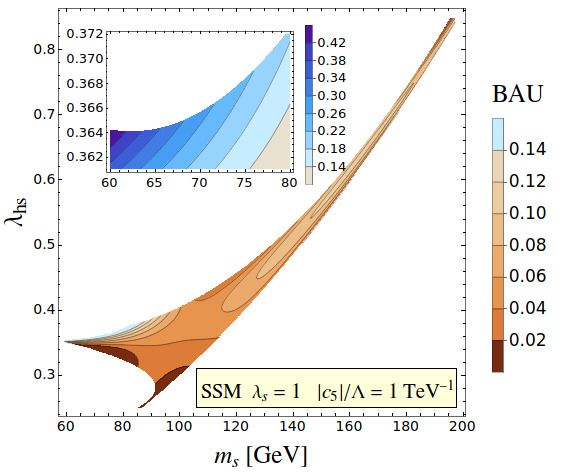}
    \quad
    \includegraphics[scale=0.52]
    {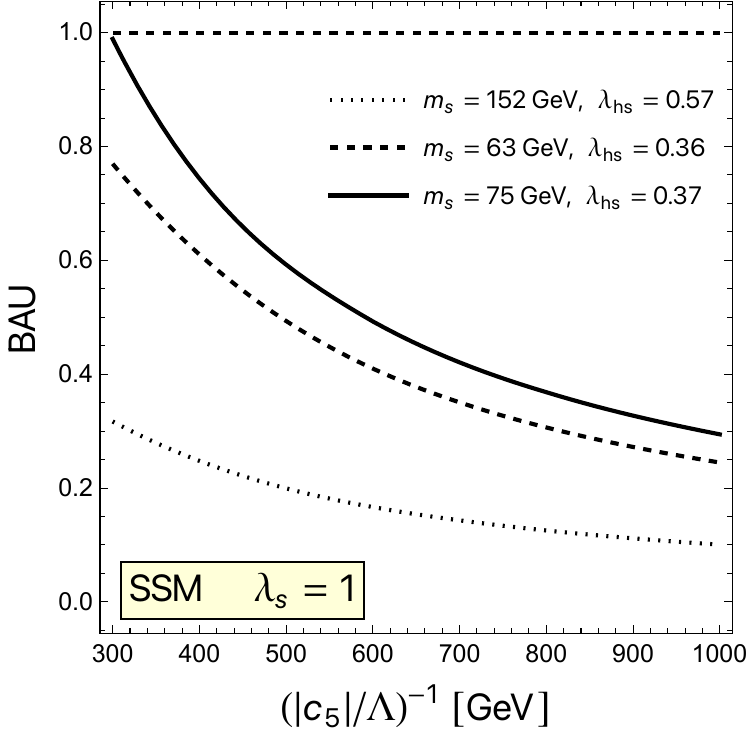}
    \caption{{\it Left panel}. Contour plot of the BAU in the $(m_s,\lambda_{hs})$ plane for the SSM with $\lambda_s=1$ augmented with the dimension-5 operator in Eq.~\eqref{eq: dim5 operator}. The scale $\Lambda$ is set to $\Lambda=1$ TeV, and the Wilson coefficient is fixed to $c_5=\pm 1$, with the sign chosen in order to realise a positive BAU. The results are normalised in terms of the observed baryon asymmetry. The inside figure is a zoom of the largest one in a specific corner. {\it Right panel}. Plot of the BAU versus the $|c_5|/\Lambda$. The dotted, dashed and thick lines refer to the three benchmark points indicated in the figure, while the horizontal dashed line is for successful baryogenesis.}
    \label{fig: BAU}
\end{figure}

The outcome of the computation of the baryon asymmetry of the universe (BAU) in the SSM with $\lambda_s=1$, modified with the dimension five operator in Eq.~\eqref{eq: dim5 operator} with  
$|c_5|/\Lambda = 1$~TeV$^{-1}$, is shown in Fig.~\ref{fig: BAU}. The results are normalised in terms of the observed baryon asymmetry, $\eta_{\rm obs}\sim 8.7\times 10^{-11}$, so that successful baryogenesis require BAU$\,=1$. In most of the parameter space, however, we find that the BAU is $\mathcal O(10^{-1})-\mathcal O(10^{-2})$. The largest values are found in the leftmost upper corner of the parameter space, where the BAU grows up to $\sim 0.4$. Differently from what one might naively expect, this is the region where the largest values for $v_w$ are found. The reason for these results is that the BAU not only depends on the wall velocity, but more generally on the static (nucleation temperature $T_n$ and VEVs $h_n$, $s_n$) and dynamical ($v_w,\,\delta_s,\,L_h,\,L_s$) parameters of the transition, that vary non-trivially across the parameter space. 

By consistently extracting their values for each point of the parameter space for which the EWPhT is strongly first-order, it appears that, for the ``augmented'' SSM with $\lambda_s=1$ and $|c_5|/\Lambda = 1$~TeV$^{-1}$, a complete baryogenesis can be hardly obtained.

The right panel of Fig.\,\ref{fig: BAU} shows the dependence of the baryon asymmetry, within the LTE approximation, on the scale $\Lambda$ (with $|c_5| = 1$). Since the BAU decreases with the latter, successful baryogenesis could be obtained for sufficiently small values of $\Lambda$ but only in a corner of the parameter space.

This result highlights the importance of including out-of-equilibrium effects, which generally lead to a reduction in wall velocity across the entire parameter space, thereby facilitating the achievement of the BAU. This will be explored in a forthcoming work \cite{in preparation}.

\section{Conclusions}
\label{sec: conclusions}

In this paper, we performed a detailed and systematic investigation of the bubble wall dynamics during a first-order EWPhT within three representative extensions of the Higgs sector of the SM: the real singlet scalar model, the real triplet model, and the inert doublet model. Our study was performed under the assumption of local thermal equilibrium, using an optimised numerical algorithm, originally introduced in previous works \cite{DeCurtis:2022hlx, DeCurtis:2023hil, DeCurtis:2024hvh}  and further refined here, to extract the key dynamical properties of the bubble wall: the velocity $v_w$, the Higgs and scalar wall thicknesses ($L_h$ and  $L_s$), and the scalar field displacement $\delta_s$. The full determination of the field and fluid profiles were crucial in the calculation of cosmological observables related to the transition, such as the EW baryogenesis. 

Our approach relied on the solution of the full set of coupled scalar equations of motion along with the hydrodynamic conservation laws. Our results confirm that, within the LTE framework, steady-state solutions are of deflagration type. Detonations could be found when out-of-equilibrium effects are included. 

An outcome of our study is the emergence of an approximate model-independence in the correlation between the bubble wall velocity and global phase transition parameters, such as the critical and nucleation temperatures. This behaviour points to possible universal features in the hydrodynamic response of the plasma during EWPhTs, and may simplify efforts to estimate bubble dynamics across a wide class of models without requiring a full microscopic analysis.

Our comparative study highlights several consistent trends across models. In particular, we observe a clear correlation between the wall velocity and the parameters of the phase transition, namely the critical temperature $T_c$ and the amount of supercooling $T_n/T_c$. We showed that this also leads to an upper bound on $T_n/T_c$. An inverse relation between wall thicknesses and the strength of the phase transition is also observed, with thinner walls associated with stronger transitions. This trend is consistent with theoretical expectations. 

Looking forward, the results presented here provide a step toward a fully out-of-equilibrium analysis, which we will pursue in a future work \cite{in preparation} by applying the complete numerical solution of the Boltzmann equation developed in \cite{DeCurtis:2022hlx, DeCurtis:2023hil, DeCurtis:2024hvh}. Given that out-of-equilibrium effects introduce additional frictional forces, our LTE-based wall velocities should be interpreted as conservative upper bounds. We also expect corrections to the other wall properties  including $\delta_s$ and 
$L_{h,s}$.

Capturing these effects will be essential for accurately assessing the viability of EW baryogenesis and the gravitational wave signatures of these transitions, that we have preliminarly studied here within the LTE approximation. \\
Overall, this work offers a robust framework for studying bubble dynamics in first-order phase transitions and sets the stage for more comprehensive kinetic treatments that are crucial for connecting theoretical models to cosmological observables.
\\
\section*{Acknowledgments}
We thank G.~Barni, A.~Guiggiani, \'A.~Gil Muyor and G.~Panico for useful discussions. The work has been funded by the European Union – Next Generation EU
through the research grant number P2022Z4P4B “SOPHYA - Sustainable Optimised PHYsics
Algorithms: fundamental physics to build an advanced society” under the program PRIN 2022
PNRR of the Italian Ministero dell’Universit\`{a} e Ricerca (MUR) and by the research grant number 20227S3M3B “Bubble Dynamics in Cosmological Phase Transitions” under the program PRIN 2022 of the Italian Ministero dell’Universit\`{a} e Ricerca (MUR). The work has been also partially supported by
ICSC – Centro Nazionale di Ricerca in High Performance Computing, Big Data and Quantum
Computing. C.Banchina and A. Conaci would like to thank the Galileo Galilei Institute for Theoretical Physics, INFN National Center for Advanced Studies, Largo E. Fermi, 2, 50125 Firenze, Italy, for hospitality and support during the completion of this work.

\section*{Appendix}

\appendix

\renewcommand{\theequation}{A.\arabic{equation}}

\setcounter{equation}{0}

\section{Field-dependent thermal masses} 
In the following  we give the expressions of the field-dependent masses and thermal masses used in the calculation of the one-loop finite temperature effective potential for the three models studied in this work.
\subsection{SSM}
\label{section: thermal masses SSM}
The field-dependent masses $m^2_i(h, s)$ for the $W$ and $Z$ bosons, the top and the Goldstones are given by
\begin{equation}
           m^2_W = \frac{g^2}{4} h^2 , \ \ \ m^2_Z = \frac{g^2 + g'^2}{4} h^2, \ \ \ m^2_t = \frac{1}{2} y^2_t h^2, \ \ \
           m^2_{\chi} = \mu_{h}^2 + \lambda_{h} h^2 + \lambda_{hs} s^2.
 \end{equation}
As for the $h$ and $s$ fields, the mass matrix reads
       \begin{equation}
        \mathcal M^2_{h s}= \begin{pmatrix}
         \mu_h^2 + 3 \lambda_h h^2 + \lambda_{hs} {s}^2 & 2 \lambda_{hs}\,  h s   \\
        2 \lambda_{hs}\, h s  & \mu_s^2 + 3 \lambda_s s^2 +  \lambda_{hs} h^2
       \end{pmatrix}.
       \end{equation}
The structure of the perturbative expansion at finite temperature requires resummation of the leading self-energy daisy diagrams. In the Parwani scheme adopted in the paper, this is achieved by shifting the tree-level masses $m^2_i \to m^2_i + \Pi_i(T)$, where $\Pi_i(T)$ denotes the thermal contribution to the zero external-momentum self-energy, that gives the thermal mass. In particular, comparing with the notation used in \eqref{pot quadratic}, we have: 
\begin{align}
  \Pi_W(T)&= \frac{11}{6}g^2 T^2, \\
  \Pi_B(T)&= \frac{11}{6}(g^2+ g'^2)  T^2,\\
  \Pi_h(T) &= \left( \frac{3 g^2 + g'^2}{16} + \frac{y^2_t}{4} + \frac{\lambda_h}{2} + \frac{\lambda_{h s}}{12} \right) T^2, \\
  \Pi_s (T)&= \left( \frac{\lambda_{hs}}{3}+ \frac{\lambda_s}{4} \right) T^2,
\label{singlet self-energy}
\end{align}
with $\Pi_h=c_h T^2$ and $\Pi_s=c_s T^2$.

\subsection{RTSM}
\label{section: thermal masses RTSM}
We report here the analogous expressions for the RTSM.
The field-dependent masses $m^2_i(h, \sigma_3)$ are obtained as     
\begin{equation}
           m^2_W = \frac{g^2}{4} h^2 + g^2 {\sigma_3}^2, \ \ \  m^2_Z = \frac{g^2 + g'^2}{4} h^2, \ \ \  m^2_t = \frac{1}{2} y^2_t h^2, 
\end{equation}      
\begin{equation}
           m^2_{\chi} = \mu_{h}^2+\lambda_{h} h^2  + \lambda_{h \sigma} {\sigma_3}^2, \ \ \ m^2_{\sigma_{1,2}} = \mu_{\sigma}^2 +\lambda_\sigma {\sigma_3}^2 +\lambda_{h \sigma} h^2 .
 \end{equation}
Concerning the fields $h$ and $\sigma_3$, the mass matrix reads
       \begin{equation}
        \mathcal M^2_{h \sigma_3}= \begin{pmatrix}
        \mu_h^2 + 3 \lambda_h h^2  + \lambda_{h \sigma} {\sigma_3}^2 & 2 \lambda_{h \sigma} h {\sigma_3} \\
        2 \lambda_{h \sigma} h{\sigma_3} &  \mu_{\sigma}^2 + 3 \lambda_\sigma {\sigma_3}^2 +\lambda_{h \sigma}h^2  
       \end{pmatrix}.
       \end{equation}
Finally, the thermal masses are obtained from 
\begin{align}
    \Pi_W(T)&= \frac{13}{6}g^2 T^2\\
\Pi_B(T)&= \frac{11}{6}(g^2 + g'^2)T^2\\
    \Pi_h(T) &= \left( \frac{3 g^2 + g'^2}{16} + \frac{y^2_t}{4} + \frac{\lambda_h}{2} + \frac{\lambda_{h\sigma}}{4} \right) T^2\\
        \Pi_\sigma (T)&= \left( \frac{5}{12}\lambda_\sigma+\frac{g^2}{2}+\frac{\lambda_{h \sigma}}{3}\right) T^2.
\end{align}

\subsection{IDM}
\label{section: thermal masses IDM}
We give here the expression of the field-dependent and thermal masses for the IDM.
Again, field-dependent masses are obtained as
\begin{equation}
           m^2_W = \frac{g^2}{4}( h^2 + h'^2), \ \ \  m^2_Z = \frac{g^2 + g'^2 }{4} ( h^2 + h'^2), \ \ \  m^2_t = \frac{1}{2} y^2_t h^2,
\end{equation}  
for the $W$, $Z$ and the top, and from the following mass matrices for the scalar fields, 
\begin{equation}
\mathcal M^2_{h h'} = \begin{pmatrix}
             \mu_1^2  + 3 \lambda_1 h^2 + \frac{\lambda_{345}}{2}h'^2 & \lambda_{345} h h' \\
            \lambda_{345} h h'  & \mu_2^2 + 3 \lambda_2 h'^2 + \frac{\lambda_{345}}{2}h^2
            \end{pmatrix},
\end{equation}
\begin{equation}
\mathcal M^2_{\chi^3 \chi^6} = \begin{pmatrix}
            \mu_1^2 +  \lambda_1 h^2 + \frac{\tilde{\lambda}_{345}}{2}h'^2 & \lambda_{5} h h' \\
            \lambda_{5} h h'  & \mu_2^2 +  \lambda_2 h'^2 + \frac{\tilde{\lambda}_{345}}{2}h^2
            \end{pmatrix},
\end{equation}

\begin{equation}
\mathcal M^2_{\chi^1 \chi^4, \chi^2 \chi^5} = \begin{pmatrix}
            \mu_1^2 + \lambda_1 h^2 + \frac{\lambda_3}{2}h'^2 & \frac{\lambda_4 + \lambda_{5}}{2} h h' \\
            \frac{\lambda_4 + \lambda_{5}}{2} h h'  & \mu_2^2 + \lambda_2 h'^2 + \frac{\lambda_3}{2}h^2
            \end{pmatrix} .
\end{equation}
The thermal contributions to the zero external momentum self-energy are
\begin{align}
  \Pi_W(T)&= 2g^2 T^2, \\
    \Pi_B(T)&= 2g'^2 T^2, \\
\Pi_h(T) &= \left(\frac{3g^2 + g'^2 }{16}  + \frac{y_t^2}{4}+ \frac{6\lambda_1 + 2\lambda_3 + \lambda_4 }{12} \right)T^2, \\
\Pi_{h'}(T)  &= \left( \frac{3g^2 + g'^2 }{16}  + \frac{6\lambda_2 + 2\lambda_3 + \lambda_4 }{12} \right) T^2.
\end{align}


\clearpage


\begin{thebibliography}{99}

\bibitem{Kirzhnits:1972ut}
D.~A.~Kirzhnits and A.~D.~Linde,
\emph{Macroscopic Consequences of the Weinberg Model},''
Phys. Lett. B \textbf{42} (1972), 471-474

\bibitem{Dolan:1973qd}
L.~Dolan and R.~Jackiw,
\emph{Symmetry Behavior at Finite Temperature},
Phys. Rev. D \textbf{9} (1974), 3320-3341

\bibitem{Weinberg:1974hy}
S.~Weinberg,
\emph{Gauge and Global Symmetries at High Temperature},
Phys. Rev. D \textbf{9} (1974), 3357-3378

\bibitem{Kajantie:1995kf}
K.~Kajantie, M.~Laine, K.~Rummukainen and M.~E.~Shaposhnikov,
\emph{The Electroweak phase transition: A Nonperturbative analysis},
Nucl. Phys. B \textbf{466} (1996), 189-258, arXiv:hep-lat/9510020 [hep-lat].

\bibitem{Kajantie:1996mn}
K.~Kajantie, M.~Laine, K.~Rummukainen and M.~E.~Shaposhnikov,
\emph{Is there a~ hot electroweak phase transition at $m_H \gtrsim m_W$?},
Phys. Rev. Lett. \textbf{77} (1996), 2887-2890, arXiv:hep-ph/9605288 [hep-ph].

\bibitem{Csikor:1998eu}
F.~Csikor, Z.~Fodor and J.~Heitger,
\emph{Endpoint of the hot electroweak phase transition},
Phys. Rev. Lett. \textbf{82} (1999), 21-24, arXiv:hep-ph/9809291 [hep-ph]].


\bibitem{LIGOScientific:2016aoc}
B.~P.~Abbott \textit{et al.} [LIGO Scientific and Virgo],
\emph{Observation of Gravitational Waves from a Binary Black Hole Merger},
Phys. Rev. Lett. \textbf{116} (2016) no.6, 061102, arXiv:1602.03837 [gr-qc].

\bibitem{LIGOScientific:2016sjg}
B.~P.~Abbott \textit{et al.} [LIGO Scientific and Virgo],
\emph{GW151226: Observation of Gravitational Waves from a 22-Solar-Mass Binary Black Hole Coalescence},
Phys. Rev. Lett. \textbf{116} (2016) no.24, 241103, arXiv:1606.04855 [gr-qc].

\bibitem{Hindmarsh:2020hop}
M.~B.~Hindmarsh, M.~L\"uben, J.~Lumma and M.~Pauly,
\emph{Phase transitions in the early universe},
SciPost Phys. Lect. Notes \textbf{24} (2021), 1,
arXiv:2008.09136 [astro-ph.CO].

\bibitem{Athron:2023xlk}
P.~Athron, C.~Bal\'azs, A.~Fowlie, L.~Morris and L.~Wu,
\emph{Cosmological phase transitions: From perturbative particle physics to gravitational waves},
Prog. Part. Nucl. Phys. \textbf{135} (2024), 104094, arXiv:2305.02357 [hep-ph].

\bibitem{Moore:1995ua}
G.~D.~Moore and T.~Prokopec,
\emph{Bubble wall velocity in a first order electroweak phase transition},
Phys. Rev. Lett. \textbf{75} (1995), 777-780, arXiv:hep-ph/9503296 [hep-ph].

\bibitem{Moore:1995si}
G.~D.~Moore and T.~Prokopec,
\emph{How fast can the wall move? A Study of the electroweak phase transition dynamics},
Phys. Rev. D \textbf{52} (1995), 7182-7204, arXiv:hep-ph/9506475 [hep-ph].

\bibitem{John:2000zq}
P.~John and M.~G.~Schmidt,
\emph{Do stops slow down electroweak bubble walls?},
Nucl. Phys. B \textbf{598} (2001), 291-305
[erratum: Nucl. Phys. B \textbf{648} (2003), 449-452], arXiv:hep-ph/0002050 [hep-ph].

\bibitem{Moore:2000wx}
G.~D.~Moore,
\emph{Electroweak bubble wall friction: Analytic results},
JHEP \textbf{03} (2000), 006, arXiv:hep-ph/0001274 [hep-ph].

\bibitem{Cline:2000nw}
J.~M.~Cline, M.~Joyce and K.~Kainulainen,
\emph{Supersymmetric electroweak baryogenesis},
JHEP \textbf{07} (2000), 018, arXiv:hep-ph/0006119 [hep-ph].

\bibitem{Bodeker:2009qy}
D.~Bodeker and G.~D.~Moore,
\emph{Can electroweak bubble walls run away?},
JCAP \textbf{05} (2009), 009, arXiv:0903.4099 [hep-ph].

\bibitem{Megevand:2009gh}
A.~Megevand and A.~D.~Sanchez,
\emph{Velocity of electroweak bubble walls},
Nucl. Phys. B \textbf{825} (2010), 151-176, arXiv:0908.3663 [hep-ph].

\bibitem{Espinosa:2010hh}
J.~R.~Espinosa, T.~Konstandin, J.~M.~No and G.~Servant,
\emph{Energy Budget of Cosmological First-order Phase Transitions},
JCAP \textbf{06} (2010), 028, arXiv:1004.4187 [hep-ph].



\bibitem{Leitao:2010yw}
L.~Leitao and A.~Megevand,
\emph{Spherical and non-spherical bubbles in cosmological phase transitions},
Nucl. Phys. B \textbf{844} (2011), 450-470, arXiv:1010.2134 [astro-ph.CO].

\bibitem{Huber:2013kj}
S.~J.~Huber and M.~Sopena,
\emph{An efficient approach to electroweak bubble velocities},
arXiv:1302.1044 [hep-ph].

\bibitem{Megevand:2013hwa}
A.~M\'egevand,
\emph{Friction forces on phase transition fronts},
JCAP \textbf{07} (2013), 045, arXiv:1303.4233 [astro-ph.CO].


\bibitem{Megevand:2013yua}
A.~Megevand and F.~A.~Membiela,
\emph{Stability of cosmological deflagration fronts},
Phys. Rev. D \textbf{89} (2014) no.10, 103507, arXiv:1311.2453 [astro-ph.CO].

\bibitem{Megevand:2014yua}
A.~Megevand and F.~A.~Membiela,
\emph{Stability of cosmological detonation fronts},
Phys. Rev. D \textbf{89} (2014) no.10, 103503, arXiv:1402.5791 [astro-ph.CO].

\bibitem{Konstandin:2014zta}
T.~Konstandin, G.~Nardini and I.~Rues,
\emph{From Boltzmann equations to steady wall velocities},
JCAP \textbf{09} (2014), 028, arXiv:1407.3132 [hep-ph].

\bibitem{Leitao:2014pda}
L.~Leitao and A.~Megevand,
\emph{Hydrodynamics of phase transition fronts and the speed of sound in the plasma},''
Nucl. Phys. B \textbf{891} (2015), 159-199, arXiv:1410.3875 [hep-ph].

\bibitem{Megevand:2014dua}
A.~Megevand, F.~A.~Membiela and A.~D.~Sanchez,
\emph{Lower bound on the electroweak wall velocity from hydrodynamic instability},
JCAP \textbf{03} (2015), 051, arXiv:1412.8064 [hep-ph].

\bibitem{Kozaczuk:2015owa}
J.~Kozaczuk,
\emph{Bubble Expansion and the Viability of Singlet-Driven Electroweak Baryogenesis},
JHEP \textbf{10} (2015), 135, arXiv:1506.04741 [hep-ph].

\bibitem{Bodeker:2017cim}
D.~Bodeker and G.~D.~Moore,
\emph{Electroweak Bubble Wall Speed Limit},
JCAP \textbf{05} (2017), 025, arXiv:1703.08215 [hep-ph].

\bibitem{Dorsch:2018pat}
G.~C.~Dorsch, S.~J.~Huber and T.~Konstandin,
\emph{Bubble wall velocities in the Standard Model and beyond},
JCAP \textbf{12} (2018), 034,
arXiv:1809.04907 [hep-ph].

\bibitem{DeCurtis:2019rxl}
S.~De Curtis, L.~Delle Rose and G.~Panico, \emph{Composite Dynamics in the Early Universe},
JHEP \textbf{12} (2019), 149,
arXiv:1909.07894 [hep-ph].

\bibitem{Cline:2020jre}
J.~M.~Cline and K.~Kainulainen,
\emph{Electroweak baryogenesis at high bubble wall velocities},''
Phys. Rev. D \textbf{101} (2020) no.6, 063525, arXiv:2001.00568 [hep-ph].

\bibitem{BarrosoMancha:2020fay}
M.~Barroso Mancha, T.~Prokopec and B.~Swiezewska,
\emph{Field-theoretic derivation of bubble-wall force},
JHEP \textbf{01} (2021), 070, arXiv:2005.10875 [hep-th].

\bibitem{Hoche:2020ysm}
S.~H\"oche, J.~Kozaczuk, A.~J.~Long, J.~Turner and Y.~Wang,
\emph{Towards an all-orders calculation of the electroweak bubble wall velocity},
JCAP \textbf{03} (2021), 009, arXiv:2007.10343 [hep-ph].

\bibitem{Laurent:2020gpg}
B.~Laurent and J.~M.~Cline,
\emph{Fluid equations for fast-moving electroweak bubble walls},
Phys. Rev. D \textbf{102} (2020) no.6, 063516, arXiv:2007.10935 [hep-ph].

\bibitem{Friedlander:2020tnq}
A.~Friedlander, I.~Banta, J.~M.~Cline and D.~Tucker-Smith,
\emph{Wall speed and shape in singlet-assisted strong electroweak phase transitions},
Phys. Rev. D \textbf{103} (2021) no.5, 055020, arXiv:2009.14295 [hep-ph].

\bibitem{Azatov:2020ufh}
A.~Azatov and M.~Vanvlasselaer,
\emph{Bubble wall velocity: heavy physics effects},
JCAP \textbf{01} (2021), 058, arXiv:2010.02590 [hep-ph].

\bibitem{Balaji:2020yrx}
S.~Balaji, M.~Spannowsky and C.~Tamarit,
\emph{Cosmological bubble friction in local equilibrium},
JCAP \textbf{03} (2021), 051, arXiv:2010.08013 [hep-ph]

\bibitem{Cai:2020djd}
R.~G.~Cai and S.~J.~Wang,
\emph{Effective picture of bubble expansion},
JCAP \textbf{03} (2021), 096, arXiv:2011.11451 [astro-ph.CO].

\bibitem{Wang:2020zlf}
X.~Wang, F.~P.~Huang and X.~Zhang,
\emph{Bubble wall velocity beyond leading-log approximation in electroweak phase transition}, arXiv:2011.12903 [hep-ph].

\bibitem{Cline:2021iff}
J.~M.~Cline, A.~Friedlander, D.~M.~He, K.~Kainulainen, B.~Laurent and D.~Tucker-Smith,
\emph{Baryogenesis and gravity waves from a UV-completed electroweak phase transition},
Phys. Rev. D \textbf{103} (2021) no.12, 123529, arXiv:2102.12490 [hep-ph].

\bibitem{Bigazzi:2021ucw}
F.~Bigazzi, A.~Caddeo, T.~Canneti and A.~L.~Cotrone,
\emph{Bubble wall velocity at strong coupling},
JHEP \textbf{08} (2021), 090, arXiv:2104.12817 [hep-ph].

\bibitem{Dorsch:2021ubz}
G.~C.~Dorsch, S.~J.~Huber and T.~Konstandin,
\emph{On the wall velocity dependence of electroweak baryogenesis},
JCAP \textbf{08} (2021), 020, arXiv:2106.06547 [hep-ph].

\bibitem{Cline:2021dkf}
J.~M.~Cline and B.~Laurent,
\emph{Electroweak baryogenesis from light fermion sources: A critical study},
Phys. Rev. D \textbf{104} (2021) no.8, 083507, arXiv:2108.04249 [hep-ph].

\bibitem{Ai:2021kak}
W.~Y.~Ai, B.~Garbrecht and C.~Tamarit,
\emph{Bubble wall velocities in local equilibrium},
JCAP \textbf{03} (2022) no.03, 015,
 arXiv:2109.13710 [hep-ph].

\bibitem{Lewicki:2021pgr}
M.~Lewicki, M.~Merchand and M.~Zych,
\emph{Electroweak bubble wall expansion: gravitational waves and baryogenesis in Standard Model-like thermal plasma},
JHEP \textbf{02} (2022), 017, arXiv:2111.02393 [astro-ph.CO].


\bibitem{Gouttenoire:2021kjv}
Y.~Gouttenoire, R.~Jinno and F.~Sala,
\emph{Friction pressure on relativistic bubble walls},
JHEP \textbf{05} (2022), 004, arXiv:2112.07686 [hep-ph].

\bibitem{Dorsch:2021nje}
G.~C.~Dorsch, S.~J.~Huber and T.~Konstandin,
\emph{A sonic boom in bubble wall friction},
JCAP \textbf{04} (2022) no.04, 010
arXiv:2112.12548 [hep-ph].

\bibitem{DeCurtis:2022hlx}
S.~De Curtis, L.~D.~Rose, A.~Guiggiani, \'A.~G.~Muyor and G.~Panico,
\emph{Bubble wall dynamics at the electroweak phase transition},
JHEP \textbf{03} (2022), 163, arXiv:2201.08220 [hep-ph].



\bibitem{Laurent:2022jrs}
B.~Laurent and J.~M.~Cline, 
\emph{First principles determination of bubble wall velocity},
Phys. Rev. D \textbf{106} (2022) no.2, 023501, arXiv:2204.13120 [hep-ph].

\bibitem{Lewicki:2022nba}
M.~Lewicki, V.~Vaskonen and H.~Veerm\"ae,
\emph{Bubble dynamics in fluids with N-body simulations},
Phys. Rev. D \textbf{106} (2022) no.10, 103501, 
arXiv:2205.05667 [astro-ph.CO].

\bibitem{Janik:2022wsx}
R.~A.~Janik, M.~Jarvinen, H.~Soltanpanahi and J.~Sonnenschein,
\emph{Perfect Fluid Hydrodynamic Picture of Domain Wall Velocities at Strong Coupling},
Phys. Rev. Lett. \textbf{129} (2022) no.8, 081601, arXiv:2205.06274 [hep-th].

\bibitem{DeCurtis:2022llw}
S.~De Curtis, L.~Delle Rose, A.~Guiggiani, \'A.~Gil Muyor and G.~Panico,
\emph{Dynamics of bubble walls at the electroweak phase transition},
EPJ Web Conf. \textbf{270} (2022), 00035, arXiv:2209.06509 [hep-ph].

\bibitem{DeCurtis:2022djw}
S.~De Curtis, L.~Delle Rose, A.~Guiggiani, \'A.~Gil Muyor and G.~Panico,
\emph{Bubble wall dynamics at the electroweak scale},
PoS \textbf{ICHEP2022}, 080

\bibitem{Ellis:2022lft}
J.~Ellis, M.~Lewicki, M.~Merchand, J.~M.~No and M.~Zych,
\emph{The scalar singlet extension of the Standard Model: gravitational waves versus baryogenesis},
JHEP \textbf{01} (2023), 093, arXiv:2210.16305 [hep-ph].

\bibitem{Jiang:2022btc}
S.~Jiang, F.~P.~Huang and X.~Wang,
\emph{Bubble wall velocity during electroweak phase transition in the inert doublet model},
Phys. Rev. D \textbf{107} (2023) no.9, 095005, arXiv:2211.13142 [hep-ph].


\bibitem{DeCurtis:2023hil}
S.~De Curtis, L.~Delle Rose, A.~Guiggiani, \'A.~Gil Muyor and G.~Panico, \emph{Collision integrals for cosmological phase transitions}, JHEP \textbf{05} (2023), 194, arXiv:2303.05846 [hep-ph].


\bibitem{Ai:2023see}
W.~Y.~Ai, B.~Laurent and J.~van de Vis,
\emph{Model-independent bubble wall velocities in local thermal equilibrium},
JCAP \textbf{07} (2023), 002, arXiv:2303.10171 [astro-ph.CO].

\bibitem{Krajewski:2023clt}
T.~Krajewski, M.~Lewicki and M.~Zych,
\emph{Hydrodynamical constraints on the bubble wall velocity},
Phys. Rev. D \textbf{108} (2023) no.10, 103523, arXiv:2303.18216 [astro-ph.CO].

\bibitem{Baldes:2023cih}
I.~Baldes, M.~Dichtl, Y.~Gouttenoire and F.~Sala,
\emph{Ultrahigh-Energy Particle Collisions and Heavy Dark Matter at Phase Transitions},
Phys. Rev. Lett. \textbf{134} (2025) no.6, 6, arXiv:2306.15555 [hep-ph].

\bibitem{Azatov:2023xem}
A.~Azatov, G.~Barni, R.~Petrossian-Byrne and M.~Vanvlasselaer,
\emph{Quantisation across bubble walls and friction},
JHEP \textbf{05} (2024), 294, arXiv:2310.06972 [hep-ph].

\bibitem{Dorsch:2023tss}
G.~C.~Dorsch and D.~A.~Pinto,
\emph{Bubble wall velocities with an extended fluid Ansatz},
JCAP \textbf{04} (2024), 027, arXiv:2312.02354 [hep-ph].

\bibitem{Sanchez-Garitaonandia:2023zqz}
M.~Sanchez-Garitaonandia and J.~van de Vis,
\emph{Prediction of the bubble wall velocity for a large jump in degrees of freedom},
Phys. Rev. D \textbf{110} (2024) no.2, 023509, arXiv:2312.09964 [hep-ph].

\bibitem{Ai:2024shx}
W.~Y.~Ai, X.~Nagels and M.~Vanvlasselaer,
\emph{Criterion for ultra-fast bubble walls: the impact of hydrodynamic obstruction},
JCAP \textbf{03} (2024), 037, arXiv:2401.05911 [hep-ph].


\bibitem{DeCurtis:2024hvh}
S.~De Curtis, L.~Delle Rose, A.~Guiggiani, \'A.~Gil Muyor and G.~Panico, \emph{Non-linearities in cosmological bubble wall dynamics},
JHEP \textbf{05} (2024), 009, arXiv:2401.13522 [hep-ph].

\bibitem{Krajewski:2024gma}
T.~Krajewski, M.~Lewicki and M.~Zych,
\emph{Bubble-wall velocity in local thermal equilibrium: hydrodynamical simulations vs analytical treatment},
JHEP \textbf{05} (2024), 011, arXiv:2402.15408 [astro-ph.CO].

\bibitem{Wang:2024wcs}
D.~W.~Wang, Q.~S.~Yan and M.~Huang,
\emph{Bubble wall velocity and gravitational wave in the minimal left-right symmetric model},
Phys. Rev. D \textbf{110} (2024) no.7, 076011, arXiv:2405.01949 [gr-qc].

\bibitem{Azatov:2024auq}
A.~Azatov, G.~Barni and R.~Petrossian-Byrne,
\emph{NLO friction in symmetry restoring phase transitions},
JHEP \textbf{12} (2024), 056, arXiv:2405.19447 [hep-ph].

\bibitem{Barni:2024lkj}
G.~Barni, S.~Blasi and M.~Vanvlasselaer,
\emph{The hydrodynamics of inverse phase transitions},
JCAP \textbf{10}, 042 (2024),
arXiv:2406.01596 [hep-ph].

\bibitem{Yuwen:2024hme}
Z.~Y.~Yuwen, J.~C.~Wang and S.~J.~Wang,
\emph{Bubble wall velocity from number density current in (non)equilibrium},
arXiv:2409.20045 [hep-ph].

\bibitem{Branchina:2024rva}
C.~Branchina, A.~Conaci, S.~De Curtis, L.~Delle Rose, A.~Guiggiani, A.~Gil Muyor and G.~Panico,
\emph{New calculation of collision integrals for cosmological phase transitions},
EPJ Web Conf. \textbf{314} (2024), 00031, arXiv:2410.00766 [hep-ph].

\bibitem{Ekstedt:2024fyq}
A.~Ekstedt, O.~Gould, J.~Hirvonen, B.~Laurent, L.~Niemi, P.~Schicho and J.~van de Vis,
\emph{How fast does the WallGo? A package for computing wall velocities in first-order phase transitions}, 
JHEP \textbf{04} (2025), 101, arXiv:2411.04970 [hep-ph].

\bibitem{Ai:2024btx}
W.~Y.~Ai, B.~Laurent and J.~van de Vis,
\emph{Bounds on the bubble wall velocity},
JHEP \textbf{02} (2025), 119, arXiv:2411.13641 [hep-ph].


\bibitem{Krajewski:2024xuz}
T.~Krajewski, M.~Lewicki, M.~Vasar, V.~Vaskonen, H.~Veerm\"ae and M.~Zych,
\emph{Thermalization effects on the dynamics of growing vacuum bubbles},
JHEP \textbf{03} (2025), 178, arXiv:2411.15094 [hep-ph].

\bibitem{Krajewski:2024zxg}
T.~Krajewski, M.~Lewicki, I.~Na\l{}\k{e}cz and M.~Zych,
\emph{Steady-state bubbles beyond local thermal equilibrium}, arXiv:2411.16580 [astro-ph.CO].

\bibitem{Dorsch:2024jjl}
G.~C.~Dorsch, T.~Konstandin, E.~Perboni and D.~A.~Pinto,
\emph{Non-singular solutions to the Boltzmann equation with a fluid Ansatz},
JCAP \textbf{04} (2025), 033, arXiv:2412.09266 [hep-ph].

\bibitem{Ramsey-Musolf:2025jyk}
M.~J.~Ramsey-Musolf and J.~Zhu,
\emph{Bubble wall velocity from Kadanoff-Baym equations: fluid dynamics and microscopic interactions}, arXiv:2504.13724 [hep-ph].

\bibitem{Ai:2025bjw}
W.~Y.~Ai, M.~Carosi, B.~Garbrecht, C.~Tamarit and M.~Vanvlasselaer,
\emph{Bubble wall dynamics from nonequilibrium quantum field theory},
arXiv:2504.13725 [hep-ph].

\bibitem{Carena:2025flp}
M.~Carena, A.~Ireland, T.~Ou and I.~R.~Wang,
\emph{The Discriminant Power of Bubble Wall Velocities: Gravitational Waves and Electroweak Baryogenesis}, arXiv:2504.17841 [hep-ph].

\bibitem{Gyulassy:1983rq}
M.~Gyulassy, K.~Kajantie, H.~Kurki-Suonio and L.~D.~McLerran,
\emph{Deflagrations and Detonations as a Mechanism of Hadron Bubble Growth in Supercooled Quark Gluon Plasma},
Nucl. Phys. B \textbf{237} (1984), 477-501

\bibitem{Parwani:1991gq}
R.~R.~Parwani,
\emph{Resummation in a hot scalar field theory},''
Phys. Rev. D \textbf{45} (1992), 4695
[erratum: Phys. Rev. D \textbf{48} (1993), 5965], arXiv:hep-ph/9204216 [hep-ph].

\bibitem{Wainwright:2011kj}
C.~L.~Wainwright,
\emph{CosmoTransitions: Computing Cosmological Phase Transition Temperatures and Bubble Profiles with Multiple Fields},
Comput. Phys. Commun. \textbf{183} (2012), 2006-2013, arXiv:1109.4189 [hep-ph].

\bibitem{Basler:2018cwe}
P.~Basler and M.~M\"uhlleitner,
\emph{BSMPT (Beyond the Standard Model Phase Transitions): A tool for the electroweak phase transition in extended Higgs sectors},
Comput. Phys. Commun. \textbf{237} (2019), 62-85, arXiv:1803.02846 [hep-ph].

\bibitem{Basler:2020nrq}
P.~Basler, M.~M\"uhlleitner and J.~M\"uller,
\emph{BSMPT v2 a tool for the electroweak phase transition and the baryon asymmetry of the universe in extended Higgs Sectors},
Comput. Phys. Commun. \textbf{269} (2021), 108124, arXiv:2007.01725 [hep-ph].

\bibitem{Basler:2024aaf}
P.~Basler, L.~Biermann, M.~M\"uhlleitner, J.~M\"uller, R.~Santos and J.~Viana,
\emph{BSMPT v3 A Tool for Phase Transitions and Primordial Gravitational Waves in Extended Higgs Sectors}, arXiv:2404.19037 [hep-ph].

\bibitem{Hindmarsh:2013xza}
M.~Hindmarsh, S.~J.~Huber, K.~Rummukainen and D.~J.~Weir,
\emph{Gravitational waves from the sound of a first order phase transition},
Phys. Rev. Lett. \textbf{112} (2014), 041301, arXiv:1304.2433 [hep-ph].

\bibitem{Hindmarsh:2015qta}
M.~Hindmarsh, S.~J.~Huber, K.~Rummukainen and D.~J.~Weir,
\emph{Numerical simulations of acoustically generated gravitational waves at a first order phase transition},
Phys. Rev. D \textbf{92} (2015) no.12, 123009, arXiv:1504.03291 [astro-ph.CO].


\bibitem{Caprini:2015zlo}
C.~Caprini, M.~Hindmarsh, S.~Huber, T.~Konstandin, J.~Kozaczuk, G.~Nardini, J.~M.~No, A.~Petiteau, P.~Schwaller and G.~Servant, \textit{et al.}, 
\emph{Science with the space-based interferometer eLISA. II: Gravitational waves from cosmological phase transitions},
JCAP \textbf{04} (2016), 001, arXiv:1512.06239 [astro-ph.CO].


\bibitem{Hindmarsh:2016lnk}
M.~Hindmarsh,
\emph{Sound shell model for acoustic gravitational wave production at a first-order phase transition in the early Universe},
Phys. Rev. Lett. \textbf{120} (2018) no.7, 071301, arXiv:1608.04735 [astro-ph.CO].

\bibitem{Hindmarsh:2017gnf}
M.~Hindmarsh, S.~J.~Huber, K.~Rummukainen and D.~J.~Weir,
\emph{Shape of the acoustic gravitational wave power spectrum from a first order phase transition},
Phys. Rev. D \textbf{96} (2017) no.10, 103520
[erratum: Phys. Rev. D \textbf{101} (2020) no.8, 089902], arXiv:1704.05871 [astro-ph.CO].

\bibitem{Weir:2017wfa}
D.~J.~Weir,
\emph{Gravitational waves from a first order electroweak phase transition: a brief review},
Phil. Trans. Roy. Soc. Lond. A \textbf{376} (2018) no.2114, 20170126
[erratum: Phil. Trans. Roy. Soc. Lond. A \textbf{381} (2023), 20230212], arXiv:1705.01783 [hep-ph].

\bibitem{Ellis:2018mja}
J.~Ellis, M.~Lewicki and J.~M.~No,
\emph{On the Maximal Strength of a First-Order Electroweak Phase Transition and its Gravitational Wave Signal},
JCAP \textbf{04} (2019), 003, arXiv:1809.08242 [hep-ph].

\bibitem{Caprini:2019egz}
C.~Caprini, M.~Chala, G.~C.~Dorsch, M.~Hindmarsh, S.~J.~Huber, T.~Konstandin, J.~Kozaczuk, G.~Nardini, J.~M.~No and K.~Rummukainen, \textit{et al.}, \emph{Detecting gravitational waves from cosmological phase transitions with LISA: an update},
JCAP \textbf{03} (2020), 024, arXiv:1910.13125 [astro-ph.CO].

\bibitem{Ellis:2020awk}
J.~Ellis, M.~Lewicki and J.~M.~No,
\emph{Gravitational waves from first-order cosmological phase transitions: lifetime of the sound wave source},
JCAP \textbf{07} (2020), 050, arXiv:2003.07360 [hep-ph].


\bibitem{Caprini:2024hue}
C.~Caprini \textit{et al.} [LISA Cosmology Working Group],
\emph{Gravitational waves from first-order phase transitions in LISA: reconstruction pipeline and physics interpretation},
JCAP \textbf{10} (2024), 020, arXiv:2403.03723 [astro-ph.CO].

\bibitem{Espinosa:2011eu}
J.~R.~Espinosa, B.~Gripaios, T.~Konstandin and F.~Riva,
\emph{Electroweak Baryogenesis in Non-minimal Composite Higgs Models},
JCAP \textbf{01} (2012), 012, arXiv:1110.2876 [hep-ph].

\bibitem{Kainulainen:2024qpm}
K.~Kainulainen and N.~Venkatesan,
\emph{Systematic moment expansion for electroweak baryogenesis},
JCAP \textbf{08} (2024), 058, arXiv:2407.13639 [hep-ph].


\bibitem{in preparation}
C.~Branchina, A.~Conaci, S.~De Curtis and L.~Delle Rose,
\emph{in preparation}

\end{thebibliography}
\end{document}